\documentclass[12pt]{article}
\usepackage{amsmath,amssymb}
\usepackage{stix}
\usepackage{dsfont}
\usepackage{mathrsfs}
\usepackage{enumerate}
\usepackage{graphicx}
\usepackage{booktabs}
\usepackage{setspace}
\usepackage{geometry}
\usepackage{natbib}
\newcommand{\pr}{\mathrm{P}}
\newcommand{\E}{\mathrm{E}}

\newcommand{\pn}{\mathbb{P}_n}

\newcommand{\bcdot}{\, \boldsymbol{\cdot} \,}
\def\independenT#1#2{\mathrel{\rlap{$#1#2$}\mkern2mu{#1#2}}}
\newcommand\independent{\protect\mathpalette{\protect\independenT}{\perp}}
\DeclareMathOperator{\var}{var}
\DeclareMathOperator{\EIF}{EIF}
\def\T{{ \mathrm{\scriptscriptstyle T} }}
\newtheorem{assumption}{Assumption}
\newtheorem{lemma}{Lemma}
\newtheorem{theorem}{Theorem}
\newtheorem{corollary}{Corollary}
\newtheorem{remark}{Remark}

\usepackage{multirow}

\newcommand{\p}{\mathbb{P}}

\newtheorem{condition}{Condition}   

\title{\bf Estimating Treatment Effects Through Intermediate-Event Pathways with Multi-State Data}
\author{Yuhao Deng$^{*,1}$, Haoyu Wei$^{*,2}$, Donglin Zeng$^3$, Rui Song$^4$, Xiao-Hua Zhou$^5$ \\
{\small 1 Fred Hutchinson Cancer Center 
   2 University of California, San Diego}\\[-4pt]
{\small 3 University of Michigan 
   4 Amazon Inc
   5 Peking University}
}
\date{}

\begin{document}
\maketitle

\begin{abstract}
During clinical trials evaluating a drug's effect on a survival endpoint, intermediate events often occur before the primary event. The treatment can affect the primary event through multiple pathways via intermediate events. With multiple intermediate events, each is a treatment-induced confounder of transitions out of the others, so we cannot identify natural or separable effects. We instead consider stochastic interventions that assign each transition the identifiable observed-history hazard of either arm. Contrasts between the resulting counterfactual cumulative incidences define transition-, event-, and path-targeted effects, identifiable without cross-world or isolation assumptions. Based on the efficient influence functions for the counterfactual cumulative incidences under hypothetical interventions, we construct multiply robust and semiparametrically efficient estimators for these targeted treatment effects. 
Our proposed framework quantifies treatment effects through each transition, on each event, and along each path.
By analyzing data from the LEADER trial, we find that liraglutide significantly reduces the risk of cardiovascular and microvascular events. The reduction in all-cause mortality is primarily mediated by its effects on expanded major adverse cardiovascular events.
\end{abstract}

\noindent%

\section{Introduction} \label{sec:intro}

\begin{figure}[!tbh]
    \centering
    \includegraphics[width=0.8\textwidth]{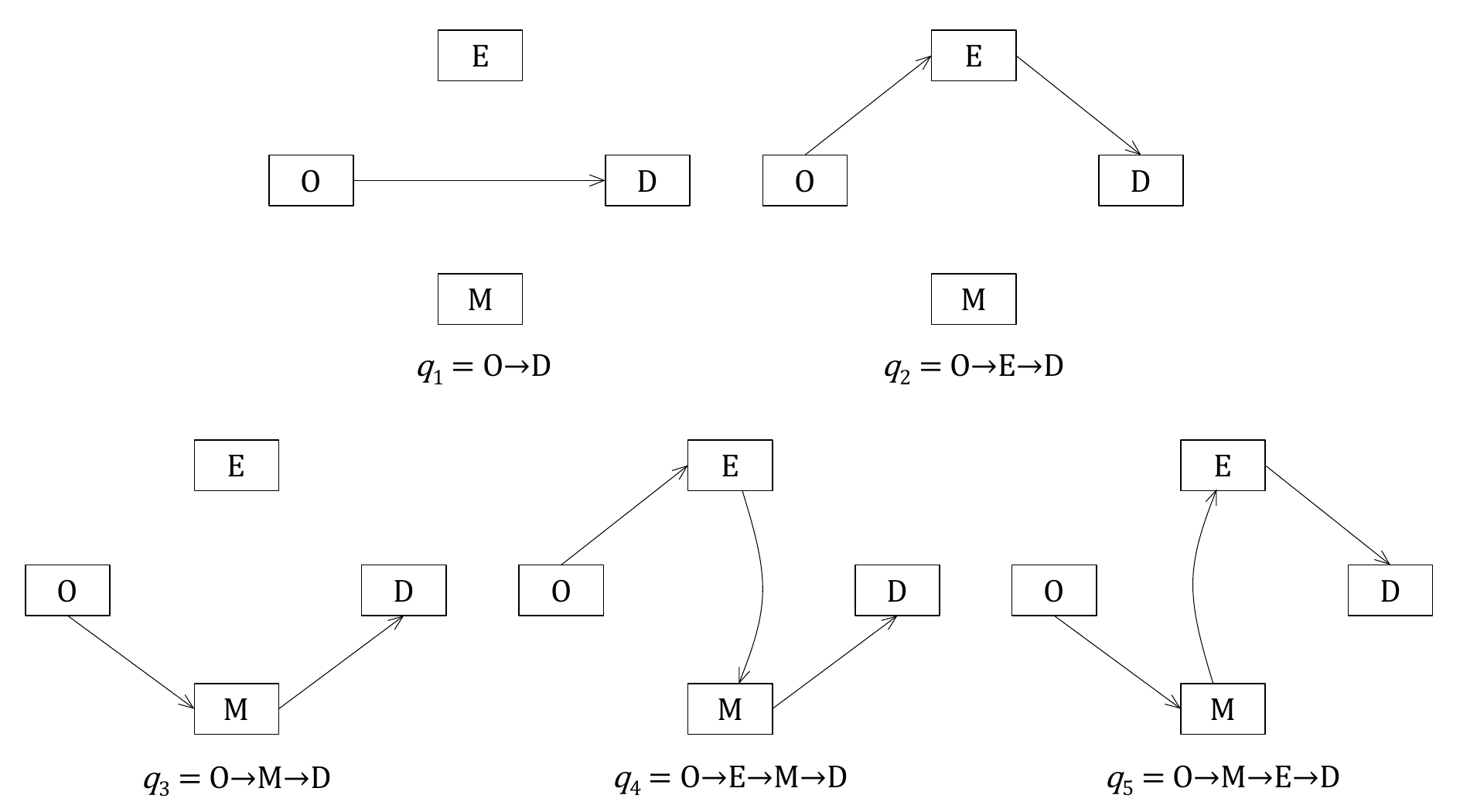}
    \caption{Five potential paths from baseline (O) to death (D) with expanded major adverse cardiovascular events (E) and microvascular events (M) as two intermediate states.}
    \label{fig:model}
\end{figure}

When evaluating the effect of a new drug on a survival endpoint, individuals may experience multiple intermediate events, thereby generating multiple transition pathways from baseline to the primary endpoint. Effects operating through each pathway provide insight into questions such as how the drug works and whether to recommend additional post-treatment interventions. As a motivational example, the LEADER trial randomized 9,340 patients with type 2 diabetes at high cardiovascular risk to liraglutide or placebo and followed them for up to five years to study whether the drug can reduce the incidence of cardiovascular events \citep{marso2013design, marso2016liraglutide}. Single-endpoint analyses found that liraglutide reduces all-cause mortality, the incidence of expanded major adverse cardiovascular events (EMACE), and the incidence of microvascular events (MVE). In this trial, a patient can die after an EMACE, after an MVE, after both in either order, or without any preceding vascular event (Figure \ref{fig:model}). Does the survival benefit arise because liraglutide lowers the hazard of a first cardiovascular event, the hazard of a first microvascular event, the hazard of death once an event has occurred, or the hazard of death without a preceding event? Analyses of single endpoints cannot separate these possibilities, because the occurrence of an event changes who is at risk of the others.

Treating the primary event and the intermediate events as distinct states yields multi-state data \citep{hougaard1999multi, andersen2002multi, putter2007tutorial}. When a single intermediate event may precede a terminal event, the data are semi-competing risks data \citep{kodell1980illness, fine2001semi, xu2010statistical}. Mediation-type frameworks, namely natural effects and separable effects, estimate direct and indirect effects on the terminal event under no-unmeasured-confounding-type assumptions \citep{huang2021causal, weir2022counterfactual, martinussen2023estimation, breum2024estimation}. However, moving from one intermediate event to multiple ones is not a routine extension. In LEADER, an EMACE changes the hazard of a subsequent MVE and vice versa, so each intermediate event is a treatment-induced confounder of the transitions out of the other. Natural effects are then not identifiable because sequential ignorability fails \citep{shpitser2016causal, miles2020semiparametric}. Separable effects posit treatment components that each act on a single event \citep{stensrud2021generalized, stensrud2022separable, robins2022interventionist}; however, identifying such components of liraglutide is clinically impossible, and the estimands cannot address effects through specific transitions.

For each transition between states, we consider a hypothetical intervention that draws the transition from the identifiable observed-history hazard of either the treated or the control arm, leaving all other transitions unaffected. This extends randomized (stochastic) interventional effects, originally proposed for mediation analysis, to multi-state data \citep{vanderweele2017mediation, vansteelandt2017interventional, lin2017interventional, lin2017mediation, diaz2020causal, hejazi2022nonparametric, valeri2023multistate}. The object intervened upon is the instantaneous risk of a transition rather than an event time \citep{lok2015defining}. Identification requires neither cross-world assumptions nor isolated treatment components. The observed-history hazard of an arm averages over the latent post-treatment states of the patients in that arm, so a regime that combines hazards from different arms describes a hypothetical multi-state law rather than an intervention on a biological mechanism; it mimics the generation of events under ``recanting twins'' of unmeasured confounders \citep{vo2026recanting}. 

Interventions on transitions translate the target questions into well-defined contrasts. In LEADER, switching the hazard of a single transition from the placebo to the liraglutide arm with all other hazards at placebo gives a transition-targeted effect; switching all transitions into a state gives an event-targeted effect; switching all transitions along a path gives a path-targeted effect. Estimating these effects reduces to estimating counterfactual cumulative incidences. This paper makes three contributions.
First, we define the hypothetical processes through their transition intensities and show that the counterfactual cumulative incidence of each state is identifiable.
Second, we derive the efficient influence function (EIF) and construct one-step estimators for the counterfactual cumulative incidences and targeted effects. Third, we establish asymptotic properties including multiple robustness and semiparametric efficiency. 
In LEADER, we find that interventions that switch the hazard of a first EMACE reduce the counterfactual incidence of death, whereas switching the death hazards themselves has no detectable effect.

\section{Stochastic Interventional Effects for Multi-State Data} \label{sec:frame}

\subsection{Notations and Estimands}

Suppose there are $K$ states, including an initial state (denoted by O), intermediate states, and terminal states. Transitions are represented by a directed graph $(\mathcal{G}, \mathcal{E})$, where $\mathcal{G}$ is the set of states and $\mathcal{E}$ the set of one-step transitions. Let $\underline{\mathcal{G}}_k$ be the set of states reachable from $k$ in one step, $\mathcal{Q}_{kg}$ the set of paths from $k$ to $g$, and $\mathcal{Q}_g$ the set of paths from O to $g$. A path $q \in \mathcal{Q}_{kg}$ is an ordered sequence of states $q = (q(0), q(1), \ldots, q(l(q)))$ with $q(0) = k$ and $q(l(q)) = g$, where $l(q)$ is the number of transitions along $q$. We assume that each state is visited at most once on any path. Figure \ref{fig:model} shows the structure of our motivating data; here $\mathcal{Q}_{\text{D}} = \{q_1, \ldots, q_5\}$.

Let $W \in \mathbb{R}^p$ be baseline covariates with support $\mathcal{W}$, and let $t^*$ be the end of the study, which by design should not exceed the maximum follow-up time. Let $X^a(t)$ be the potential state at time $t$ if treatment $a \in \{0, 1\}$ is assigned at baseline, $\overline{X}^a(t) = (X^a(s): 0 \leq s \leq t)$ the potential trajectory, and
\begin{equation*} 
    \mathrm{d} \Lambda_{kg}^{a}(t \mid W,\overline{X}^{a}(t)) := \pr \{X^{a}(t + \mathrm{d} t)=g \mid X^a(t)=k, W,\overline{X}^{a}(t)\}
\end{equation*}
the potential transition hazard.

The intervention acts on transitions rather than on event times. Let $\overline{a} = (a_{kg}: (k,g)\in\mathcal{E})$ be a vector of transition-specific levels, where $a_{kg} = 1$ means that the transition from $k$ to $g$ is drawn from the hazard of the treated arm and $a_{kg} = 0$ means that it is drawn from the hazard of the control arm. Given a hypothetical regime $\overline{a}$, let $X^{\overline{a}}(t)$ be the state at time $t$ and $\overline{X}^{\overline{a}}(t) = (X^{\overline{a}}(s): 0 \leq s \leq t)$ the hypothetical trajectory. The hypothetical state process can be generated from transition hazards. Given history $(W,\overline{X}^{\overline{a}}(t)) = (W, \overline{x}(t))$ with $x(t)=k$, we assume the transition to $g$ is drawn from the hazard
\begin{equation} \label{int_haz}
\mathrm{d} \Lambda_{kg}^{\overline{a}}(t \mid W,\overline{x}(t)) := \pr \{X^{\overline{a}}(t + \mathrm{d} t)=g \mid X^{\overline{a}}(t)=k,W,\overline{X}^{\overline{a}}(t) = \overline{x}(t)\} = \mathrm{d} \Lambda_{kg}^{a_{kg}}(t \mid W,\overline{x}(t)),
\end{equation}
the potential hazard under treatment $a_{kg}$ evaluated at the hypothetical history $\overline{x}(t)$.
In words, $\overline{a}$ only decides which arm each transition borrows its hazard from; the hazard is evaluated at the history generated under $\overline{a}$ itself.
Two features of this definition matter. First, because it operates on hazards, the intervention applies only to individuals at risk: a transition out of $k$ is drawn only while the individual is in $k$. Interventions terminate once the individual reaches a terminal state. Second, the hazard on the right-hand side conditions only on the observed history. If there are treatment-induced confounders $\overline{L}^{a}(t)$, possibly unmeasured, it marginalizes over them,
\begin{equation}\label{draw_uc}
\mathrm{d} \Lambda_{kg}^{a}(t \mid W,\overline{x}(t)) = \int_{\overline{l}(t)} \mathrm{d} \Lambda_{kg}^{a}(t \mid W,\overline{l}(t),\overline{x}(t)) \pr \{\overline{L}^{a}(t)=\overline{l}(t) \mid W, \overline{X}^{a}(t) = \overline{x}(t)\} \mathrm{d} \overline{l}(t).
\end{equation}
Equation \eqref{draw_uc} is not an additional assumption but the innovation theorem \citep[Charter II]{andersen2012statistical}: the borrowed hazard averages over the latent post-treatment states of the arm it is borrowed from, so it is a functional of a single arm's law, and no cross-world assumption is needed.

Given $W = w$, the hypothetical trajectory $\overline{X}^{\overline{a}}$ is the marked point process on $\mathcal{G}$ that starts in the initial state and whose compensator for a jump from $k$ to $g$ is
\begin{equation} \label{eq:gen}
\mathds{1}\{X^{\overline{a}}(t-) = k\} \, \mathrm{d}\Lambda_{kg}^{a_{kg}}\{t \mid w, \overline{X}^{\overline{a}}(t-)\}, \quad (k,g) \in \mathcal{E}.
\end{equation}
Equation \eqref{eq:gen} determines the joint law of all transition counting processes $\{N_{kg}^{\overline{a}}: (k,g)\in\mathcal{E}\}$ uniquely. Equivalently, on entry into state $k$ at time $u$ with history $\overline{x}(u)$, one draws independent latent times for every $g \in \underline{\mathcal{G}}_k$ from the hazards $\mathrm{d}\Lambda_{kg}^{a_{kg}}(\cdot \mid w, \overline{x}(u))$ on $(u, t^*]$, moves to the state with the earliest time, and repeats. The construction is therefore a stochastic intervention on observed-history transition intensities. Supplementary Material B develops the methodology of stochastic interventions that condition on observed history and a latent frailty.

The primary object of interest is the time to reach a state $j \in \mathcal{G}$ (typically a terminal event such as death), $T_j^{\overline{a}} = \inf\{t \in [0, t^*]: X^{\overline{a}}(t) = j \}$, with $T_j^{\overline{a}} = \infty$ if $j$ is not visited by $t^*$. The primary estimand is the counterfactual cumulative incidence function (CIF) of state $j$,
\begin{equation}\label{target_par_all}
    F_j^{\overline{a}}(t) := \pr ( T_j^{\overline{a}} \leq t ) = \pr(j \in \overline{X}^{\overline{a}}(t)), \quad t \in [0, t^*],
\end{equation}
the probability that $j$ is reached by time $t$ under the intervention. Since multiple paths may lead to $j$, we let $F_j^{\overline{a}}(t;q)$ denote the sub-incidence of $j$ along the path $q \in \mathcal{Q}_j$, so that
\begin{equation}\label{par_relationship}
    F_j^{\overline{a}}(t) = \sum_{q \in \mathcal{Q}_{j}} F_j^{\overline{a}}(t; q).
\end{equation}
Since each state is visited at most once, the trajectory makes at most $K-1$ jumps and its sequence of states before reaching $j$ is a unique element of $\mathcal{Q}_j$; these events are disjoint, and their union is $\{T_j^{\overline{a}}\le t\}$.

\begin{figure}[!tb]
\centering
\includegraphics[width=0.5\textwidth]{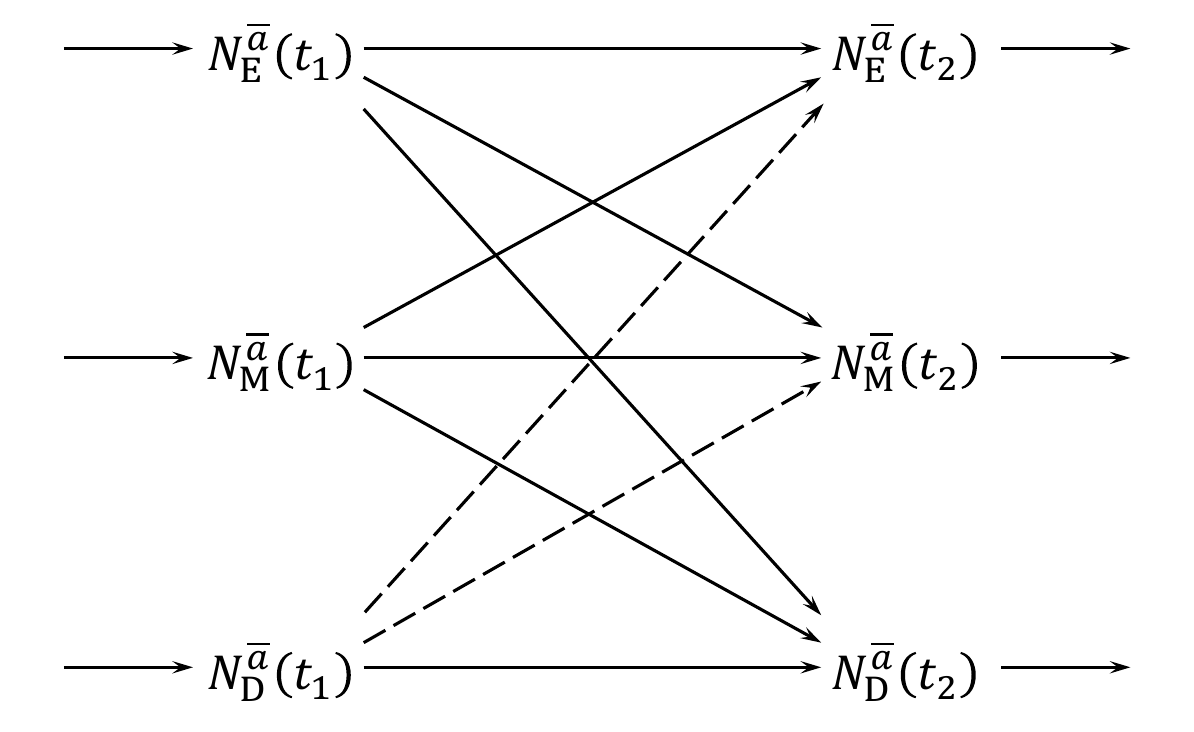}
\caption{Causal graphs of counterfactual counting processes. Solid arrows represent stochastic effects on the counting processes of E, M, and D. Dashed arrows represent deterministic effects, enforcing that $N_{\text{E}}^{\overline{a}}(t+s) = N_{\text{E}}^{\overline{a}}(t)$ and $N_{\text{M}}^{\overline{a}}(t+s) = N_{\text{M}}^{\overline{a}}(t)$ for all $s>0$ if $N_{\text{D}}^{\overline{a}}(t) = 1$ at some $t>0$.} \label{fig:dag_int}
\end{figure}

For the structure in Figure \ref{fig:model}, the intervention vector has seven components, $$\overline{a} = (a_{\text{OE}}, a_{\text{OM}}, a_{\text{OD}}, a_{\text{EM}}, a_{\text{ME}}, a_{\text{ED}}, a_{\text{MD}}).$$ For example, $\overline{a} = (1,0,0,0,0,1,0)$ draws the transitions along $\text{O}\to\text{E}\to\text{D}$ from the treated-arm hazards and all other transitions from the control-arm hazards. Figure \ref{fig:dag_int} shows the causal graph of the counterfactual event-counting processes $N_{\text{E}}^{\overline{a}}(t)$, $N_{\text{M}}^{\overline{a}}(t)$, and $N_{\text{D}}^{\overline{a}}(t)$, each the maximum of the counting processes of the transitions into that event, so that $F_{\text{D}}^{\overline{a}}(t) = \pr\{N_{\text{D}}^{\overline{a}}(t)=1\}$.

\subsection{Identifiability}

By \eqref{int_haz}, the hazards of the hypothetical process are potential hazards, and the counterfactual CIF is a functional of them: the compensator \eqref{eq:gen} determines the law of the process \citep{jacod1975multivariate}, and Theorem \ref{thm1} gives the functional explicitly.
It therefore suffices to identify the potential hazards. Let $T_{\text{c}}^a$ be the potential censoring time under treatment $a$, $N^a_{\text{c}}(t) = \mathds{1}(T_{\text{c}}^a \leq t)$, $\mathcal{F}_t^{a} = \sigma\{W, \overline{X}^{a}(t), N_{\text{c}}^{a}(s): s \leq t\}$, and $\mathcal{H}_t^{a} = \sigma\{W, \overline{X}^{a}(t)\}$. 
We make the following standard assumptions in causal inference for survival data.

\begin{assumption}[Ignorability] \label{ass:ign}
$A \independent \{(\overline{X}^{a}(t^*), T_{\text{c}}^{a}): a = 0, 1\} \mid W$.
\end{assumption}

\begin{assumption}[Random censoring]\label{ass:ran}
For $a \in \{0,1\}$ and every $(k,g) \in \mathcal{E}$, on the event $\{T_{\text{c}}^a \geq t\}$ the $\mathcal{F}^{a}$-intensity of the transition counting process $N_{kg}^{a}$ given $A = a$ equals its $\mathcal{H}^{a}$-intensity $\mathds{1}\{X^{a}(t-) = k\}\, \mathrm{d}\Lambda_{kg}^{a}\{t \mid W, \overline{X}^{a}(t-)\}$.
\end{assumption}

\begin{assumption}[Positivity] \label{ass:pos}
(i) There exists a constant $\epsilon > 0$ such that $\epsilon \leq \pr(A=1 \mid W) \leq 1 - \epsilon$. (ii) For $a=0,1$ and $t \leq t^*$, $\pr\{T_{\text{c}} \geq t \mid A=a, W, \overline{X}(t-)\} \geq \epsilon$ on the support of $(W, \overline{X}(t-))$ given $A = a$. (iii) Given $W$, the laws of the potential trajectories $\overline{X}^{1}(t^*)$ and $\overline{X}^{0}(t^*)$ are mutually absolutely continuous.
\end{assumption}

Ignorability excludes unmeasured confounding between treatment assignment and the joint collection of potential trajectories and censoring times. It holds by design in a randomized trial. Assumption \ref{ass:ran} is the usual independent-censoring condition for event-history data: being uncensored at $t$ carries no information about the next transition, while censoring itself may depend on the history. 
The first two parts of Assumption \ref{ass:pos} ensure that every observed-history hazard in arm $a$ is estimable; the last part ensures that the law of the hypothetical trajectory under any $\overline{a}$ is absolutely continuous with respect to the law of each potential trajectory, so every history generated by $\overline{a}$ lies in the support of the arm from which each hazard is borrowed.

Let $\overline{X}(t) = ( X(s): 0 \leq s \leq t )$ be the observed trajectory, with $X(t) = \text{c}$ for $t \geq T_{\text{c}}$ if the individual is censored at $T_{\text{c}}$. 
Let $\Delta_k = 1$ if state $k$ is visited before the end of the study, and let $T_k$ be the time of the first visit to $k$ if $\Delta_k = 1$ and $T_k = T_{\text{c}}\wedge t^*$ otherwise.

\begin{assumption}[Consistency] \label{ass:con}
$T_{\text{c}}^A = T_{\text{c}}$ and $\overline{X}^{A}(T_{\text{c}}^A\wedge t^*) = \overline{X}(T_{\text{c}}\wedge t^*)$.
\end{assumption}

Under Assumptions \ref{ass:ign}--\ref{ass:con}, the potential transition hazards are identified as
\[
\mathrm{d} \Lambda_{kg}^a (t \mid w, \overline{x}(t)) = \pr \{t < T_g < t+\mathrm{d} t, \Delta_g = 1 \mid A=a, W=w, \overline{X}(t) = \overline{x}(t), X(t) = k\},
\]
and the counterfactual CIF follows.

\begin{theorem} \label{thm1}
    Under Assumptions \ref{ass:ign}--\ref{ass:con}, the counterfactual cumulative incidence $F_j^{\overline{a}}(t)$ of any state $j \in \mathcal{G}$ is identified as
    \begin{align}
        F_j^{\overline{a}}(t) =  \sum_{q\in\mathcal{Q}_{j}}\int_{\mathcal{W}}\int_0^t\int_0^{t_{l(q)}}\cdots\int_0^{t_2} \prod_{r=1}^{l(q)} \mathrm{d} \pr_r(t_r \mid a_{q(r-1), \bcdot},w,q[\overline{t}_r]) \mathrm{d} \pr(w), \label{eq:id}
    \end{align}
where the iterated integrals run over the strictly ordered transition times $0 < t_1 < t_2 < \cdots < t_{l(q)} \leq t$, $q[\overline{t}_r]$ is the history of the path $q = (q(0), q(1), \ldots, q(r-1))$ before time $t_r$ with transition times $\overline{t}_r = \{t_s: s=1,\ldots,r-1\}$ and $x(t_r) = q(r-1)$, and $\pr(w)$ is the distribution function of covariates. Here, $\mathrm{d} \pr_r(t_r \mid a_{q(r-1), \bcdot},w,q[\overline{t}_r])$ is the probability that the next transition is to $q(r)$ and occurs in $[t_r, t_r + \mathrm{d}t_r)$ given the history $q[\overline{t}_r]$,
    \begin{equation} \label{eq:dens}
    \begin{aligned}
    \mathrm{d} \pr_r(t_r \mid a_{q(r-1), \bcdot},w,q[\overline{t}_r]) &= \exp\left\{-\sum_{ k \in \underline{\mathcal{G}}_{q(r-1)}} \Lambda_{q(r-1),k}^{a_{q(r-1),k}}(t_r \mid w,q[\overline{t}_r])\right\} \mathrm{d} \Lambda_{q(r-1),q(r)}^{a_{q(r-1),q(r)}}(t_r \mid w,q[\overline{t}_r]).
    \end{aligned}
    \end{equation}
\end{theorem}

The proof is in Supplementary Material E.1. In \eqref{eq:dens}, $\Lambda_{q(r-1),k}^{a}(t_r \mid w, q[\overline{t}_r])$ is the hazard of the transition $(q(r-1),k)$ accumulated during the sojourn in $q(r-1)$, that is, over $(t_{r-1}, t_r]$; equivalently, $\mathrm{d}\Lambda_{kg}^a(t \mid w, \overline{x}(t)) = 0$ whenever $x(t) \neq k$. If all components of $\overline{a}$ equal $a\in\{0,1\}$, the counterfactual CIF coincides with the potential CIF $F_j^a(t) = \pr(j \in \overline{X}^a(t))$ under treatment $a$, so $F_j^{\overline{1}}(t) = F_j^1(t)$ and $F_j^{\overline{0}}(t) = F_j^0(t)$ with $\overline{1} = (1, \ldots, 1)$ and $\overline{0} = (0, \ldots, 0)$, and the total effect $\tau_j(t) = F_j^1(t) - F_j^0(t)$ is a contrast between two such regimes.

\subsection{Transition-, Event-, and Path-Targeted Effects} \label{sec:targeted}

\begin{figure}[!tb]
\centering
\includegraphics[width=0.85\textwidth]{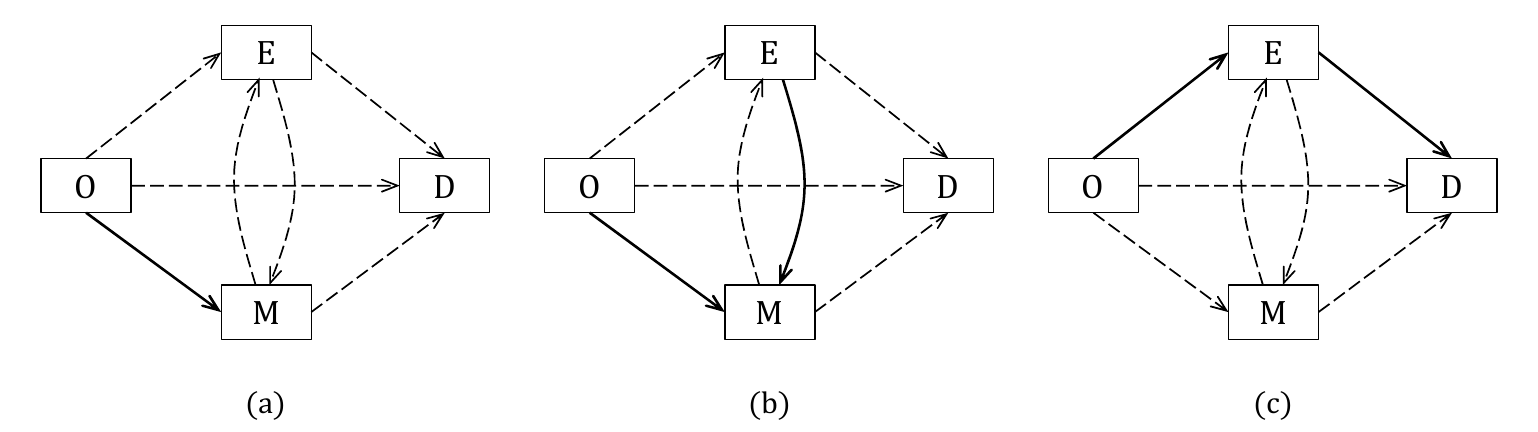}
\caption{Three targeted interventions with the all-control reference. Solid arrows represent transitions assigned the treated-arm hazard, and dashed arrows represent transitions assigned the control-arm hazard. (a) Transition-targeted intervention on O$\to$E. (b) Event-targeted intervention on M. (c) Path-targeted intervention on O$\to$E$\to$D.} \label{fig:tes}
\end{figure}

We define treatment effects as contrasts between two intervention vectors. For a set $\mathcal{S} \subseteq \mathcal{E}$ of transitions and a reference level $b \in \{0, 1\}$, let $\overline{a}(\mathcal{S}; b)$ denote the vector with $a_{kg} = 1 - b$ for $(k,g) \in \mathcal{S}$ and $a_{kg} = b$ otherwise, and define
\begin{equation} \label{eq:switch}
\tau_j^{\mathcal{S}}(t; b) = (1 - 2b) \left\{ F_j^{\overline{a}(\mathcal{S}; b)}(t) - F_j^{\overline{b}}(t) \right\},
\end{equation}
the effect on the cumulative incidence of $j$ of switching the transitions in $\mathcal{S}$ from the control-arm to the treated-arm hazards, with all other transitions held at the reference level $b$. With $b = 0$, the reference regime is all-control; with $b = 1$, it is all-treated. We call $\tau_j^{\mathcal{S}}$ a \textit{transition-targeted effect} when $\mathcal{S} = \{(k,g)\}$, an \textit{event-targeted effect} for event $g$ when $\mathcal{S}$ contains all transitions into $g$, and a \textit{path-targeted effect} for path $q$ when $\mathcal{S} = \{(q(r-1), q(r)): r = 1, \ldots, l(q)\}$. Figure \ref{fig:tes} illustrates the three types with the all-control reference. In general, $\tau_j^{\mathcal{S}}(t; 0) \neq \tau_j^{\mathcal{S}}(t; 1)$ because the effect of switching the transitions in $\mathcal{S}$ depends on the hazards of the transitions outside $\mathcal{S}$. An order-averaged (Shapley) version, defined in Supplementary Material A.3, decomposes the total effect into a sum of transition-targeted or of path-targeted effects \citep{shapley1953value}.

\begin{remark}[Illness--death model] \label{rem:id}
Consider an illness--death model with $\overline{a} = (a_{\text{OI}},a_{\text{OD}},a_{\text{ID}})$ for the transitions from the initial state to illness, from the initial state to death, and from illness to death. The total effect on death is decomposed as
\[
F_{\text{D}}^1(t) - F_{\text{D}}^0(t) = \{F_{\text{D}}^{(0,1,1)}(t) - F_{\text{D}}^{\overline{0}}(t)\} + \{F_{\text{D}}^{\overline{1}}(t) - F_{\text{D}}^{(0,1,1)}(t)\},
\]
where the first term is an event-targeted effect on death (an interventional direct effect). This decomposition coincides with the natural effects under sequential ignorability and with the separable effects under the dismissible components condition \citep{deng2024direct, breum2024estimation}. Alternatively,  
\[
F_{\text{D}}^1(t) - F_{\text{D}}^0(t) = \{F_{\text{D}}^{(1,0,0)}(t) - F_{\text{D}}^{\overline{0}}(t)\} + \{F_{\text{D}}^{\overline{1}}(t) - F_{\text{D}}^{(1,0,0)}(t)\},
\]
where the first term is a transition-targeted effect from the initial state to illness. There are two path-targeted effects on death with the all-control reference: $F_{\text{D}}^{(1,0,1)}(t) - F_{\text{D}}^{\overline{0}}(t)$ for $\text{O}\to\text{I}\to\text{D}$ and $F_{\text{D}}^{(0,1,0)}(t) - F_{\text{D}}^{\overline{0}}(t)$ for $\text{O}\to\text{D}$; they do not add up to the total effect. 
\end{remark}

\section{Estimation and Inference} \label{sec:estim}

\subsection{Efficient Influence Function}

We observe $n$ independent and identically distributed copies of $O = (A, W, \overline{X}(t^*))$, where the trajectory $\overline{X}(t^*)$ can equivalently be written as $(T_k,\Delta_k: k\in\mathcal{G})$. We write $\overline{X}(s-)$ for the observed history on $[0,s)$ and $\overline{x}(s-)$ for a generic such history, with $x(s-)$ its current state. For $(k,g)\in\mathcal{E}$, let $N_{kg}(s)$ count the observed transitions from $k$ to $g$ by time $s$, let $Y_k(s) = \mathds{1}\{X(s-) = k\}$ indicate being observed, uncensored, in state $k$ just before $s$, and let
\[
\mathrm{d} M_{kg}(s; a, w, \overline{x}(s-)) = \mathrm{d} N_{kg}(s) - Y_k(s)\, \mathrm{d}\Lambda_{kg}^{a}(s \mid w, \overline{x}(s-))
\]
be the transition-specific martingale increment. We derive the efficient influence function (EIF) of $F_j^{\overline{a}}(t)$ in the nonparametric model, in which the transition hazards may depend on the entire history arbitrarily.

Let $F_j^{\overline{a}}(t \mid w; \overline{x}(s-)) = \pr\{T_j^{\overline{a}} \leq t \mid W = w, \overline{X}^{\overline{a}}(s-) = \overline{x}(s-)\}$ be the counterfactual CIF of $j$ given that the hypothetical trajectory has followed $\overline{x}(s-)$ up to time $s$; it is 1 if $j$ is in $\overline{x}(s-)$ and is otherwise calculated from Equation \eqref{eq:id} starting from $\overline{x}(s-)$. Writing $\overline{x}(s-) \oplus (g,s)$ for the history extended by a transition to $g$ at time $s$,
we define
\begin{equation}\label{eq:delta}
\Delta_{kg}^{\overline{a}}(t \mid w; \overline{x}(s-)) = F_j^{\overline{a}}(t \mid w; \overline{x}(s-) \oplus (g,s)) - F_j^{\overline{a}}(t \mid w; \overline{x}(s-)), \quad x(s-) = k,
\end{equation}
the change in the counterfactual CIF of $j$ caused by moving a unit of hazard mass from $k$ to $g$ at time $s$. We assume that the cumulative hazards are absolutely continuous in $t$ and write $\lambda_{kg}^{a}(t \mid w, \overline{x}(t)) = \frac{\mathrm{d}}{\mathrm{d}t}\Lambda_{kg}^{a}(t \mid w, \overline{x}(t))$ for the rate, so that the transition probability $\mathrm{d}\pr_r$ in \eqref{eq:dens} has the density $$p_r(t_r \mid a_{q(r-1), \bcdot}, w, q[\overline{t}_r]) = \exp\left\{-\sum_{k \in \underline{\mathcal{G}}_{q(r-1)}} \Lambda_{q(r-1),k}^{a_{q(r-1),k}}(t_r \mid w, q[\overline{t}_r])\right\} \lambda_{q(r-1),q(r)}^{a_{q(r-1),q(r)}}(t_r \mid w, q[\overline{t}_r])$$ with respect to Lebesgue measure on $t_r$. For a history $\overline{x}(s-) = q[\overline{t}_r]$ consisting of $r-1$ transitions along a path $q$ at times $t_1 < \cdots < t_{r-1} < s$, let
\begin{equation}\label{eq:hist_dens}
p^{\overline{a}}(s, \overline{x}(s-) \mid w) = \prod_{\ell=1}^{r-1} p_\ell(t_\ell \mid a_{q(\ell-1),\bcdot}, w, q[\overline{t}_\ell]) \exp\left\{-\sum_{g\in\underline{\mathcal{G}}_{q(r-1)}} \Lambda_{q(r-1),g}^{a_{q(r-1),g}}(s \mid w, q[\overline{t}_r])\right\}
\end{equation}
be the density of the hypothetical history $\overline{X}^{\overline{a}}(s-)$ given $W = w$, with respect to the Lebesgue measure on the transition times $(t_1, \ldots, t_{r-1})$ and the counting measure on the sequence of visited states; the last factor is the probability of no transition out of $q(r-1)$ during $(t_{r-1}, s)$. Let $p^{a}(s, \overline{x}(s-) \mid w)$ be the same quantity with every component of $\overline{a}$ set to $a$, which is the density of the potential trajectory $\overline{X}^{a}(s-)$, and let $G^{a}(s \mid w, \overline{x}(s-)) = \pr\{T_{\text{c}} \geq s \mid A = a, W = w, \overline{X}^{a}(s-) = \overline{x}(s-)\}$ be the probability of remaining uncensored. Given $x(s-)=k$,
\begin{equation}\label{eq:omega}
\omega_k^{\overline{a}}(s; a, w, \overline{x}(s-)) = \frac{p^{\overline{a}}(s, \overline{x}(s-) \mid w)}{p^{a}(s, \overline{x}(s-) \mid w)\, G^{a}(s \mid w, \overline{x}(s-))}
\end{equation}
is the density of the hypothetical history relative to the observed, uncensored history in arm $a$.

\begin{assumption}[Overlap of hypothetical and observed histories] \label{ass:overlap}
There is a constant $C < \infty$ such that $\omega_k^{\overline{a}}(s; a_{kg}, w, \overline{x}(s-)) \leq C$ for every $(k,g) \in \mathcal{E}$, $s \in [0, t^*]$, and every history $\overline{x}(s-)$ with $x(s-) = k$ in the support of $\overline{X}^{\overline{a}}(s-)$ given $W = w$.
\end{assumption}

Assumption \ref{ass:overlap} reduces to the censoring part of Assumption \ref{ass:pos} when all components of $\overline{a}$ coincide. It holds under Assumption \ref{ass:pos} together with (i) bounded hazard ratios between arms, $\lambda_{kg}^{1}(s \mid w, \overline{x}(s-)) / \lambda_{kg}^{0}(s \mid w, \overline{x}(s-)) \in [1/C', C']$ for every transition and history, and (ii) cumulative hazards $\Lambda_{kg}^{a}(t^* \mid w, \overline{x})$ bounded uniformly in $w$ and the history. In words, the treatment may shift the risk of a transition but cannot make a possible transition impossible.

\begin{lemma}[EIF in the nonparametric model] \label{thm_eif_np}
Under Assumptions \ref{ass:ign}--\ref{ass:overlap}, the efficient influence function of $F_j^{\overline{a}}(t)$ in the nonparametric model is
\begin{equation}\label{eif_np}
\begin{aligned}
\EIF\{F_j^{\overline{a}}(t)\} &= F_j^{\overline{a}}(t \mid W) - F_j^{\overline{a}}(t) + \sum_{(k,g)\in\mathcal{E}} \frac{\mathds{1}(A = a_{kg})}{\pr(A = a_{kg} \mid W)} \\
&\quad \times \int_0^t \omega_k^{\overline{a}}(s; a_{kg}, W, \overline{X}(s-))\, \Delta_{kg}^{\overline{a}}(t \mid W; \overline{X}(s-))\, \mathrm{d}M_{kg}(s; a_{kg}, W, \overline{X}(s-)).
\end{aligned}
\end{equation}
\end{lemma}

In particular, for a history $q[\overline{t}_r]$ in which the unit entered $k = q(r-1)$ after transitions at times $t_1 < \cdots < t_{r-1} < s$,
\begin{equation}\label{eq:omega_prod}
\begin{aligned}
\omega_k^{\overline{a}}(s; a, w, q[\overline{t}_r]) &= \frac{1}{G^{a}(s \mid w, q[\overline{t}_r])} \prod_{\ell=1}^{r-1} \frac{\lambda_{q(\ell-1),q(\ell)}^{a_{q(\ell-1),q(\ell)}}(t_\ell \mid w, q[\overline{t}_\ell])}{\lambda_{q(\ell-1),q(\ell)}^{a}(t_\ell \mid w, q[\overline{t}_\ell])} \\
&\quad \times \prod_{\ell=1}^{r} \exp\Bigg[-\sum_{g' \in \underline{\mathcal{G}}_{q(\ell-1)}} \left\{\Lambda_{q(\ell-1),g'}^{a_{q(\ell-1),g'}} - \Lambda_{q(\ell-1),g'}^{a}\right\}(t_\ell \mid w, q[\overline{t}_\ell])\Bigg],
\end{aligned}
\end{equation}
where the cumulative hazards are accumulated over the sojourns. The weight therefore involves only the transitions upstream of and out of $k$ that $\overline{a}$ borrows from the other arm. The exponential factors are survival ratios and require only cumulative hazards. The proof of Lemma \ref{thm_eif_np} is in Supplementary Material E.2. Each transition contributes one weighted martingale residual, computed from the units in the arm $a_{kg}$ that the intervention assigns to this transition while they are observed in state $k$. The weight $\omega_k^{\overline{a}}$ reweights these units from the observed law of histories in arm $a_{kg}$ to the hypothetical law under $\overline{a}$, and $\Delta_{kg}^{\overline{a}}$ is the sensitivity of the g-formula \eqref{eq:id} to the hazard of $(k,g)$ at time $s$.

The Aalen--Johansen estimator and the clock-forward Cox models that are standard in multi-state analysis impose the Markov property, under which the hazard depends on the past only through the current state \citep{putter2007tutorial, andersen2012statistical}. More generally, one may assume that the hazards depend on the history only through a finite-valued summary, such as the set of visited states or their order. In such a submodel, the EIF is the projection of \eqref{eif_np} onto the smaller tangent space, and the efficiency bound can be smaller than in the nonparametric model. Supplementary Material A.2 derives the EIF under the extended Markov property.

\subsection{Estimation} \label{sec:est_eif}

The EIF involves the transition hazards, censoring hazards, propensity score, and transition hazard ratios between arms.
For each transition $(k,g) \in \mathcal{E}$, we fit a Cox model with treatment as a covariate,
\begin{equation}\label{eq:cox}
\mathrm{d} \Lambda_{kg}^a (t \mid w, \overline{x}(t-)) = \mathrm{d} \Lambda_{0,kg}(t) \exp\left\{\alpha_{kg} a + \beta_{kg}^{\T} w + \gamma_{kg}^{\T} z_{kg}(t) + a \left(\delta_{kg}^{\T} w + \zeta_{kg}^{\T} z_{kg}(t)\right)\right\},
\end{equation}
subject to $x(t-) = k$.
Here $\Lambda_{0,kg}$ is an unspecified baseline cumulative hazard on the scale of time since randomization, namely that of a unit in state $k$ with $a=0$, $w=0$ and $z_{kg}(t)=0$, shared by both arms and all histories ending in $k$ (transitions into the same state may also share it, with the source state entering through $z_{kg}$).
Besides,
$z_{kg}(t)$ is a bounded vector of summaries of the history before $t$, such as indicators of the states visited, their order, or functions of the time since entry into $k$, and $\theta_{kg} = (\alpha_{kg}, \beta_{kg}, \gamma_{kg}, \delta_{kg}, \zeta_{kg})$ collects the regression coefficients. Write $R_{kg}(t; a, w, \overline{x}(t-); \theta_{kg})$ for the exponential factor in \eqref{eq:cox}, the relative risk. The ratio of rates between arms in \eqref{eq:omega_prod} is
\[
\frac{\lambda_{kg}^{1}(t \mid w, \overline{x}(t-))}{\lambda_{kg}^{0}(t \mid w, \overline{x}(t-))} = \frac{R_{kg}(t; 1, w, \overline{x}(t-); \theta_{kg})}{R_{kg}(t; 0, w, \overline{x}(t-); \theta_{kg})} = \exp\left\{\alpha_{kg} + \delta_{kg}^{\T} w + \zeta_{kg}^{\T} z_{kg}(t)\right\},
\]
a function of the coefficients only. Therefore, the weight requires only regression coefficients, cumulative hazards, and the censoring model. For each transition, $\theta_{kg}$ is estimated by maximizing the partial likelihood over the units of both arms at risk in $k$, and the baseline cumulative hazard is estimated by the Breslow estimator \citep{breslow1972disussion, breslow1974covariance}
\[
\widehat\Lambda_{0,kg}(t) = \sum_{i=1}^{n} \int_0^t \frac{\mathrm{d}N_{kg,i}(s)}{\sum_{i'=1}^{n} Y_{k,i'}(s)\, R_{kg}(s; A_{i'}, W_{i'}, \overline{X}_{i'}(s-); \widehat\theta_{kg})},
\]
where the subscript $i$ indexes units, so that $\widehat\Lambda_{kg}^{a}(t \mid w, \overline{x}(t-)) = \int_0^t R_{kg}(s; a, w, \overline{x}(s-); \widehat\theta_{kg})\, \mathrm{d}\widehat\Lambda_{0,kg}(s)$. The fitted ratio of rates between arms is $\exp\{\widehat\alpha_{kg} + \widehat\delta_{kg}^{\T} w + \widehat\zeta_{kg}^{\T} z_{kg}(t)\}$. The censoring hazards are fitted with arm-specific Cox models, and the propensity score is fitted by logistic regression.

Plugging the fitted cumulative hazards into  \eqref{eq:id} and \eqref{eq:dens} gives the fitted transition probabilities $\mathrm{d}\widehat{\pr}_r$ along each path, the sub-incidences $\widehat{F}_j^{\overline{a}}(t \mid W; q)$, the conditional CIF $\widehat{F}_j^{\overline{a}}(t \mid W) = \sum_{q\in\mathcal{Q}_j}\widehat{F}_j^{\overline{a}}(t \mid W;q)$, and the \textit{plug-in} estimator $\widehat{F}_j^{\overline{a}}(t) = \pn\{\widehat{F}_j^{\overline{a}}(t \mid W)\}$, where $\pn$ denotes the empirical average. The same formula starting from a fixed history $\overline{x}(s-)$ gives $\widehat{F}_j^{\overline{a}}(t \mid W; \overline{x}(s-))$ and hence the fitted increments $\widehat\Delta_{kg}^{\overline{a}}$, which we compute by the backward recursion of Supplementary Material E.2. The plug-in estimator is biased if the working models are misspecified. We therefore update it with the fitted debiasing term $\widehat\varphi_j^{\overline{a}}(t;O)$, obtained by substituting into the martingale terms of \eqref{eif_np} the fitted propensity score, the fitted cumulative hazards in the martingale increments and in $\widehat\Delta_{kg}^{\overline{a}}$, and the fitted weight $\widehat\omega_k^{\overline{a}}$, built from the fitted cumulative hazards, the fitted ratio of rates between arms, and the fitted censoring survival. The \textit{one-step} estimator is
\begin{equation}\label{cond_2_est_eq}
    \widetilde{F}_j^{\overline{a}}(t) = \pn \left[ \widehat{F}_j^{\overline{a}}(t \mid W) + \widehat\varphi_j^{\overline{a}}(t; O) \right].
\end{equation}
It corrects the first-order bias of the plug-in estimator and solves the estimating equation $\pn [\widehat{\EIF}\{F_j^{\overline{a}}(t)\}] := \pn [ \widehat{F}_j^{\overline{a}}(t \mid W) + \widehat\varphi_j^{\overline{a}}(t; O) - \widetilde{F}_j^{\overline{a}}(t)] = 0$. We calculate the standard error as $(n^{-1} \pn [\widehat{\EIF}\{F_j^{\overline{a}}(t)\}^2])^{1/2}$.

\subsection{Asymptotic Properties} \label{sec:asy_pro}

For a state $k \in \mathcal{G}$, let $\mathcal{U}_k = \{(k', g') \in \mathcal{E}: k' = k \text{ or } \mathcal{Q}_{k'k} \neq \emptyset\}$ be the set of transitions out of $k$ and out of the states from which $k$ can be reached. By \eqref{eq:omega_prod}, the weight of a transition out of $k$ involves the hazards of the transitions in $\mathcal{U}_k$ that $\overline{a}$ borrows from an arm other than $a_{kg}$.

\begin{theorem}[Multiple robustness] \label{thm_multi_robust}
Suppose that the assumptions in Lemma \ref{thm_eif_np} hold and that the fitted propensity score, censoring hazards, and transition hazards converge to limits, which may be misspecified. Then $\widetilde{F}_j^{\overline{a}}(t)$ is consistent if either
\begin{enumerate}[(1)]
    \item the limit of $\widehat\Lambda_{kg}^{a_{kg}}$ is correct for every $(k,g) \in \mathcal{E}$; or
    \item the propensity score and the censoring hazards are correctly specified and, for every transition $(k,g)$ whose fitted hazard $\widehat\Lambda_{kg}^{a_{kg}}$ is misspecified, the hazards of every transition $(k', g') \in \mathcal{U}_k$ with $a_{k'g'} \neq a_{kg}$ are correctly specified in both arms.
\end{enumerate}
Condition (2) holds whenever a single transition hazard is misspecified under any $\overline{a}$, and whenever all components of $\overline{a}$ are equal.
\end{theorem}

\begin{theorem}[Semiparametric efficiency] \label{thm_asy_normal}
Suppose that the assumptions in Lemma \ref{thm_eif_np} and the regularity conditions in Supplementary Material A.1 hold. Then $\widetilde{F}_j^{\overline{a}}(\cdot)$ weakly converges to a Gaussian process, with
\begin{align*}
\sqrt{n} \{\widetilde{F}_j^{\overline{a}}(t) - F_j^{\overline{a}}(t)\} \rightsquigarrow \mathcal{N}\left(0, \ \E[\EIF\{F_j^{\overline{a}}(t)\}]^2\right)
\end{align*}
for $t \in [0,t^*]$. The asymptotic variance attains the semiparametric efficiency bound for $F_j^{\overline{a}}(t)$ in the nonparametric model.
\end{theorem}

The proofs of Theorems \ref{thm_multi_robust} and \ref{thm_asy_normal} are in Supplementary Material E.5 and E.6. The regularity conditions require the fitted nuisance functions to be correctly specified, to converge at rates such that the product of the error of the weight and the error of the cumulative hazards is $o_p(n^{-1/2})$, and to lie in a Donsker class. The Cox model with a bounded baseline hazard and time-varying covariates of bounded variation is Donsker \citep{zeng2007maximum}, and it converges at the $n^{-1/2}$ rate when correctly specified, so the regularity conditions hold for the working models. Pointwise confidence intervals follow from the normal approximation. Flexible machine-learning estimators of the nuisance functions generally violate the Donsker condition; in that case, the one-step estimator should be combined with cross-fitting \citep{chernozhukov2018double}, under which Condition S1(3) can be dropped and Condition S1(2) applies to the cross-fitted estimators.

\subsection{Inference for Targeted Effects} \label{sec:inference}

For a target state $j \in \mathcal{G}$, a general contrast between two intervention vectors is $\tau_j(t;\overline{a},\overline{a}^*) := F_j^{\overline{a}}(t) - F_j^{\overline{a}^*}(t)$.
Its one-step estimator is $\widetilde\tau_j(t;\overline{a},\overline{a}^*) = \widetilde{F}_j^{\overline{a}}(t) - \widetilde{F}_j^{\overline{a}^*}(t)$. The Shapley order-averaged effects (Supplementary Material A.3) are linear combinations of such contrasts, and their estimators and influence functions are the same linear combinations of the one-step estimators and the EIFs. Inference follows from Theorem \ref{thm_asy_normal} and the linearity of influence functions.

\begin{corollary}\label{cor_inference_path}
    Suppose that the conditions in Theorem \ref{thm_asy_normal} hold. Then for $t \in [0,t^*]$,
    \[
        \sqrt{n} \{\widetilde{\tau}_j(t;\overline{a},\overline{a}^*) - \tau_j(t;\overline{a},\overline{a}^*)\} \rightsquigarrow \mathcal{N}\left(0, \ \E[\EIF\{F_j^{\overline{a}}(t)\}-\EIF\{F_j^{\overline{a}^*}(t)\}]^2\right).
    \]
\end{corollary}

As a summary for follow-up, we use the difference in the event-specific restricted mean time lost up to a horizon $t^*$,
\begin{equation} \label{teststat}
\mathcal{T}_j(\overline{a},\overline{a}^*) = \int_0^{t^*} \{\widetilde{F}_j^{\overline{a}}(s) - \widetilde{F}_j^{\overline{a}^*}(s)\} \mathrm{d}s.
\end{equation}
We test the transition-, event-, and path-targeted effects based on the restricted mean time lost.

\begin{corollary}\label{cor_inference_test}
    Suppose that the conditions in Theorem \ref{thm_asy_normal} hold. 
    Under the null hypothesis $H_0: \tau_j(t;\overline{a},\overline{a}^*) = 0$ for all $t\in[0,t^*]$,
    $\sqrt{n} \mathcal{T}_j(\overline{a},\overline{a}^*) \rightsquigarrow \mathcal{N}(0, \sigma_j^2(\overline{a},\overline{a}^*))$, where 
    \[
    \sigma_j^2(\overline{a},\overline{a}^*) = \var\left(\int_0^{t^*} [\EIF\{F_j^{\overline{a}}(s)\} - \EIF\{F_j^{\overline{a}^*}(s)\}] \mathrm{d}s\right).
    \]
\end{corollary}

Statements about the whole curve, such as $H_0$ above, require simultaneous bands, for which we use the multiplier bootstrap. Let $\widehat{\psi}_i(t)$ be the estimated influence function of the contrast for unit $i$ and $\widehat{\varsigma}(t)$ the standard error of the contrast at $t$. With independent standard normal $\xi_1, \ldots, \xi_n$, let $c_{0.95}$ be the 95\% quantile of $\sup_{t \in \mathcal{T}} |n^{-1}\sum_i \xi_i \widehat{\psi}_i(t)| / \widehat{\varsigma}(t)$ over a finite grid $\mathcal{T}$. The band $\widetilde{\tau}_j(t) \pm c_{0.95}\, \widehat{\varsigma}(t)$, $t \in \mathcal{T}$, has asymptotic simultaneous coverage 95\% under the conditions of Theorem \ref{thm_asy_normal} \citep{van2013weak}.

Supplementary Material C reports a simulation study with two intermediate events and a terminal event under a mixed intervention, comparing the one-step and plug-in estimators when all models are correctly specified, when the propensity score is misspecified, and when the hazard model of an intermediate event omits the indicator of a preceding event. The one-step estimator is unbiased in all cases, as Theorem \ref{thm_multi_robust} implies, whereas the plug-in estimator is biased under the misspecified hazard model.

\section{Application to the LEADER Trial} \label{sec:real_data}

\subsection{Data and Transition Structure} \label{sec:leader_data}

Vascular events are leading causes of death in type 2 diabetes \citep{stratton2000association, advance2008intensive}. They are classified as expanded major adverse cardiovascular events (EMACE: myocardial infarction, stroke, unstable angina pectoris, heart failure, coronary revascularization, and cardiovascular death) and microvascular events (MVE: nephropathy and retinopathy) \citep{emerging2010diabetes, forbes2013mechanisms}. The LEADER trial randomized 9,340 patients with type 2 diabetes at high cardiovascular risk to liraglutide (4,668) or placebo (4,672) at 410 sites in 32 countries \citep{marso2013design, marso2016liraglutide}. Supplementary Material D.1 provides details on data pre-processing. Of the 828 deaths, 325 were fatal first EMACE, 172 were cardiovascular (CV) deaths after a nonfatal EMACE, and 331 were non-CV deaths. Follow-up ranged from about 42 to 60 months since randomization, with a median of 46 months among survivors. We used $t^* = 54$ months as the end of the study. 
Supplementary Material D.2 shows the potential cumulative incidences of and total effects on first EMACE, MVE, and death. Single-endpoint analyses indicate that liraglutide reduces the risks of EMACE, MVE, and all-cause death, but they do not show whether the reduction in death operates through the risks of vascular events or through the risks of death after such events.

\begin{figure}[!tb]
\centering
\includegraphics[width=0.9\textwidth]{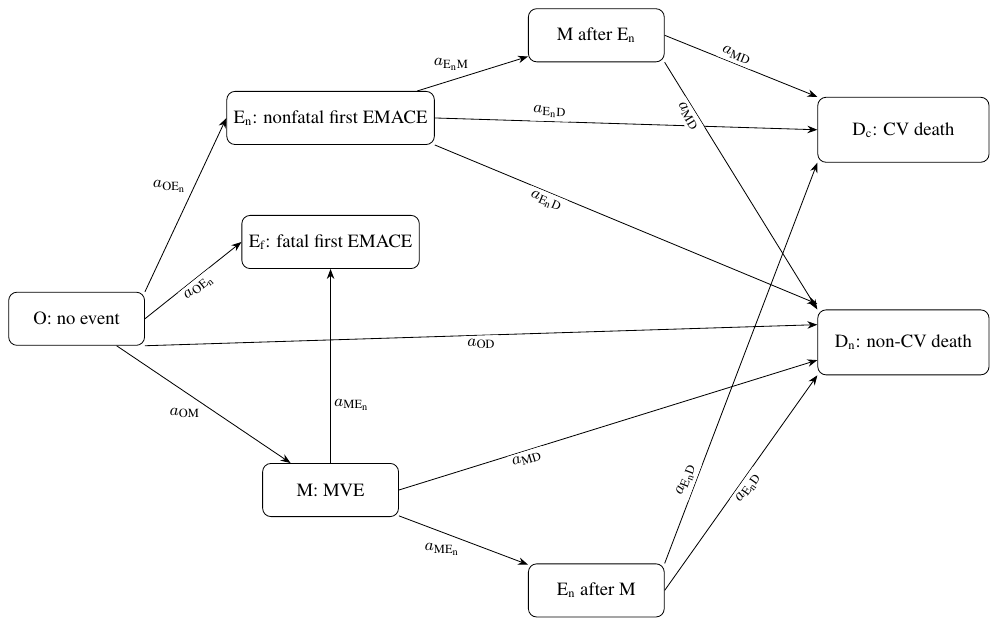}
\caption{State structure of the LEADER analysis. Each state is entered at most once (first occurrences). Transient states: no event (O), nonfatal first EMACE (E$_{\text{n}}$), first MVE (M), and the two histories with both events, ``M after E$_{\text{n}}$'' and ``E$_{\text{n}}$ after M''. Absorbing states: fatal first EMACE (E$_{\text{f}}$), that is, a CV death without a preceding nonfatal EMACE, which by the trial's definition is a first EMACE; CV death after a nonfatal EMACE (D$_{\text{c}}$); and non-CV death (D$_{\text{n}}$). Each arrow is labeled by the component of $\overline{a}$ that sets its hazard.} \label{fig:leader_model}
\end{figure}

Figure \ref{fig:leader_model} shows the state structure in our analysis. Since EMACE includes fatal and nonfatal events, we split the first EMACE into two states: nonfatal first EMACE and fatal first EMACE. To distinguish the causes of death, we split death other than fatal first EMACE into two states: CV death and non-CV death. The first intermediate states are nonfatal first EMACE (E$_{\text{n}}$) and first MVE (M). The second intermediate states are ``M after E$_{\text{n}}$'' and ``E$_{\text{n}}$ after M''. The absorbing states are fatal first EMACE (E$_{\text{f}}$), CV death after a nonfatal EMACE (D$_{\text{c}}$), and non-CV death (D$_{\text{n}}$). Transitions into E$_{\text{f}}$ and E$_{\text{n}}$ from the same state share the same intervention, while transitions into D$_{\text{c}}$ and D$_{\text{n}}$ from the same state share the same intervention. The cumulative incidence of first EMACE counts entry into E$_{\text{n}}$ and E$_{\text{f}}$, and the cumulative incidence of all-cause death counts entry into E$_{\text{f}}$, D$_{\text{c}}$, and D$_{\text{n}}$.

We adjust for age, sex, BMI, HbA1c, diabetes duration, CV risk group, and insulin-naive status as covariates. The missing covariates of 32 patients are imputed by nearest neighbors. Although treatment was randomized 1:1, we estimate the propensity score using logistic regression, which may improve efficiency, as confirmed in simulation. The hazards of nonfatal EMACE, MVE, CV death (together with fatal first EMACE), and non-CV death are each fitted by a Cox model \eqref{eq:cox}, pooling the arms with treatment interactions, with $z_{kg}(t)$ consisting of indicators of the states visited and the log-transformed time since the last transition, capped at 12 months.

\subsection{Results}

\begin{figure}[!tb]
\centering
\includegraphics[width=0.96\textwidth]{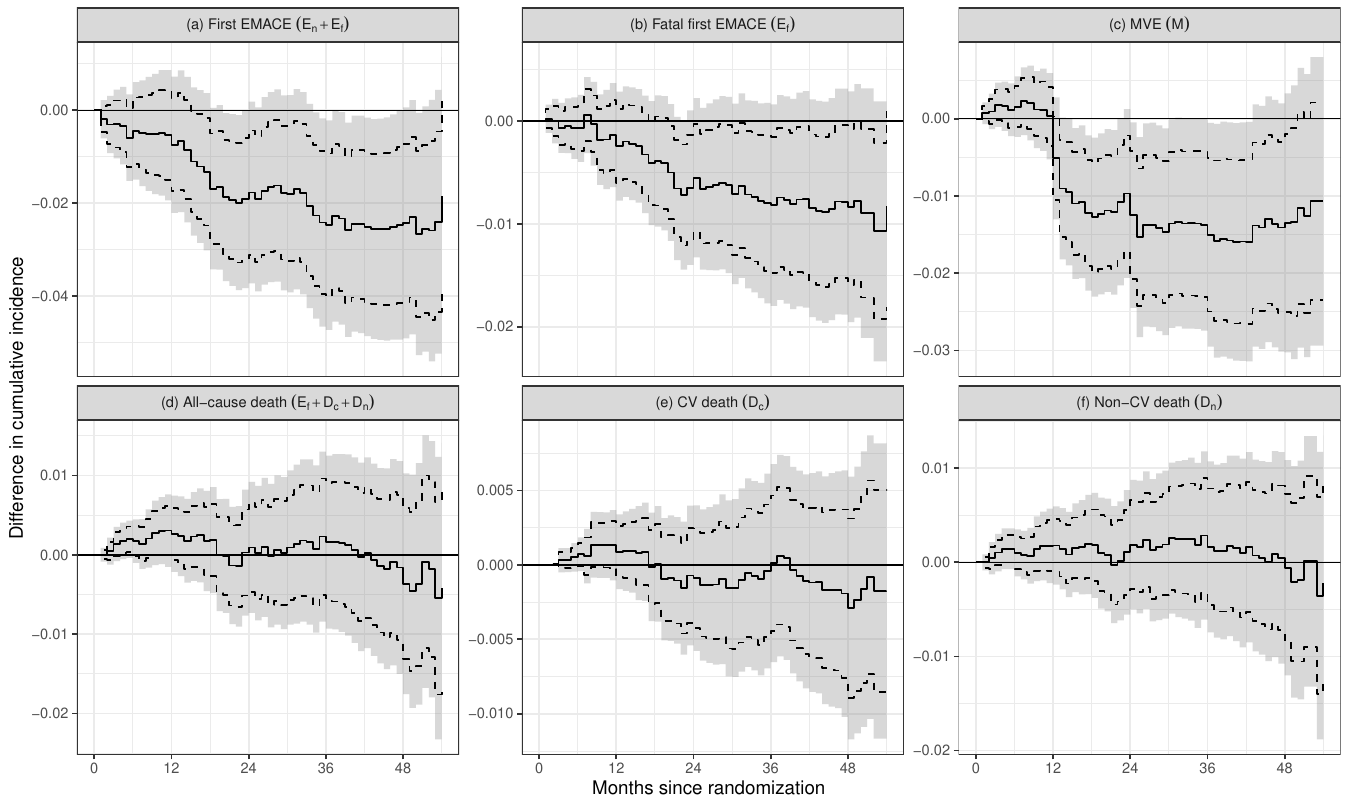}
\caption{Event-targeted effects in LEADER: difference in the counterfactual cumulative incidence between the regime in which the hazards of the designated transitions are taken from the liraglutide arm, with all other hazards at placebo, and the all-placebo regime. (a) First EMACE (E$_{\text{n}}$ or E$_{\text{f}}$) and (b) fatal first EMACE (E$_{\text{f}}$), switching the transitions into E$_{\text{n}}$ and E$_{\text{f}}$; (c) MVE, switching the transitions into M; (d) all-cause death (E$_{\text{f}}$, D$_{\text{c}}$, or D$_{\text{n}}$), (e) CV death after a nonfatal EMACE (D$_{\text{c}}$), and (f) non-CV death (D$_{\text{n}}$), switching the transitions into D$_{\text{c}}$ and D$_{\text{n}}$. Solid: estimate; dashed: pointwise 95\% confidence intervals; shaded: 95\% simultaneous confidence band.} \label{fig:ate1}
\end{figure}

Figure \ref{fig:ate1} shows event-targeted effects over time for different choices of the primary endpoint $j$ in each panel. We consider three groups of event-targeted effects: on first EMACE, on MVE, and on death. For EMACE, we intervene in transitions into first EMACE (E$_{\text{n}}$ and E$_{\text{f}}$) from all-placebo, and present the cumulative incidences of first EMACE and fatal first EMACE. The cumulative incidence of first EMACE is lowered, with 0.90 months of first EMACE-free time gained by 54 months (SE 0.30, $P = 0.003$). The cumulative incidence of fatal first EMACE is also lowered, with 0.29 months of fatal first EMACE-free time gained (SE 0.12, $P = 0.017$). For MVE, we intervene in transitions into MVE (M), and present the cumulative incidence of MVE. Switching the hazards of the transitions into M gained 0.53 months of MVE-free time (SE 0.19, $P = 0.004$). For death, we intervene in transitions into D$_{\text{c}}$ and D$_{\text{n}}$, that is, death other than fatal first EMACE. The cumulative incidences have no detectable change on all-cause death ($-0.02$ months, SE 0.13, $P = 0.86$), CV death after a nonfatal EMACE (0.03 months, SE 0.08, $P = 0.71$), or non-CV death ($-0.06$ months, SE 0.11, $P = 0.61$).

Now we take all-cause death as the primary endpoint ($j = \{\text{E}_{\text{f}}, \text{D}_{\text{c}}, \text{D}_{\text{n}}\}$). The total effect on all-cause death is 0.36 months of survival gained by 54 months (SE 0.18, $P = 0.043$). Table \ref{tab:transpath} reports the transition- and path-targeted effects on all-cause death. For each transition-targeted effect, we intervene in the transitions indexed by the same $\overline{a}$ component in Figure \ref{fig:leader_model}. The only transition whose switch has a detectable effect with either reference is $(\text{O}, \text{E})$, the transition to first EMACE from the initial state. For path-targeted effects, we collapse transitions into all-cause death into five paths, consistent with Figure \ref{fig:model}. In particular, death due to fatal first EMACE is attributed to O$\to$E$\to$D or O$\to$M$\to$E$\to$D, depending on whether MVE occurred before first EMACE. Supplementary Material D.3 shows the path-targeted effects along these five paths over time.

\begin{table}[!tb]
\centering
\caption{Transition- and path-targeted effects on all-cause death in LEADER: survival time gained by 54 months (months; standard error in parentheses) when the hazards of the designated transitions are switched from placebo to liraglutide, with all other hazards at placebo (placebo reference) or at liraglutide (liraglutide reference), and Shapley order-averaged effects. Labels are the components of $\overline{a}$ in Figure \ref{fig:leader_model}: E includes E$_{\text{n}}$ and E$_{\text{f}}$, while D includes D$_{\text{c}}$ and D$_{\text{n}}$; E$\to$D in the path-targeted effects includes E$_{\text{f}}$.} \label{tab:transpath}
\begin{tabular}{lcccccc}
\toprule
 & \multicolumn{2}{c}{Placebo reference} & \multicolumn{2}{c}{Liraglutide reference} & \multicolumn{2}{c}{Shapley} \\
\cmidrule(lr){2-3} \cmidrule(lr){4-5} \cmidrule(lr){6-7}
 & Estimate & $P$ & Estimate & $P$ & Estimate & $P$ \\
\midrule
\multicolumn{7}{l}{\textit{Transition-targeted}} \\
(O, E) & 0.364 (0.121) & 0.003 & 0.345 (0.121) & 0.004 & 0.354 (0.121) & 0.003 \\
(O, M) & 0.018 (0.010) & 0.073 & 0.021 (0.012) & 0.069 & 0.019 (0.010) & 0.048 \\
(O, D) & $-$0.087 (0.085) & 0.307 & $-$0.092 (0.088) & 0.297 & $-$0.089 (0.087) & 0.302 \\
(E, M) & 0.011 (0.010) & 0.269 & $-$0.001 (0.007) & 0.846 & 0.004 (0.007) & 0.552 \\
(M, E) & 0.024 (0.028) & 0.392 & 0.017 (0.024) & 0.464 & 0.021 (0.026) & 0.422 \\
(E, D) & 0.116 (0.092) & 0.206 & 0.098 (0.082) & 0.234 & 0.107 (0.087) & 0.218 \\
(M, D) & $-$0.053 (0.040) & 0.188 & $-$0.057 (0.035) & 0.108 & $-$0.055 (0.037) & 0.139 \\
\midrule
\multicolumn{7}{l}{\textit{Path-targeted}} \\
O$\to$D & $-$0.087 (0.085) & 0.307 & $-$0.092 (0.088) & 0.297 & $-$0.090 (0.087) & 0.300 \\
O$\to$E$\to$D & 0.463 (0.149) & 0.002 & 0.460 (0.149) & 0.002 & 0.228 (0.073) & 0.002 \\
O$\to$M$\to$D & $-$0.022 (0.040) & 0.583 & $-$0.049 (0.037) & 0.190 & $-$0.010 (0.019) & 0.622 \\
O$\to$E$\to$M$\to$D & 0.304 (0.127) & 0.017 & 0.303 (0.126) & 0.017 & 0.151 (0.064) & 0.018 \\
O$\to$M$\to$E$\to$D & 0.146 (0.096) & 0.127 & 0.148 (0.087) & 0.088 & 0.081 (0.051) & 0.109 \\
\midrule
Total effect & \multicolumn{4}{c}{} & 0.361 (0.178) & 0.043 \\
\bottomrule
\end{tabular}
\end{table}

The Shapley order-averaged effects decompose the total effect on death into either the sum of transition-targeted effects or the sum of path-targeted effects. For transition-targeted effects, $(\text{O}, \text{E})$ accounts for 0.35 months (SE 0.12, $P = 0.003$), while the transitions into death (excluding fatal first EMACE) contribute nothing detectable. Among the path-targeted effects, O$\to$E$\to$D is the largest (0.23 months, SE 0.07, $P = 0.002$), followed by O$\to$E$\to$M$\to$D (0.15 months, SE 0.06, $P = 0.018$). Since a common transition O$\to$E lies in both paths, intervention in this transition would contribute to both path-targeted effects simultaneously. Shapley order-averaged effects allocate this interventional effect to each path in fraction, so the decomposed path-targeted effects are lower than the ones taking all-placebo (or all-liraglutide) as a reference. Supplementary Material D.4 compares hypothetical cumulative incidences under different treatment regimes (switching treatment once an event occurs) and finds that continuing to take the drug can increase survival time.

\subsection{Sensitivity Analyses} \label{sec:sens}

Supplementary Material D.5 re-estimates the event-targeted effects using the EIF-based estimator under the extended Markov model; the estimates and standard errors are practically identical to those in the main analysis. In Supplementary Material D.6, we perform a sensitivity analysis in which the intervention is applied conditionally on a shared gamma frailty across all transitions of a patient, as well as on the observed history. With the variance $\nu$ of the frailty treated as fixed, the potential cumulative incidences and the total effect do not depend on $\nu$, whereas the mixed regimes do. The maximum marginal likelihood estimation yields $\widehat\nu = 0.63$. Varying $\nu$ from 0 to 0.75 changes the transition-, event-, and path-targeted effects by at most 0.03 months in terms of time gained by 54 months, and none of the conclusions change. 

Two secondary analyses are performed. In Supplementary Material D.7, we split the first-occurrence event into nine mutually exclusive types so that transition-, event-, and path-targeted effects coincide. Among the individual events, the transition-targeted effects on nephropathy (0.55 months of nephropathy-free time gained by 54 months, SE 0.15, $P < 0.001$) and first-occurrence CV death (i.e., fatal first EMACE; 0.29 months, SE 0.12, $P = 0.017$) are the only significant ones at the 0.05 level. In Supplementary Material D.8, we merge fatal first EMACE into CV death and allow different interventions for transitions into CV death and non-CV death; therefore, there are 10 transitions to intervene on. Switching the hazards of the transitions into CV death gained 0.33 months of survival by 54 months (SE 0.14, $P = 0.018$), most of it through fatal first EMACE (0.31 months, SE 0.12, $P = 0.011$), whereas switching the hazards of the transitions into non-CV death had no detectable effect ($-0.06$ months, SE 0.11, $P = 0.60$).

\section{Discussion} \label{sec:discus}

The estimands of this paper are defined by stochastic interventions on transition hazards. They are well defined and identifiable without cross-world assumptions, whatever the treatment-induced or unmeasured confounding of transitions, because the estimands are functionals of the observed-history hazards of each arm. This generality has a price. A regime that combines hazards from different arms describes a hypothetical multi-state law, not an intervention on a biological mechanism. We have therefore framed the transition-, event-, and path-targeted effects as contrasts between counterfactual multi-state laws. If there are no treatment-induced confounding of transitions given the observed history and no carry-over of earlier treatment, regimes in which all transitions out of a state receive the same level would correspond to dynamic treatment regimes that switch treatment on entry into a state; trials with re-randomization at the occurrence of an event could test this.

The approach differs from related frameworks in what is intervened upon. Interventional direct and indirect effects with one or more mediators \citep{vansteelandt2017interventional, lin2017interventional, diaz2020causal, vansteelandt2019Mediation} replace the distribution of a mediator by a random draw from its distribution under another treatment level. Here the objects replaced are the intensities of an event-history process, so the hypothetical process must respect the order of events, the absorbing states, and the dependence of each hazard on the history. Path-specific effects defined by edge interventions \citep{avin2005identifiability, shpitser2016causal} require the absence of recanting witnesses and cross-world independence, which fail when each intermediate event confounds the transitions out of the other. Our path-targeted effects need neither, but they answer a different question. A limitation is that interventional effects are not sharp \citep{miles2022on}: a null transition-targeted effect does not imply that the treatment has no effect on that transition in a mechanistic sense, because changes in the distribution of latent states among those at risk can offset a genuine effect.

Because the time origins of being at risk in intermediate states are not aligned, the weight in the EIF is a density ratio along the history, and its estimation requires some restriction on how the hazards differ between arms. We use Cox models with treatment as a covariate, which make the hazard ratio between arms a regression function and leave the dependence on the history free, including continuous dependence on the time since the last transition. Two alternatives are (1) kernel smoothing of arm-specific baseline hazards, and (2) restricting the hazards to depend on the history only through a finite-valued summary (Supplementary Material A.2), under which the projected EIF involves only cumulative hazards and occupancy probabilities and can be more efficient. Extended Markovity produces a lower semiparametric efficiency bound, but the EIF-based estimator under this submodel loses multiple robustness.

\section*{Acknowledgements}
We thank Kajsa Kvist from Novo Nordisk A/S for discussions about the LEADER trial.

\section*{Data Availability Statement}
The data that support the findings of this study are available from Novo Nordisk A/S. Restrictions apply to the availability of these data, which were used under license for this study.

\section*{Supplementary Material}
The online Supplementary Material includes (A) efficient influence functions, regularity conditions, and Shapley order-averaged effects, (B) the frailty-indexed sensitivity analysis, (C) simulation studies, (D) additional data analysis results, including the sensitivity and secondary analyses, and (E) proofs of theoretical results.

\bibliographystyle{apalike}
\bibliography{ref}

\begin{thebibliography}{}

\bibitem[{ADVANCE Collaborative Group}, 2008]{advance2008intensive}
{ADVANCE Collaborative Group} (2008).
\newblock Intensive blood glucose control and vascular outcomes in patients
  with type 2 diabetes.
\newblock {\em New England Journal of Medicine}, 358(24):2560--2572.

\bibitem[Andersen et~al., 2012]{andersen2012statistical}
Andersen, P.~K., Borgan, O., Gill, R.~D., and Keiding, N. (2012).
\newblock {\em Statistical models based on counting processes}.
\newblock Springer Science \& Business Media.

\bibitem[Andersen and Keiding, 2002]{andersen2002multi}
Andersen, P.~K. and Keiding, N. (2002).
\newblock Multi-state models for event history analysis.
\newblock {\em Statistical Methods in Medical Research}, 11(2):91--115.

\bibitem[Avin et~al., 2005]{avin2005identifiability}
Avin, C., Shpitser, I., and Pearl, J. (2005).
\newblock Identifiability of path-specific effects.
\newblock In {\em IJCAI International Joint Conference on Artificial
  Intelligence}, pages 357--363.

\bibitem[Bickel et~al., 1993]{bickel1993efficient}
Bickel, P.~J., Klaassen, C.~A., Bickel, P.~J., Ritov, Y., Klaassen, J.,
  Wellner, J.~A., and Ritov, Y. (1993).
\newblock {\em Efficient and adaptive estimation for semiparametric models},
  volume~4.
\newblock Springer.

\bibitem[Br{\'e}maud, 1981]{bremaud1981point}
Br{\'e}maud, P. (1981).
\newblock {\em Point Processes and Queues: Martingale Dynamics}.
\newblock Springer, New York.

\bibitem[Breslow, 1972]{breslow1972disussion}
Breslow, N. (1972).
\newblock Disussion of regression models and life-tables by {D. R. Cox}.
\newblock {\em Journal of the Royal Statistical Society Series B: Statistical
  Methodology}, 34:216--217.

\bibitem[Breslow, 1974]{breslow1974covariance}
Breslow, N. (1974).
\newblock Covariance analysis of censored survival data.
\newblock {\em Biometrics}, 30(1):89--99.

\bibitem[Breum et~al., 2024]{breum2024estimation}
Breum, M.~S., Munch, A., Gerds, T.~A., and Martinussen, T. (2024).
\newblock Estimation of separable direct and indirect effects in a
  continuous-time illness--death model.
\newblock {\em Lifetime Data Analysis}, 30(1):143--180.

\bibitem[Chernozhukov et~al., 2018]{chernozhukov2018double}
Chernozhukov, V., Chetverikov, D., Demirer, M., Duflo, E., Hansen, C., Newey,
  W., and Robins, J. (2018).
\newblock Double/debiased machine learning for treatment and structural
  parameters.
\newblock {\em The Econometrics Journal}, 21(1):C1--C68.

\bibitem[Deng et~al., 2024]{deng2024direct}
Deng, Y., Wang, Y., and Zhou, X.-H. (2024).
\newblock Direct and indirect treatment effects in the presence of
  semicompeting risks.
\newblock {\em Biometrics}, 80(2):ujae032.

\bibitem[D{\'\i}az and Hejazi, 2020]{diaz2020causal}
D{\'\i}az, I. and Hejazi, N.~S. (2020).
\newblock Causal mediation analysis for stochastic interventions.
\newblock {\em Journal of the Royal Statistical Society Series B: Statistical
  Methodology}, 82(3):661--683.

\bibitem[{Emerging Risk Factors Collaboration} et~al.,
  2010]{emerging2010diabetes}
{Emerging Risk Factors Collaboration} et~al. (2010).
\newblock Diabetes mellitus, fasting blood glucose concentration, and risk of
  vascular disease: a collaborative meta-analysis of 102 prospective studies.
\newblock {\em The lancet}, 375(9733):2215--2222.

\bibitem[Fine et~al., 2001]{fine2001semi}
Fine, J.~P., Jiang, H., and Chappell, R. (2001).
\newblock On semi-competing risks data.
\newblock {\em Biometrika}, 88(4):907--919.

\bibitem[Forbes and Cooper, 2013]{forbes2013mechanisms}
Forbes, J.~M. and Cooper, M.~E. (2013).
\newblock Mechanisms of diabetic complications.
\newblock {\em Physiological Reviews}, 93(1):137--188.

\bibitem[Ha et~al., 2020]{ha2020frailty}
Ha, I.~D., Xiang, L., Peng, M., Jeong, J.-H., and Lee, Y. (2020).
\newblock Frailty modelling approaches for semi-competing risks data.
\newblock {\em Lifetime Data Analysis}, 26:109--133.

\bibitem[Hejazi et~al., 2022]{hejazi2022nonparametric}
Hejazi, N.~S., Rudolph, K.~E., Van Der~Laan, M.~J., and Díaz, I. (2022).
\newblock Nonparametric causal mediation analysis for stochastic interventional
  (in)direct effects.
\newblock {\em Biostatistics}, 24(3):686--707.

\bibitem[Hougaard, 1999]{hougaard1999multi}
Hougaard, P. (1999).
\newblock Multi-state models: a review.
\newblock {\em Lifetime Data Analysis}, 5:239--264.

\bibitem[Huang, 2021]{huang2021causal}
Huang, Y.-T. (2021).
\newblock Causal mediation of semicompeting risks.
\newblock {\em Biometrics}, 77(4):1143--1154.

\bibitem[Jacod, 1975]{jacod1975multivariate}
Jacod, J. (1975).
\newblock Multivariate point processes: predictable projection,
  {R}adon--{N}ikodym derivatives, representation of martingales.
\newblock {\em Zeitschrift f{\"u}r Wahrscheinlichkeitstheorie und verwandte
  Gebiete}, 31(3):235--253.

\bibitem[Kodell and Nelson, 1980]{kodell1980illness}
Kodell, R. and Nelson, C. (1980).
\newblock An illness--death model for the study of the carcinogenic process
  using survival/sacrifice data.
\newblock {\em Biometrics}, 36(2):267--277.

\bibitem[Li et~al., 2020]{li2020shared}
Li, J., Zhang, Y., Bakoyannis, G., and Gao, S. (2020).
\newblock On shared gamma-frailty conditional {Markov} model for semicompeting
  risks data.
\newblock {\em Statistics in Medicine}, 39(23):3042--3058.

\bibitem[Lin and VanderWeele, 2017]{lin2017interventional}
Lin, S.-H. and VanderWeele, T. (2017).
\newblock Interventional approach for path-specific effects.
\newblock {\em Journal of Causal Inference}, 5(1):20150027.

\bibitem[Lin et~al., 2017]{lin2017mediation}
Lin, S.-H., Young, J.~G., Logan, R., and VanderWeele, T.~J. (2017).
\newblock Mediation analysis for a survival outcome with time-varying
  exposures, mediators, and confounders.
\newblock {\em Statistics in Medicine}, 36(26):4153--4166.

\bibitem[Lok, 2015]{lok2015defining}
Lok, J.~J. (2015).
\newblock Defining and estimating causal direct and indirect effects when
  setting the mediator to specific values is not feasible.
\newblock {\em Statistics in Medicine}, 35(22):4008--4020.

\bibitem[Marso et~al., 2016]{marso2016liraglutide}
Marso, S.~P., Daniels, G.~H., Brown-Frandsen, K., Kristensen, P., Mann, J.~F.,
  Nauck, M.~A., Nissen, S.~E., Pocock, S., Poulter, N.~R., Ravn, L.~S., et~al.
  (2016).
\newblock Liraglutide and cardiovascular outcomes in type 2 diabetes.
\newblock {\em New England Journal of Medicine}, 375(4):311--322.

\bibitem[Marso et~al., 2013]{marso2013design}
Marso, S.~P., Poulter, N.~R., Nissen, S.~E., Nauck, M.~A., Zinman, B., Daniels,
  G.~H., Pocock, S., Steinberg, W.~M., Bergenstal, R.~M., Mann, J.~F., et~al.
  (2013).
\newblock Design of the liraglutide effect and action in diabetes: evaluation
  of cardiovascular outcome results {(LEADER)} trial.
\newblock {\em American Heart Journal}, 166(5):823--830.

\bibitem[Martinussen and Stensrud, 2023]{martinussen2023estimation}
Martinussen, T. and Stensrud, M.~J. (2023).
\newblock Estimation of separable direct and indirect effects in continuous
  time.
\newblock {\em Biometrics}, 79(1):127--139.

\bibitem[Miles, 2023]{miles2022on}
Miles, C.~H. (2023).
\newblock On the causal interpretation of randomised interventional indirect
  effects.
\newblock {\em Journal of the Royal Statistical Society Series B: Statistical
  Methodology}, 85(4):1154--1172.

\bibitem[Miles et~al., 2020]{miles2020semiparametric}
Miles, C.~H., Shpitser, I., Kanki, P., Meloni, S., and {Tchetgen Tchetgen},
  E.~J. (2020).
\newblock On semiparametric estimation of a path-specific effect in the
  presence of mediator-outcome confounding.
\newblock {\em Biometrika}, 107(1):159--172.

\bibitem[Putter et~al., 2007]{putter2007tutorial}
Putter, H., Fiocco, M., and Geskus, R.~B. (2007).
\newblock Tutorial in biostatistics: competing risks and multi-state models.
\newblock {\em Statistics in Medicine}, 26(11):2389--2430.

\bibitem[Robins et~al., 2022]{robins2022interventionist}
Robins, J.~M., Richardson, T.~S., and Shpitser, I. (2022).
\newblock An interventionist approach to mediation analysis.
\newblock In {\em Probabilistic and causal inference: the works of Judea
  Pearl}, pages 713--764. ACM.

\bibitem[Shapley, 1953]{shapley1953value}
Shapley, L.~S. (1953).
\newblock A value for n-person games.
\newblock In Kuhn, H.~W. and Tucker, A.~W., editors, {\em Contributions to the
  Theory of Games II}, pages 307--317. Princeton University Press.

\bibitem[Shpitser and Tchetgen~Tchetgen, 2016]{shpitser2016causal}
Shpitser, I. and Tchetgen~Tchetgen, E. (2016).
\newblock Causal inference with a graphical hierarchy of interventions.
\newblock {\em Annals of Statistics}, 44(6):2433.

\bibitem[Stensrud et~al., 2021]{stensrud2021generalized}
Stensrud, M.~J., Hern{\'a}n, M.~A., Tchetgen~Tchetgen, E.~J., Robins, J.~M.,
  Didelez, V., and Young, J.~G. (2021).
\newblock A generalized theory of separable effects in competing event
  settings.
\newblock {\em Lifetime Data Analysis}, 27(4):588--631.

\bibitem[Stensrud et~al., 2022]{stensrud2022separable}
Stensrud, M.~J., Young, J.~G., Didelez, V., Robins, J.~M., and Hern{\'a}n,
  M.~A. (2022).
\newblock Separable effects for causal inference in the presence of competing
  events.
\newblock {\em Journal of the American Statistical Association},
  117(537):175--183.

\bibitem[Stratton et~al., 2000]{stratton2000association}
Stratton, I.~M., Adler, A.~I., Neil, H. A.~W., Matthews, D.~R., Manley, S.~E.,
  Cull, C.~A., Hadden, D., Turner, R.~C., and Holman, R.~R. (2000).
\newblock Association of glycaemia with macrovascular and microvascular
  complications of type 2 diabetes ({UKPDS 35}): prospective observational
  study.
\newblock {\em BMJ}, 321(7258):405--412.

\bibitem[Valeri et~al., 2023]{valeri2023multistate}
Valeri, L., Proust-Lima, C., Fan, W., Chen, J.~T., and Jacqmin-Gadda, H.
  (2023).
\newblock A multistate approach for the study of interventions on an
  intermediate time-to-event in health disparities research.
\newblock {\em Statistical Methods in Medical Research}, 32(8):1445--1460.

\bibitem[van~der vaart and Wellner, 2013]{van2013weak}
van~der vaart, A. and Wellner, J. (2013).
\newblock {\em Weak Convergence and Empirical Processes: With Applications to
  Statistics}.
\newblock Springer Series in Statistics. Springer New York.

\bibitem[Van~der Vaart, 2000]{van2000asymptotic}
Van~der Vaart, A.~W. (2000).
\newblock {\em Asymptotic statistics}, volume~3.
\newblock Cambridge university press.

\bibitem[VanderWeele and Tchetgen~Tchetgen, 2017]{vanderweele2017mediation}
VanderWeele, T.~J. and Tchetgen~Tchetgen, E.~J. (2017).
\newblock Mediation analysis with time varying exposures and mediators.
\newblock {\em Journal of the Royal Statistical Society Series B: Statistical
  Methodology}, 79(3):917--938.

\bibitem[Vansteelandt and Daniel, 2017]{vansteelandt2017interventional}
Vansteelandt, S. and Daniel, R.~M. (2017).
\newblock Interventional effects for mediation analysis with multiple
  mediators.
\newblock {\em Epidemiology}, 28(2):258--265.

\bibitem[Vansteelandt et~al., 2019]{vansteelandt2019Mediation}
Vansteelandt, S., Linder, M., Vandenberghe, S., Steen, J., and Madsen, J.
  (2019).
\newblock Mediation analysis of time‐to‐event endpoints accounting for
  repeatedly measured mediators subject to time‐varying confounding.
\newblock {\em Statistics in Medicine}, 38(24):4828 -- 4840.

\bibitem[Vo et~al., 2026]{vo2026recanting}
Vo, T.-T., Williams, N., Liu, R., Rudolph, K.~E., and D{\'\i}az, I. (2026).
\newblock Recanting twins: Addressing intermediate confounding in mediation
  analysis.
\newblock {\em Statistics in Medicine}, 45(3-5):e70432.

\bibitem[Weir et~al., 2022]{weir2022counterfactual}
Weir, I.~R., Rider, J.~R., and Trinquart, L. (2022).
\newblock Counterfactual mediation analysis in the multistate model framework
  for surrogate and clinical time-to-event outcomes in randomized controlled
  trials.
\newblock {\em Pharmaceutical Statistics}, 21(1):163--175.

\bibitem[Xu et~al., 2010]{xu2010statistical}
Xu, J., Kalbfleisch, J.~D., and Tai, B. (2010).
\newblock Statistical analysis of illness--death processes and semicompeting
  risks data.
\newblock {\em Biometrics}, 66(3):716--725.

\bibitem[Zeng and Lin, 2007]{zeng2007maximum}
Zeng, D. and Lin, D. (2007).
\newblock Maximum likelihood estimation in semiparametric regression models
  with censored data.
\newblock {\em Journal of the Royal Statistical Society Series B: Statistical
  Methodology}, 69(4):507--564.

\end{thebibliography}

\newpage

\appendix                       
\setcounter{table}{0}
\setcounter{figure}{0}
\setcounter{equation}{0}
\setcounter{theorem}{0}
\setcounter{condition}{0}
\setcounter{assumption}{0}
\setcounter{lemma}{0}
\setcounter{remark}{0}
\renewcommand{\thetable}{S\arabic{table}}
\renewcommand{\thefigure}{S\arabic{figure}}
\renewcommand{\theequation}{S\arabic{equation}}
\renewcommand{\thetheorem}{S\arabic{theorem}}
\renewcommand{\thecondition}{S\arabic{condition}}
\renewcommand{\theassumption}{S\arabic{assumption}}
\renewcommand{\thelemma}{S\arabic{lemma}}
\renewcommand{\theremark}{S\arabic{remark}}
\allowdisplaybreaks[4]

\begin{center}
{\large\bf Supplementary Material for ``Estimating Treatment Effects Through
Intermediate-Event Pathways with Multi-State Data''}
\end{center}

\bigskip

\section{Efficient influence functions and regularity conditions} \label{sec:eif}

\subsection{Efficient influence function in the nonparametric model} \label{sec:np}

Under Assumptions 1--5, Lemma 1 states that the efficient influence function of $F_j^{\overline{a}}(t)$ in the nonparametric model is
\begin{equation}\label{eif_np_s}
\begin{aligned}
\psi(O) &= F_j^{\overline{a}}(t \mid W) - F_j^{\overline{a}}(t) + \sum_{(k,g)\in\mathcal{E}} \frac{\mathds{1}(A = a_{kg})}{\pr(A = a_{kg} \mid W)} \\
&\quad \times \int_0^t \omega_k^{\overline{a}}(s; a_{kg}, W, \overline{X}(s-))\, \Delta_{kg}^{\overline{a}}(t \mid W; \overline{X}(s-))\, \mathrm{d}M_{kg}(s; a_{kg}, W, \overline{X}(s-)),
\end{aligned}
\end{equation}
where the weight $\omega_k^{\overline{a}}$ is the density of the hypothetical history relative to the observed, uncensored history in arm $a_{kg}$; by Equation (13), it is a product of ratios of rates between arms at the unit's earlier transition times, of survival ratios over the sojourns, and of the inverse censoring survival.

The one-step estimator is
\begin{equation}\label{cond_2_est}
\begin{aligned}
\widetilde{F}_j^{\overline{a}}(t) &= \pn \Bigg[ \widehat{F}_j^{\overline{a}}(t \mid W) + \sum_{(k,g)\in\mathcal{E}} \frac{\mathds{1}(A = a_{kg})}{\widehat{\pr}(A = a_{kg} \mid W)} \\
&\qquad\quad \times \int_0^t \widehat\omega_k^{\overline{a}}(s; a_{kg}, W, \overline{X}(s-))\, \widehat\Delta_{kg}^{\overline{a}}(t \mid W; \overline{X}(s-))\, \mathrm{d}\widehat{M}_{kg}(s; a_{kg}, W, \overline{X}(s-)) \Bigg],
\end{aligned}
\end{equation}
where the hats denote quantities computed from the fitted propensity score, censoring survival, transition hazards, and hazard ratios between arms, and $\mathrm{d}\widehat{M}_{kg}(s; a, W, \overline{X}(s-)) = \mathrm{d}N_{kg}(s) - Y_k(s)\, \mathrm{d}\widehat\Lambda_{kg}^{a}(s \mid W, \overline{X}(s-))$. The fitted weight $\widehat\omega_k^{\overline{a}}$ is Equation (13) evaluated at the fitted hazard ratios, cumulative hazards, and censoring survival.

Let $\mathbb{P}(h) = \int h(o) f(o) \mathrm{d}o$ be the measure with respect to the true data-generating process $f(o)$. The following conditions ensure that the estimator converges to the target estimand at the parametric rate.

\begin{condition}[Regularity conditions for asymptotic normality] \label{cond2}
(1) The fitted propensity score, cumulative hazards, and weights are $L_2$-consistent over $[0,t^*]$: for $a \in \{0,1\}$, every $(k,g)\in\mathcal{E} \cup (\mathcal{G}\times\{\text{c}\})$, and every $(k,g) \in \mathcal{E}$ respectively,
\begin{align*}
    \mathbb{P} \left\{ \widehat{\pr}(A=a \mid W) - \pr(A=a \mid W) \right\}^2 &= o_p(1), \\
    \sup_{t} \sup_{\overline{x}(t)} \mathbb{P} \left\{ \widehat\Lambda_{kg}^a(t \mid W, \overline{x}(t)) - \Lambda_{kg}^a(t \mid W, \overline{x}(t)) \right\}^2 &= o_p(1), \\
    \sup_{s} \sup_{\overline{x}(s-)} \mathbb{P} \left\{ \widehat\omega_k^{\overline{a}}(s; a_{kg}, W, \overline{x}(s-)) - \omega_k^{\overline{a}}(s; a_{kg}, W, \overline{x}(s-)) \right\}^2 &= o_p(1),
\end{align*}
and the fitted weights are bounded with probability tending to one.
(2) The second-order remainder is negligible: for every $(k,g)\in\mathcal{E}$,
\[
\sup_{t \leq t^*} \left| \mathbb{P}\, \E^{\overline{a}}\left[ \int_0^t \widehat{e}_{kg}(s; W, \overline{X}^{\overline{a}}(s-))\, Y_k^{\overline{a}}(s)\, \mathrm{d}\left\{\widehat\Lambda_{kg}^{a_{kg}} - \Lambda_{kg}^{a_{kg}}\right\}(s \mid W, \overline{X}^{\overline{a}}(s-)) \Bigm| W \right] \right| = o_p(n^{-1/2}),
\]
where $\E^{\overline{a}}(\cdot \mid W)$ is the expectation under the law of the true hypothetical process given $W$, and $\widehat{e}_{kg} = \{1 - \widehat{r}_{kg}\, \widehat\rho_k / \rho_k\}\, \widehat\Delta_{kg}^{\overline{a}}$. Here $\widehat{r}_{kg}$ is the ratio of the true to the fitted propensity score and censoring survival, and $\widehat\rho_k / \rho_k$ is the ratio of the fitted to the true density ratio $p^{\overline{a}} / p^{a_{kg}}$ of the history. The term $\widehat{e}_{kg}$ is the coefficient in the exact bias identity proved in Section E.5. It is a bounded factor times the errors of the fitted propensity score and censoring survival and of the fitted hazards and hazard ratios, in both arms, of the transitions in $\mathcal{U}_k$ (Section 3.3) that $\overline{a}$ borrows from an arm other than $a_{kg}$. It does not involve the error of $\widehat\Lambda_{kg}^{a_{kg}}$, and it vanishes when all components of $\overline{a}$ are equal and the propensity score and censoring survival are correctly specified.
(3) The function $\{\widehat{F}_j^{\overline{a}}(t \mid W) + \widehat\varphi_j^{\overline{a}}(t; O): t\in[0,t^*]\}$ belongs to a Donsker class.
\end{condition}

Condition S1(2) is a product-rate condition. If the true and fitted cumulative hazards are absolutely continuous in $s$, with rates $\lambda_{kg}^{a}$ and $\widehat\lambda_{kg}^{a}$, the Cauchy--Schwarz inequality bounds each term by a constant times $\|\widehat{e}_{kg}\|\,\|\widehat\lambda_{kg}^{a} - \lambda_{kg}^{a}\|$ in $L_2(\mathbb{P} \times \mathrm{d}s)$. It then suffices that this product be $o_p(n^{-1/2})$, for instance that all fitted models converge at rate $o_p(n^{-1/4})$. This bound does not apply to Breslow-type estimators, whose error is not small in total variation \citep{breslow1972disussion,breslow1974covariance}.
For them, Condition S1(2) holds under three conditions: (i) $\widehat\Lambda_{kg}^{a}(s \mid w, \overline{x}) - \Lambda_{kg}^{a}(s \mid w, \overline{x}) = (\pn - \mathbb{P})\ell_{s,w,\overline{x}} + o_p(n^{-1/2})$ uniformly in $(s, w, \overline{x})$, where the influence processes $s \mapsto \ell_{s,w,\overline{x}}(O)$ have uniformly bounded variation and form a Donsker class; (ii) $\widehat{e}_{kg}$ has uniformly bounded variation in $s$ over each sojourn; and (iii) $\sup|\widehat{e}_{kg}| = o_p(1)$. Integrating by parts over the sojourn in $k$, each term equals $(\pn - \mathbb{P})\widehat{f} + o_p(n^{-1/2})$ with $\widehat{f}(o) = \mathbb{P}\,\E^{\overline{a}}[\int_0^t \widehat{e}_{kg}(s; W, \overline{X}^{\overline{a}}(s-))\, Y_k^{\overline{a}}(s)\, \mathrm{d}_s \ell_{s,W,\overline{X}^{\overline{a}}(s-)}(o) \mid W]$. Since $\|\widehat{f}\|_{L_2} = o_p(1)$, the term is $o_p(n^{-1/2})$ \citep[Lemma 19.24]{van2000asymptotic}. The fitted weight has bounded variation in $s$: the hazard ratios enter at the fixed earlier transition times, and the dependence on $s$ is through cumulative hazards only. Correctly specified Cox models with Breslow estimators converge at rate $O_p(n^{-1/2})$ and satisfy the Donsker condition \citep{zeng2007maximum}, as does the logistic model. Under the Cox model of Section 3.2, the hazard ratios in the weight are functions of the regression coefficients and also converge at the $n^{-1/2}$ rate. Condition S1(3), required uniformly in $t$, also justifies the multiplier-bootstrap simultaneous bands. It holds for the Cox and logistic working models but generally not for flexible machine-learning estimators. With such estimators, cross-fitting should be used; Condition S1(3) is then not needed, and Conditions S1(1)--(2) are required of the cross-fitted estimators.

\paragraph{Estimating the weight with Cox working models.}
Lemma 1 is a statement about the nonparametric model, and Assumption 5 is a condition on the true distribution rather than on the working models. By Equation (13), the weight $\omega_k^{\overline{a}}$ has two kinds of factors. The exponential factors involve only cumulative hazards accumulated over each sojourn, so they can be estimated using Breslow-type estimators. The remaining factors are ratios of rates between arms at the observed earlier transition times. The factor of an earlier transition equals one if that transition is assigned the same arm as $(k, g)$, so these factors matter exactly when the intervention borrows the hazards of an earlier transition and of $(k, g)$ from different arms. Arm-specific Breslow-type estimators cannot estimate such a ratio because the jump times in the numerator and denominator are almost surely different. There are two ways around this.
\begin{itemize}
\item[(i)] \textit{Common baseline hazard.} This is the route taken in Section 3.2. The Cox model (14) with treatment as a covariate gives the hazard ratio $\exp\{\alpha_{kg} + \delta_{kg}^{\T} w + \zeta_{kg}^{\T} z_{kg}(s)\}$. The weight $\omega_k^{\overline{a}}$ then requires only regression coefficients, cumulative hazards, and the censoring model, all estimable at the $n^{-1/2}$ rate. The cost is proportional baseline hazards between arms for every transition, which arm-specific models do not impose.
\item[(ii)] \textit{Smoothing.} Kernel smoothing of the arm-specific Breslow increments, $\widehat{\lambda}_{0,kg}^{a}(s) = \int K_b(s - u) \, \mathrm{d}\widehat{\Lambda}_{0,kg}^{a}(u)$, estimates the baseline ratio without the proportionality restriction. It converges at a nonparametric rate, with boundary bias near $0$ and $t^*$; boundary-corrected kernels should be used when early or late transition times are common. A slower rate for the weight is tolerable when the cumulative hazards themselves are estimated at the $n^{-1/2}$ rate.
\end{itemize}
A third route, developed in Section \ref{sec:extmarkov}, restricts the hazards to depend on the history only through a finite partition of the histories, as under the Markov property. The projected EIF then replaces $\omega_k^{\overline{a}}$ by a ratio of occupancy probabilities that involves only cumulative hazards. Neither a model for the hazard ratio between arms nor smoothing is needed, and the estimator is more efficient in the submodel. The price is the partition as a modeling assumption, which excludes continuous dependence on past transition times such as the time since entry into the current state.

\subsection{Extended Markov model} \label{sec:extmarkov}

This section gives the EIF in a semiparametric submodel in which the hazards depend on the past only through a finite-valued summary of the history, as under the Markov property imposed by the clock-forward Cox models \citep{putter2007tutorial, andersen2012statistical}. We write $h(\overline{x}(t))$ for the class of histories that share the same value of the summary as $\overline{x}(t)$.

\begin{assumption}[Extended Markovness] \label{ass:extmarkov}
There is a partition of the possible histories into finitely many classes, with $h(\overline{x}(t))$ denoting the class containing $\overline{x}(t)$, such that for every $(k,g)\in\mathcal{E}$ and $a \in \{0,1\}$,
\begin{align*}
    \mathrm{d}\Lambda_{kg}^{a}(t \mid w, \overline{x}(t)) &= \mathrm{d}\Lambda_{kg}^{a}(t \mid w, h(\overline{x}(t))),
\end{align*}
and $\pr \{\overline{X}(t) \in h(\overline{x}(t)), T_{\text{c}} \geq t \mid A=a, W\} > \epsilon$ for some constant $\epsilon>0$ on the support of $(t, \overline{x}(t))$ under the hypothetical trajectory $\overline{X}^{\overline{a}}$.
\end{assumption}

Within each class, the transition hazards are unrestricted functions of calendar time $t$ and $w$; across classes they may differ arbitrarily. Assumption \ref{ass:extmarkov} is thus a Markov property on the enlarged state space formed by the classes. Since the hazard of $(k,g)$ is evaluated only at histories with current state $k$, we take each class to consist of histories with the same current state, without loss of generality. Three partitions are of particular interest: (i) \textit{visited states}, under which two histories are in the same class if the same set of states has been visited, so that in Figure 1 the hazard of death depends on whether neither, only E, only M, or both events have occurred; (ii) \textit{ordered visits}, under which the hazard of death may also depend on the order of E and M; and (iii) \textit{visited states and sojourn time}, under which the class additionally records the time since the last transition, discretized into a fixed set of intervals. Partition (iii) accommodates semi-Markov dependence such as elevated mortality shortly after a cardiovascular event. Under Assumption \ref{ass:extmarkov}, the model is a submodel of the nonparametric model. When Assumption 5 also holds, the new EIF is the projection of the nonparametric EIF onto the smaller tangent space. It replaces the history-specific weight $\omega_k^{\overline{a}} \Delta_{kg}^{\overline{a}}$ by its conditional mean among the uncensored units of arm $a_{kg}$ that are in state $k$ with a history in the same class. For a class $c$, let
\begin{align*}
P_k^{\overline{a}}(s \mid w; c) &= \pr\{X^{\overline{a}}(s-) = k, \overline{X}^{\overline{a}}(s-) \in c \mid W = w\}, \\
D_{kg}^{\overline{a}}(t \mid w; c, s) &= \E\{\Delta_{kg}^{\overline{a}}(t \mid w; \overline{X}^{\overline{a}}(s-)) \mid X^{\overline{a}}(s-) = k, \overline{X}^{\overline{a}}(s-) \in c, W = w\}, \\
\pi_k^{a}(s \mid w; c) &= \pr\{X(s-) = k, \overline{X}(s-) \in c \mid A = a, W = w\}
\end{align*}
be the hypothetical probability of occupying state $k$ with a history in class $c$ just before $s$, the average change in the counterfactual CIF caused by a transition from $k$ to $g$ at $s$ over such histories, and the probability of being observed, uncensored, in state $k$ with a history in class $c$ in arm $a$. By Assumption \ref{ass:extmarkov}, $\pi_k^{a}(s \mid w; c) > \epsilon$ whenever $P_k^{\overline{a}}(s \mid w; c) > 0$.

\begin{lemma}[EIF under extended Markovness] \label{lem:extmarkov}
Under Assumptions 1--4 and Assumption \ref{ass:extmarkov}, the efficient influence function of $F_j^{\overline{a}}(t)$ in the semiparametric model defined by Assumption \ref{ass:extmarkov} is
\begin{equation}\label{eif_h}
\begin{aligned}
\EIF\{F_j^{\overline{a}}(t)\} &= F_j^{\overline{a}}(t \mid W) - F_j^{\overline{a}}(t) + \sum_{(k,g)\in\mathcal{E}} \frac{\mathds{1}(A = a_{kg})}{\pr(A = a_{kg} \mid W)} \\
&\quad \times \int_0^t \frac{P_k^{\overline{a}}(s \mid W; h(\overline{X}(s-)))\, D_{kg}^{\overline{a}}(t \mid W; h(\overline{X}(s-)), s)}{\pi_k^{a_{kg}}(s \mid W; h(\overline{X}(s-)))}\, \mathrm{d}M_{kg}(s; a_{kg}, W, h(\overline{X}(s-))).
\end{aligned}
\end{equation}
Equivalently, organized by path,
\begin{equation} \label{eif1}
    \begin{aligned}
    \EIF\{F_j^{\overline{a}}(t)\} &= \sum_{q\in\mathcal{Q}_j} \int_0^t\int_0^{t_{l(q)}}\cdots\int_0^{t_2} \prod_{r= 1}^{l(q)} \mathrm{d} \pr_{r}(t_{r} \mid a_{q(r-1),\bcdot},W,q[\overline{t}_{r}]) \\
    &\quad \times \sum_{r = 1}^{l(q)} \Bigg[ \frac{\mathrm{d} \eta_{q(r-1),q(r)}^{a_{q(r-1),q(r)}}(t_r \mid W,q[\overline{t}_r])}{\mathrm{d} \Lambda_{q(r-1),q(r)}^{a_{q(r-1),q(r)}}(t_r \mid W,q[\overline{t}_r]) } - \sum_{k\in\underline{\mathcal{G}}_{q(r-1)}} \eta_{q(r-1),k}^{a_{q(r-1),k}}(t_r \mid W,q[\overline{t}_r]) \Bigg] \\
    &\quad  + F_j^{\overline{a}}(t \mid W) - F_j^{\overline{a}}(t),
    \end{aligned}
\end{equation}
where
\begin{equation} \label{eq:mart}
\begin{aligned}
    \eta_{q(r-1),k}^{a_{q(r-1),k}}(t \mid W,\overline{x}(t)) & = \frac{\mathds{1}(A=a_{q(r-1),k})}{\pr (A=a_{q(r-1),k}\mid W)} \\
    &\quad \times \int_{t_{r-1}}^t \frac{\mathrm{d} M_{q(r-1),k}(s;a_{q(r-1),k}, W, \overline{x}(s-))}{\pr \{T_{k}\wedge T_{\text{c}}\geq s, \overline{X}(s-) \in h(\overline{x}(s-)) \mid A=a_{q(r-1),k}, W\}},
\end{aligned}
\end{equation}
with $t_{r-1}$ the entry time into the current state $q(r-1)$ recorded in $\overline{x}(t)$ ($t_0 = 0$), and $M_{q(r-1),k}(t; \penalty 0 a_{q(r-1),k}, W, \overline{x}(t))$ the martingale of the transition from $q(r-1)$ to $k$ among individuals whose observed history lies in the same class as $\overline{x}(\cdot)$,
\[
M_{q(r-1),k}(t;a, W, \overline{x}(t)) = \int_0^t \mathds{1}\{\overline{X}(s-)\in h(\overline{x}(s-))\} \left\{\mathrm{d}N_{k}(s) - Y_{\bcdot k}(s) \mathrm{d}\Lambda_{q(r-1),k}^{a}(s\mid W,\overline{x}(s-)) \right\},
\]
with $N_{k}(t) = \Delta_k \mathds{1} \{T_k \leq t\}$ counting entry into $k$ and $Y_{\bcdot k}(t) = \mathds{1} \{T_k \wedge T_{\text{c}} \geq t\}$ indicating being at risk of entering $k$. Because each class records the current state, the denominator equals $\pi_{q(r-1)}^{a}(s \mid W; h(\overline{x}(s-)))$.
\end{lemma}

Lemma \ref{lem:extmarkov} requires Assumption \ref{ass:extmarkov} in place of Assumption 5: the class positivity bounds the weight in \eqref{eif_h}, and no density ratio appears. The hazard $\mathrm{d} \Lambda_{q(r-1),q(r)}^{a_{q(r-1),q(r)}}(t_r \mid W,q[\overline{t}_r])$ in the denominator of \eqref{eif1} cancels with the same factor in $\mathrm{d}\pr_{r}$, so the EIF \eqref{eif_h} involves the hazards only through cumulative hazards and the occupancy probabilities $P_k^{\overline{a}}$ and $\pi_k^{a}$. When both assumptions hold, the nonparametric EIF \eqref{eif_np_s} remains an influence function in the restricted model. The one-step estimator of the main text therefore remains consistent and asymptotically normal there, but \eqref{eif_h} has a smaller variance unless $\omega_k^{\overline{a}} \Delta_{kg}^{\overline{a}}$ is constant within classes. The proof is in Section \ref{sec:proof_lemmaS1}.

\paragraph{Estimation and inference.}
A one-step estimator can be built on \eqref{eif_h} exactly as in Section 3.2. Each transition hazard is fitted by a Cox model with the class indicators as time-varying covariates, separately by arm or pooled with treatment interactions. The fitted cumulative hazards give the plug-in $\widehat{F}_j^{\overline{a}}(t \mid W)$, the hypothetical class occupancies $\widehat{P}_k^{\overline{a}}(s \mid W; c)$, and the class averages $\widehat{D}_{kg}^{\overline{a}}(t \mid W; c, s)$ of the fitted increments under the fitted hypothetical law. The observed occupancies $\widehat\pi_k^{a}(s \mid W; c)$ are computed from the fitted transition and censoring hazards by treating censoring as an additional absorbing state. The estimator is
\begin{equation}\label{cond_h_est}
\begin{aligned}
\widetilde{F}_{j,h}^{\overline{a}}(t) &= \pn \Bigg[ \widehat{F}_j^{\overline{a}}(t \mid W) + \sum_{(k,g)\in\mathcal{E}} \frac{\mathds{1}(A = a_{kg})}{\widehat{\pr}(A = a_{kg} \mid W)} \\
&\qquad\quad \times \int_0^t \frac{\widehat{P}_k^{\overline{a}}(s \mid W; c_s)\, \widehat{D}_{kg}^{\overline{a}}(t \mid W; c_s, s)}{\widehat\pi_k^{a_{kg}}(s \mid W; c_s)}\, \mathrm{d}\widehat{M}_{kg}(s; a_{kg}, W, c_s) \Bigg],
\end{aligned}
\end{equation}
with $c_s = h(\overline{X}(s-))$ and $\mathrm{d}\widehat{M}_{kg}(s; a, W, c) = \mathrm{d}N_{kg}(s) - Y_k(s)\, \mathrm{d}\widehat\Lambda_{kg}^{a}(s \mid W, c)$. Because the class of a history can change with $s$ under partitions such as (iii), all class indicators are evaluated at the left limit at each $s$. All quantities are functionals of cumulative hazards, so the estimator is computable from Breslow estimators without a between-arm proportionality restriction or smoothing.

\begin{condition}[Regularity conditions for \eqref{cond_h_est}] \label{condh}
(1) The fitted propensity score and censoring survival, the fitted class-specific cumulative hazards, the fitted occupancies $\widehat{P}_k^{\overline{a}}$ and $\widehat\pi_k^{a}$, and the fitted class averages $\widehat{D}_{kg}^{\overline{a}}$ are $L_2$-consistent uniformly over $s, t \leq t^*$ and the finitely many classes, and $\widehat\pi_k^{a}(s \mid W; c) > \epsilon/2$ with probability tending to one whenever $P_k^{\overline{a}}(s \mid W; c) > 0$.
(2) The second-order remainder is negligible: for every $(k,g) \in \mathcal{E}$ and class $c$, with $a = a_{kg}$,
\[
\sup_{t \leq t^*} \left| \mathbb{P} \int_0^t \left\{ P_k^{\overline{a}}\widetilde{D}_{kg} - \widehat{P}_k^{\overline{a}}\widehat{D}_{kg}\widehat\rho_{kg} \right\}(s \mid W; c)\, \mathrm{d}\left\{\widehat\Lambda_{kg}^{a} - \Lambda_{kg}^{a}\right\}(s \mid W, c) \right| = o_p(n^{-1/2}),
\]
where $\widetilde{D}_{kg}$ is the class average of the fitted increment under the true hypothetical law and $\widehat\rho_{kg} = \pr(A = a \mid W)\, \pi_k^{a} / \{\widehat\pr(A = a \mid W)\, \widehat\pi_k^{a}\}$. The expression in braces is the coefficient in the exact bias identity proved in Section \ref{sec:proof_thmS1}. It is a bounded factor times the errors of the fitted propensity score, censoring survival, occupancies, and class averages, so the condition is a product-rate condition between these and the fitted cumulative hazard of $(k,g)$.
(3) The function $\{\widehat{F}_j^{\overline{a}}(t \mid W) + \widehat\varphi_{j,h}^{\overline{a}}(t; O): t \in [0, t^*]\}$, with $\widehat\varphi_{j,h}^{\overline{a}}$ the debiasing term of \eqref{cond_h_est}, belongs to a Donsker class.
\end{condition}

With Cox models and Breslow estimators, the fitted occupancies and class averages are smooth functionals of the fitted cumulative hazards and inherit their $n^{-1/2}$ rate and Donsker property, and the integration-by-parts argument shows that Condition \ref{condh}(2) holds when the working models are correctly specified. Cross-fitting can replace Condition \ref{condh}(3).

\begin{theorem}[Asymptotic normality and efficiency in the extended Markov model] \label{thmS1}
Suppose that Assumptions 1--4, Assumption \ref{ass:extmarkov}, and Condition \ref{condh} hold. Then $\widetilde{F}_{j,h}^{\overline{a}}(\cdot)$ weakly converges to a Gaussian process on $[0, t^*]$, with
\[
\sqrt{n} \{\widetilde{F}_{j,h}^{\overline{a}}(t) - F_j^{\overline{a}}(t)\} \rightsquigarrow \mathcal{N}\left(0, \ \E[\psi_h(O)]^2\right),
\]
where $\psi_h$ is the right-hand side of \eqref{eif_h}. The asymptotic variance attains the semiparametric efficiency bound for $F_j^{\overline{a}}(t)$ in the model defined by Assumption \ref{ass:extmarkov}, and is no larger than the variance in Theorem 3.
\end{theorem}

The proof is in Section \ref{sec:proof_thmS1}. The estimator \eqref{cond_h_est} is more efficient than that of the main text whenever the extended Markov restriction holds. However, the partition becomes part of the working model, and we cannot make a multiple robustness claim for \eqref{cond_h_est}.

\paragraph{Illness--death example.}
Let $\mathcal{G} = \{\text{O}, \text{I}, \text{D}\}$, $\mathcal{E} = \{(\text{O},\text{I}), (\text{O},\text{D}), (\text{I},\text{D})\}$, $\overline{a} = (a_{\text{OI}}, a_{\text{OD}}, a_{\text{ID}})$, and suppose that the hazards are Markov, $\mathrm{d}\Lambda_{kg}^{a}(t \mid w, \overline{x}(t)) = \mathrm{d}\Lambda_{kg}^{a}(t \mid w)$, and that censoring depends on $w$ only, with $G^{a}(s \mid w) = \pr(T_{\text{c}} \geq s \mid A = a, W = w)$. Write
\[
P_{\text{O}}^{\overline{a}}(s \mid w) = \exp\{-\Lambda_{\text{OI}}^{a_{\text{OI}}}(s \mid w) - \Lambda_{\text{OD}}^{a_{\text{OD}}}(s \mid w)\}, \quad P_{\text{O}}^{a}(s \mid w) = \exp\{-\Lambda_{\text{OI}}^{a}(s \mid w) - \Lambda_{\text{OD}}^{a}(s \mid w)\}
\]
for the probabilities of remaining in O under the intervention and in arm $a$, and let $F_{\text{D}}^{\overline{a}}(t \mid w; k, s)$ be the conditional probability of death by time $t$ given occupancy of state $k$ at time $s$ under the intervention:
\begin{align*}
F_{\text{D}}^{\overline{a}}(t \mid w; \text{D}, s) &= 1, \quad
F_{\text{D}}^{\overline{a}}(t \mid w; \text{I}, s) = 1 - \exp\{-\Lambda_{\text{ID}}^{a_{\text{ID}}}(t \mid w) + \Lambda_{\text{ID}}^{a_{\text{ID}}}(s \mid w)\}, \\
F_{\text{D}}^{\overline{a}}(t \mid w; \text{O}, s) &= \int_s^t \frac{P_{\text{O}}^{\overline{a}}(u \mid w)}{P_{\text{O}}^{\overline{a}}(s \mid w)} \left\{ \mathrm{d}\Lambda_{\text{OD}}^{a_{\text{OD}}}(u \mid w) + F_{\text{D}}^{\overline{a}}(t \mid w; \text{I}, u) \, \mathrm{d}\Lambda_{\text{OI}}^{a_{\text{OI}}}(u \mid w) \right\}.
\end{align*}
Because the hazards are Markov, the counterfactual CIF given a history depends on the history only through the current state, so $\Delta_{kg}^{\overline{a}}(t \mid w; \overline{x}(s-)) = F_{\text{D}}^{\overline{a}}(t \mid w; g, s) - F_{\text{D}}^{\overline{a}}(t \mid w; k, s)$. A history in state O at time $s$ is determined by $s$, and a history in state I by the entry time $u < s$. From Equation (13), with $a = a_{\text{ID}}$ for state I,
\[
\omega_{\text{O}}^{\overline{a}}(s; a, w) = \frac{P_{\text{O}}^{\overline{a}}(s \mid w)}{P_{\text{O}}^{a}(s \mid w)\, G^{a}(s \mid w)}, \quad
\omega_{\text{I}}^{\overline{a}}(s; a, w, u) = \frac{P_{\text{O}}^{\overline{a}}(u \mid w)}{P_{\text{O}}^{a}(u \mid w)\, G^{a}(s \mid w)} \cdot \frac{\lambda_{\text{OI}}^{a_{\text{OI}}}(u \mid w)}{\lambda_{\text{OI}}^{a}(u \mid w)}.
\]
The survival factor $\exp\{-\Lambda_{\text{ID}}^{a}(s \mid w) + \Lambda_{\text{ID}}^{a}(u \mid w)\}$ for the sojourn in I has canceled, because $(\text{I},\text{D})$ is drawn from arm $a_{\text{ID}}$ under both laws. What remains is the ratio of the densities of entering I at $u$. It contains the ratio of the rates of $(\text{O},\text{I})$ between arms at the unit's own entry time whenever $a_{\text{OI}} \neq a_{\text{ID}}$. On the observed-data side, let $N_{\text{OI}}(s) = \Delta_{\text{I}} \mathds{1}(T_{\text{I}} \leq s)$, $N_{\text{OD}}(s) = (1 - \Delta_{\text{I}}) \Delta_{\text{D}} \mathds{1}(T_{\text{D}} \leq s)$, and $N_{\text{ID}}(s) = \Delta_{\text{I}} \Delta_{\text{D}} \mathds{1}(T_{\text{D}} \leq s)$ count the three transitions. Let $Y_{\text{O}}(s) = \mathds{1}(T_{\text{I}} \wedge T_{\text{D}} \wedge T_{\text{c}} \geq s)$ and $Y_{\text{I}}(s) = \Delta_{\text{I}} \mathds{1}(T_{\text{I}} < s \leq T_{\text{D}} \wedge T_{\text{c}})$ indicate being observed in O and in I, and let $M_{kg}(s; a, w) = N_{kg}(s) - \int_0^s Y_k(u) \, \mathrm{d}\Lambda_{kg}^{a}(u \mid w)$ be the transition-specific martingales. Then the nonparametric EIF \eqref{eif_np_s} reads
\begin{equation} \label{eif_id}
\begin{aligned}
\psi(O) &= F_{\text{D}}^{\overline{a}}(t \mid W) - F_{\text{D}}^{\overline{a}}(t) + \sum_{(k,g) \in \mathcal{E}} \frac{\mathds{1}(A = a_{kg})}{\pr(A = a_{kg} \mid W)} \\
&\quad \times \int_0^t \omega_k^{\overline{a}}(s; a_{kg}, W, T_{\text{I}}) \left\{ F_{\text{D}}^{\overline{a}}(t \mid W; g, s) - F_{\text{D}}^{\overline{a}}(t \mid W; k, s) \right\} \mathrm{d}M_{kg}(s; a_{kg}, W),
\end{aligned}
\end{equation}
where $\omega_{\text{O}}^{\overline{a}}$ does not depend on $T_{\text{I}}$. When all components of $\overline{a}$ equal $a$, every weight reduces to $1/G^{a}(s \mid W)$, and \eqref{eif_id} is the EIF of the potential CIF $F_{\text{D}}^{a}(t)$. Under the Markov property, take the classes in Assumption \ref{ass:extmarkov} to be the current states. Let $\pi_k^{a}(s \mid w) = \pr\{Y_k(s) = 1 \mid A = a, W = w\}$ be the probability of being observed in state $k$ at time $s$ in arm $a$, and let $P_{\text{I}}^{\overline{a}}(s \mid w) = \int_0^s P_{\text{O}}^{\overline{a}}(u \mid w) \exp\{-\Lambda_{\text{ID}}^{a_{\text{ID}}}(s \mid w) + \Lambda_{\text{ID}}^{a_{\text{ID}}}(u \mid w)\} \, \mathrm{d}\Lambda_{\text{OI}}^{a_{\text{OI}}}(u \mid w)$ be the occupancy of I under the intervention. Then \eqref{eif_h} reduces to
\begin{equation} \label{eif_id_h}
\begin{aligned}
\psi_h(O) &= F_{\text{D}}^{\overline{a}}(t \mid W) - F_{\text{D}}^{\overline{a}}(t) + \sum_{(k,g) \in \mathcal{E}} \frac{\mathds{1}(A = a_{kg})}{\pr(A = a_{kg} \mid W)} \\
&\quad \times \int_0^t \frac{P_k^{\overline{a}}(s \mid W)}{\pi_k^{a_{kg}}(s \mid W)} \left\{ F_{\text{D}}^{\overline{a}}(t \mid W; g, s) - F_{\text{D}}^{\overline{a}}(t \mid W; k, s) \right\} \mathrm{d}M_{kg}(s; a_{kg}, W).
\end{aligned}
\end{equation}
In \eqref{eif_id}, the history-specific weight $\omega_{\text{I}}^{\overline{a}}(s; a, W, T_{\text{I}})$ depends on the unit's own entry time through the ratio of rates of $(\text{O}, \text{I})$ between arms. In \eqref{eif_id_h} it is replaced by $P_{\text{I}}^{\overline{a}}(s \mid W) / \pi_{\text{I}}^{a}(s \mid W)$, its average over the entry times of the units of arm $a$ in state I at $s$. The two weights coincide when $a_{\text{OI}} = a_{\text{OD}} = a_{\text{ID}}$. The derivations are in Section \ref{sec:idm}.

\subsection{Shapley order-averaged effects} \label{sec:shapley}

The targeted effects of Section 2.3 depend on the reference regime: switching the transitions in $\mathcal{S}$ from the control-arm to the treated-arm hazards has a different effect when the other transitions are at their control-arm hazards ($b = 0$) than when they are at their treated-arm hazards ($b = 1$). The transition-targeted effects with either reference do not add up to the total effect. The Shapley value \citep{shapley1953value} removes the dependence on the reference by averaging the incremental effect of a set of transitions over all orders in which the transitions can be switched.

Partition the transitions into $K$ disjoint groups $\mathcal{E} = \mathcal{S}_1 \cup \cdots \cup \mathcal{S}_K$, the \textit{players}. For a coalition $\mathcal{C} \subseteq \{1, \ldots, K\}$, let $\overline{a}(\mathcal{C})$ be the intervention vector that assigns the treated-arm hazard to every transition in $\bigcup_{k \in \mathcal{C}} \mathcal{S}_k$ and the control-arm hazard to all other transitions, and let $v_j(t; \mathcal{C}) = F_j^{\overline{a}(\mathcal{C})}(t)$. The Shapley order-averaged effect of player $k$ on the cumulative incidence of $j$ is
\begin{equation} \label{eq:shapley}
\phi_j^{k}(t) = \sum_{\mathcal{C} \subseteq \{1, \ldots, K\} \setminus \{k\}} \frac{|\mathcal{C}|!\,(K - |\mathcal{C}| - 1)!}{K!} \left\{ v_j(t; \mathcal{C} \cup \{k\}) - v_j(t; \mathcal{C}) \right\},
\end{equation}
the average, over the $K!$ orders in which the players can be switched from control to treated, of the change in the cumulative incidence when player $k$ is switched. Each term in \eqref{eq:shapley} is a contrast of the form $\tau_j(t; \overline{a}, \overline{a}^*)$ of Section 3.4. The Shapley effects satisfy
\begin{equation} \label{eq:shapley_sum}
\sum_{k=1}^{K} \phi_j^{k}(t) = F_j^{\overline{1}}(t) - F_j^{\overline{0}}(t),
\end{equation}
so they decompose the total effect exactly. The same definition applies to the restricted mean time lost, $\int_0^{t^*} F_j^{\overline{a}}(s)\, \mathrm{d}s$, in place of $F_j^{\overline{a}}(t)$.

Because $\phi_j^{k}(t)$ is a linear combination of counterfactual cumulative incidences under the $2^K$ regimes $\overline{a}(\mathcal{C})$, its one-step estimator is the same linear combination of the one-step estimators $\widetilde{F}_j^{\overline{a}(\mathcal{C})}(t)$, and its influence function is the same linear combination of the efficient influence functions of Lemma 1. The standard error is the standard deviation of the estimated influence function divided by $\sqrt{n}$, and the $P$-value for $H_0: \phi_j^{k}(t) = 0$, or for the corresponding restricted-mean contrast, follows from the normal approximation, as in Corollaries 1 and 2. In LEADER, the seven labels require $2^7 = 128$ regimes, from which both the transition- and the path-targeted Shapley effects are computed.

\section{Frailty-indexed sensitivity analysis} \label{sec:density_est}

\subsection{Estimand} \label{sec:fr_estimand}

The estimands of the main text apply the intervention to the observed-history hazards, which average over latent heterogeneity among patients with the same observed history. To examine how the conclusions change if the intervention is instead applied conditionally on a latent patient-level variable, we use a shared frailty \citep{xu2010statistical, ha2020frailty, li2020shared}. Let $\xi$ be a frailty, independent of $(A, W)$, with a gamma distribution with mean one, variance $\nu$, and density $\phi_\nu$, shared by all transitions of a patient. The transition hazard given the history and the frailty is
\[
    \mathrm{d}\Lambda_{kg}^{a}(t \mid w,\overline{x}(t), \xi) = \xi \, \mathrm{d}\Lambda_{kg}^{a}(t \mid w,\overline{x}(t); \nu),
\]
where $\Lambda_{kg}^{a}(\cdot \mid w,\overline{x}; \nu)$ is a conditional cumulative hazard, with rate $\lambda_{kg}^{a}(\cdot \mid w, \overline{x}; \nu)$, that differs from the observed-history hazard $\Lambda_{kg}^{a}(\cdot \mid w, \overline{x})$ of the main text unless $\nu = 0$. The frailty enters the event hazards only; the censoring hazards and the propensity score are those of the main text. The intervention draws each transition from the conditional hazard of the arm assigned to it, at the patient's frailty: given $\xi$, the hypothetical process has the intensities of Equation (3) with $\xi \, \mathrm{d}\Lambda_{kg}^{a_{kg}}(\cdot \mid w, \overline{x}; \nu)$ in place of $\mathrm{d}\Lambda_{kg}^{a_{kg}}(\cdot \mid w, \overline{x})$. Let $F_j^{\overline{a}}(t \mid w, \xi; \nu)$ be its cumulative incidence of $j$ given $(W, \xi) = (w, \xi)$, given by Theorem 1 with the conditional hazards. The frailty-indexed estimand is
\begin{equation}\label{eq:fr_estimand}
    F_j^{\overline{a}}(t; \nu) = \int_{\mathcal{W}} \int_0^\infty F_j^{\overline{a}}(t \mid w, \xi; \nu)\, \phi_\nu(\xi)\, \mathrm{d}\xi \, \mathrm{d}\pr(w).
\end{equation}
Coupling the arms through the same frailty keeps a patient's frailty fixed across arm assignments. Throughout, $\nu$ is treated as fixed. Each $\nu$ defines a different estimand, which at $\nu = 0$ is that of the main text. The parameter $\nu$ indexes how much latent heterogeneity the intervention holds fixed; it does not adjust for confounding.

\subsection{Identification} \label{sec:fr_ident}

For $n = 0, 1, \ldots$ and $L \geq 0$, let $\mathcal{L}_\nu^{(n)}(L) = \int_0^\infty \xi^{n} \mathrm{e}^{-\xi L} \phi_\nu(\xi)\, \mathrm{d}\xi$, and let
\[
    m_\nu(n, L) = \frac{\mathcal{L}_\nu^{(n+1)}(L)}{\mathcal{L}_\nu^{(n)}(L)}, \quad \kappa_\nu(n, L) = \frac{\partial}{\partial L} \log m_\nu(n, L),
\]
so that $m_\nu(n, L)$ is the posterior mean of the frailty of a patient who has made $n$ transitions and accumulated the conditional hazard $L$; $m_\nu$ is decreasing in $L$, so $\kappa_\nu < 0$. For a history $\overline{x}$ with $N(s-)$ transitions before $s$, let
\begin{equation}\label{eq:fr_cumtot}
    \Lambda^{a}_{\bullet}(s \mid w, \overline{x}; \nu) = \int_0^{s} \sum_{g \in \underline{\mathcal{G}}_{x(u-)}} \lambda_{x(u-), g}^{a}(u \mid w, \overline{x}(u-); \nu)\, \mathrm{d}u
\end{equation}
be the total conditional hazard accumulated along $\overline{x}$ in arm $a$, and write $\mu^{a}(s \mid w, \overline{x}) = m_\nu\{N(s-), \penalty 0  \Lambda^{a}_{\bullet}(s \mid w, \overline{x}; \nu)\}$ and $\kappa^{a}(s \mid w, \overline{x}) = \kappa_\nu\{N(s-), \Lambda^{a}_{\bullet}(s \mid w, \overline{x}; \nu)\}$. Because the frailty is shared by all transitions and independent of $(A, W)$, the observed-history hazard in arm $a$ is the conditional hazard times the posterior mean of the frailty given the history,
\begin{equation}\label{eq:fr_demix}
    \lambda_{kg}^{a}(s \mid w, \overline{x}(s-)) = \mu^{a}(s \mid w, \overline{x})\, \lambda_{kg}^{a}(s \mid w, \overline{x}(s-); \nu).
\end{equation}
Since $\mu^{a}(s \mid w, \overline{x})$ depends on the conditional hazards only through their integral \eqref{eq:fr_cumtot} up to $s$, Equation \eqref{eq:fr_demix} is a Volterra equation for $\lambda^{a}(\cdot; \nu)$ along each history, with a unique solution for every $\nu$. The conditional hazards are therefore a one-to-one transformation of the observed-history hazards. With $\nu$ fixed, the frailty model restricts neither the observed law nor the potential hazards of the main text. Hence $F_j^{\overline{a}}(t; \nu)$ is identified under Assumptions 1--4 strengthened to hold conditionally on $\xi$, that is, randomization and censoring may not depend on $\xi$ beyond $W$ and the observed history.

Integrating the frailty out of the hypothetical process shows that $F_j^{\overline{a}}(t; \nu)$ is the cumulative incidence of a process of the form of Equation (3) whose history-dependent hazards are
\begin{equation}\label{eq:fr_remix}
    \lambda_{kg}^{\overline{a}}(s \mid w, \overline{x}(s-); \nu) = \frac{\mu^{\overline{a}}(s \mid w, \overline{x})}{\mu^{a_{kg}}(s \mid w, \overline{x})}\, \lambda_{kg}^{a_{kg}}(s \mid w, \overline{x}(s-)),
\end{equation}
where $\mu^{\overline{a}}(s \mid w, \overline{x}) = m_\nu\{N(s-), \Lambda^{\overline{a}}_{\bullet}(s \mid w, \overline{x}; \nu)\}$ and $\Lambda^{\overline{a}}_{\bullet}$ is \eqref{eq:fr_cumtot} with the conditional hazard of arm $a_{x(u-), g}$ for each transition. The frailty thus modifies the stochastic intervention of the main text by the factor $\mu^{\overline{a}} / \mu^{a_{kg}}$, the ratio of the posterior means of the frailty of a patient with history $\overline{x}$ under the regime and in arm $a_{kg}$. When all components of $\overline{a}$ equal $a$, the two accumulated hazards coincide, the factor equals one, and $F_j^{\overline{a}}(t; \nu) = F_j^{a}(t)$ for every $\nu$. The potential cumulative incidences and the total effect therefore do not depend on $\nu$. Only the mixed regimes, and hence the transition-, event-, and path-targeted effects, are affected by the frailty.

\subsection{Efficient influence function and one-step estimator} \label{sec:fr_eif}

Since the frailty model with fixed $\nu$ is a reparameterization of the nonparametric model, $F_j^{\overline{a}}(t; \nu)$ has a unique influence function there. It is obtained by the argument of Lemma 1 with one additional term. By \eqref{eq:fr_demix}, a perturbation of the observed-history hazard of a transition at time $u$ changes the conditional hazards of all later transitions along the same history. Let $p^{\overline{a}}_\xi(s, \overline{x}(s-) \mid w)$ and $p^{a}_\xi(s, \overline{x}(s-) \mid w)$ be the densities of the history under the hypothetical process given $\xi$ and under arm $a$ given $\xi$, computed as in Equation (10) with the hazards $\xi \lambda^{a_{kg}}_{kg}(\cdot; \nu)$ and $\xi \lambda^{a}_{kg}(\cdot; \nu)$. Let $\Delta_{kg}(t \mid w, \xi; \overline{x}(s-))$ be the increment of $F_j^{\overline{a}}(t \mid w, \xi; \nu)$ caused by a transition from $k$ to $g$ at $s$, as in Equation (9). Let
\begin{equation}\label{eq:fr_prop}
    \Pi^{a}(u, v \mid w, \overline{x}) = \exp\left\{ -\int_u^v \kappa^{a}(r \mid w, \overline{x}) \sum_{g \in \underline{\mathcal{G}}_{x(r-)}} \lambda^{a}_{x(r-), g}(r \mid w, \overline{x}(r-); \nu)\, \mathrm{d}r \right\} \geq 1,
\end{equation}
which propagates a change of the conditional hazards at time $u$ to the de-mixing factor at time $v$ along $\overline{x}$; within a sojourn preceded by $n$ transitions, $\Pi^{a}(u, v) = m_\nu\{n, \Lambda^{a}_{\bullet}(u)\} / m_\nu\{n, \Lambda^{a}_{\bullet}(v)\}$. Define the expected future contribution of the de-mixing,
\begin{equation}\label{eq:fr_gamma}
\begin{aligned}
    \Gamma^{a}(s \mid w, \xi; \overline{x}) &= \E^{\overline{a}}_{\xi}\Bigg[ \sum_{(k',g'): a_{k'g'} = a} \int_s^t \kappa^{a}(v \mid w, \overline{X})\, \Pi^{a}(s, v \mid w, \overline{X})\, \Delta_{k'g'}(t \mid w, \xi; \overline{X}(v-)) \\
    &\qquad\qquad \times Y_{k'}^{\overline{a}}(v)\, \xi \lambda^{a}_{k'g'}(v \mid w, \overline{X}(v-); \nu)\, \mathrm{d}v \Biggm| \overline{X}(s-) = \overline{x}, W = w, \xi \Bigg],
\end{aligned}
\end{equation}
the expectation being over the hypothetical process given $\xi$ continued from $\overline{x}$ at time $s$.

\begin{lemma}[EIF of the frailty-indexed estimand] \label{lem:frailty}
Suppose that Assumptions 1--5 hold conditionally on $\xi$, that $\xi \independent (A, W)$, and that $\nu$ is fixed. The efficient influence function of $F_j^{\overline{a}}(t; \nu)$ in the nonparametric model is
\begin{equation}\label{eq:fr_eif}
\begin{aligned}
    \psi_\nu(O) &= F_j^{\overline{a}}(t \mid W; \nu) - F_j^{\overline{a}}(t; \nu) \\
    &\quad + \sum_{a \in \{0, 1\}} \frac{\mathds{1}(A = a)}{\pr(A = a \mid W)} \sum_{(k,g) \in \mathcal{E}} \int_0^t H^{a}_{kg}(s \mid W; \overline{X}(s-))\, \mathrm{d}M_{kg}(s; a, W, \overline{X}(s-)),
\end{aligned}
\end{equation}
where $F_j^{\overline{a}}(t \mid w; \nu) = \int_0^\infty F_j^{\overline{a}}(t \mid w, \xi; \nu) \phi_\nu(\xi)\, \mathrm{d}\xi$, $M_{kg}(s; a, W, \overline{X}(s-))$ is the martingale, whose compensator is the observed-history hazard $\Lambda_{kg}^{a}(\cdot \mid W, \overline{X}(s-))$, and
\begin{equation}\label{eq:fr_H}
    H^{a}_{kg}(s \mid w; \overline{x}) = \frac{\int_0^\infty \left\{ \mathds{1}(a_{kg} = a)\, \xi\, \Delta_{kg}(t \mid w, \xi; \overline{x}) - \Gamma^{a}(s \mid w, \xi; \overline{x}) \right\} p^{\overline{a}}_\xi(s, \overline{x} \mid w)\, \phi_\nu(\xi)\, \mathrm{d}\xi}{G^{a}(s \mid w, \overline{x}) \int_0^\infty \xi\, p^{a}_\xi(s, \overline{x} \mid w)\, \phi_\nu(\xi)\, \mathrm{d}\xi}.
\end{equation}
\end{lemma}

The proof is in Section \ref{sec:proof_frailty}. The denominator of \eqref{eq:fr_H} equals $G^{a}\, \mu^{a}\, p^{a}$, with $p^{a}$ the observed-history density of arm $a$. Hence $H^{a}_{kg}$ is the weight $\omega_k^{\overline{a}} \Delta_{kg}^{\overline{a}}$ of Lemma 1 with two replacements: the density ratio $p^{\overline{a}} / p^{a}$ becomes a ratio of frailty mixtures, and the increment $\Delta_{kg}^{\overline{a}}$ becomes the mixture-weighted conditional increment $\xi \Delta_{kg}(\cdot \mid \xi)$ minus the de-mixing contribution $\Gamma^{a}$. Three features differ from Lemma 1. First, the residual of a transition $(k, g)$ receives contributions from both arms, not only from $a_{kg}$. In arm $a \neq a_{kg}$ the direct term vanishes, but $\Gamma^{a}$ is nonzero whenever a transition that can still occur after $s$ is borrowed from arm $a$, because the hazard of $(k, g)$ in arm $a$ enters the de-mixing of that transition. Second, $\Gamma^{a}$ is an expectation over the future of the hypothetical process, so the weight at time $s$ depends on the hazards after $s$ through $\kappa^{a}$ and $\Pi^{a}$ as well as through $\Delta$. Third, as $\nu \to 0$, $m_\nu \to 1$, $\kappa_\nu \to 0$, $\Gamma^{a} \to 0$, and \eqref{eq:fr_eif} reduces to Lemma 1. When all components of $\overline{a}$ equal $a$, the estimand is $F_j^{a}(t)$ by Section \ref{sec:fr_ident}, and \eqref{eq:fr_eif} coincides with its EIF for every $\nu$ by the uniqueness of the influence function in the nonparametric model.

The one-step estimator is
\begin{equation}\label{eq:fr_onestep}
\begin{aligned}
    \widetilde{F}_j^{\overline{a}}(t; \nu) &= \pn\bigg[ \widehat{F}_j^{\overline{a}}(t \mid W; \nu) \\
    &\qquad + \sum_{a \in \{0, 1\}} \frac{\mathds{1}(A = a)}{\widehat{\pr}(A = a \mid W)} \sum_{(k,g) \in \mathcal{E}} \int_0^t \widehat{H}^{a}_{kg}(s \mid W; \overline{X}(s-))\, \mathrm{d}\widehat{M}_{kg}(s; a, W, \overline{X}(s-)) \bigg],
\end{aligned}
\end{equation}
where the nuisance functions are the conditional hazards $\widehat\lambda^{a}_{kg}(\cdot; \nu)$ for both arms, fitted by a frailty working model with $\nu$ fixed, the observed-history hazards $\widehat\mu^{a} \widehat\lambda^{a}_{kg}(\cdot; \nu)$ implied by \eqref{eq:fr_demix}, which define the martingale residuals, and the fitted censoring survival and propensity score. For the gamma frailty, $\mathcal{L}_\nu^{(n)}(L) \propto (1/\nu + L)^{-(1/\nu + n)}$, so that $m_\nu(n, L) = (1/\nu + n)/(1/\nu + L)$ and $\kappa_\nu(n, L) = -1/(1/\nu + L)$. Hence the posterior mean of the frailty is $\mu^{a}(s \mid w, \overline{x}) = \{1/\nu + N(s-)\}/\{1/\nu + \Lambda^{a}_{\bullet}(s \mid w, \overline{x}; \nu)\}$, and the propagator \eqref{eq:fr_prop} is $\Pi^{a}(u, v \mid w, \overline{x}) = \{1/\nu + \Lambda^{a}_{\bullet}(v \mid w, \overline{x}; \nu)\}/\{1/\nu + \Lambda^{a}_{\bullet}(u \mid w, \overline{x}; \nu)\}$, also across transitions. The product $\kappa^{a}(v \mid w, \overline{x})\, \Pi^{a}(s, v \mid w, \overline{x}) = -1/\{1/\nu + \Lambda^{a}_{\bullet}(s \mid w, \overline{x}; \nu)\}$ does not depend on $v$ or on the history after $s$. Therefore \eqref{eq:fr_gamma} reduces to
\begin{equation}\label{eq:fr_gamma_gamma}
\begin{aligned}
    \Gamma^{a}(s \mid w, \xi; \overline{x}) &= -\frac{R^{a}(s \mid w, \xi; \overline{x})}{1/\nu + \Lambda^{a}_{\bullet}(s \mid w, \overline{x}; \nu)}, \\
    R^{a}(s \mid w, \xi; \overline{x}) &= \E^{\overline{a}}_{\xi}\Bigg[ \sum_{(k',g'): a_{k'g'} = a} \int_s^t \Delta_{k'g'}(t \mid w, \xi; \overline{X}(v-))\, \\
    &\qquad\qquad\qquad\qquad Y_{k'}^{\overline{a}}(v)\, \xi \lambda^{a}_{k'g'}(v \mid w, \overline{X}(v-); \nu)\, \mathrm{d}v \Biggm| \overline{X}(s-) = \overline{x}, W = w, \xi \Bigg],
\end{aligned}
\end{equation}
where $R^{a}$ is the expected future change in the counterfactual cumulative incidence that is compensated through the transitions borrowed from arm $a$, and $\Lambda^{a}_{\bullet}(s \mid w, \overline{x}; \nu)$ is accumulated along the patient's own history before $s$. Suppose that the conditional hazards depend on the history only through a finite partition of the histories, as in Section \ref{sec:extmarkov}. Then $R^{a}$ depends on $\overline{x}$ only through its class, and it is computed by the backward recursion that gives $F_j^{\overline{a}}(t \mid w, \xi; \overline{x})$, with $\Delta_{k'g'}\, \xi \lambda^{a}_{k'g'}$ as a running reward; no augmentation of the state is needed. The integrals over $\xi$ in $\widehat{F}_j^{\overline{a}}(t \mid W; \nu)$ and $\widehat{H}^{a}_{kg}$ are computed by Gauss--Laguerre quadrature for the gamma distribution. At each node $\xi_q$, $\widehat{F}_j^{\overline{a}}(t \mid W, \xi_q; \nu)$, $\widehat\Delta_{kg}(\cdot \mid \xi_q)$, and $\widehat{R}^{a}(\cdot \mid \xi_q)$ are obtained from backward recursions with the hazards $\xi_q \widehat\lambda(\cdot; \nu)$, and $\widehat{p}^{\overline{a}}_{\xi_q}$ and $\widehat{p}^{a}_{\xi_q}$ from the product formula along the patient's own history. Under the analogue of Condition \ref{cond2}, with the product-rate condition applying to the errors of the fitted $\widehat\mu^{a}$ and $\widehat{R}^{a}$ as well as of the fitted weights, the argument of Theorem 3 gives $\sqrt{n}\{\widetilde{F}_j^{\overline{a}}(t; \nu) - F_j^{\overline{a}}(t; \nu)\} \rightsquigarrow \mathcal{N}(0, \E[\psi_\nu(O)]^2)$. The variance attains the nonparametric efficiency bound for the functional \eqref{eq:fr_estimand}. For a constant regime, \eqref{eq:fr_onestep} has the same influence function as the one-step estimator of $F_j^{a}(t)$ in the main text, so the two are asymptotically equivalent for every $\nu$. The plug-in estimator $\pn \widehat{F}_j^{\overline{a}}(t \mid W; \nu)$, by contrast, depends on $\nu$ through the frailty working model. If $\nu$ is estimated rather than fixed, the estimand changes with $\widehat\nu$, and the variability of $\widehat\nu$ is not accounted for.

\section{Simulation studies} \label{sec:simu}

We evaluate the one-step estimator of Section 3.2 in the structure of Figure 1, with two intermediate events, $\text{m}_1$ and $\text{m}_2$, and a terminal event $\text{d}$. Let $W = (W_1,W_2,W_3)^{\T}$ be three independent $\operatorname{Bernoulli}(0.5)$ covariates. The potential transition hazards under treatment $a \in \{0,1\}$ are proportional hazards given the covariates and the visited states, with a baseline common to the two arms and a treatment effect that may depend on the covariates and on the history, as in the working model (14):
\begin{align*}
\lambda_{\text{m}_1}^a(t \mid W,T_{\text{m}_2}^a) &= (a_1 t + b_1) \exp(\alpha_1 a + W^{\T}\beta_1^a) \cdot (\delta_{21}^a)^{\mathds{1}\{T_{\text{m}_2}^a<t\}}, \\
\lambda_{\text{m}_2}^a(t \mid W,T_{\text{m}_1}^a) &= (a_2 t + b_2) \exp(\alpha_2 a + W^{\T}\beta_2^a) \cdot (\delta_{12}^a)^{\mathds{1}\{T_{\text{m}_1}^a<t\}}, \\
\lambda_{\text{d}}^a(t \mid W,T_{\text{m}_1}^a,T_{\text{m}_2}^a) &= (a_3 t + b_3) \exp(\alpha_3 a + W^{\T} \beta_3^a) \cdot (\delta_{13}^a)^{\mathds{1}\{T_{\text{m}_1}^a<t\}} (\delta_{23}^a)^{\mathds{1}\{T_{\text{m}_2}^a<t\}},
\end{align*}
where the hazard of each transition is accumulated from the entry into the current state, and
\begin{align*}
&a_1 = 0.02, \ b_1 = 0.05, \ \alpha_1 = -0.4, \ \beta_1^1 = (0.2,0.2,-0.2)^{\T}, \ \beta_1^0 = (0.1,0.1,0.0)^{\T}, \ \delta_{21}^1 = \delta_{21}^0 = 1.2, \\
&a_2 = 0.02, \ b_2 = 0.02, \ \alpha_2 = 0, \ \beta_2^1 = \beta_2^0 = (0.2,-0.2,-0.2)^{\T}, \ \delta_{12}^1 = 1.5, \ \delta_{12}^0 = 0.5, \\
&a_3 = 0.015, \ b_3 = 0.01, \ \alpha_3 = 0.5, \ \beta_3^1 = (0.1,0.1,0.1)^{\T}, \ \beta_3^0 = (0.1,0.0,0.1)^{\T}, \\
& \delta_{13}^1 = \delta_{13}^0 = 1, \ \delta_{23}^1 = 1.5, \ \delta_{23}^0 = 0.5.
\end{align*}
The treatment lowers the hazard of $\text{m}_1$ and raises that of $\text{d}$; it does not change the hazard of $\text{m}_2$ from the initial state but reverses the effect of a preceding $\text{m}_1$ on it ($\delta_{12}^a$), and likewise reverses the effect of a preceding $\text{m}_2$ on the hazard of death ($\delta_{23}^a$). The hazard of censoring is
\[
\lambda_{\text{c}}^a(t \mid W) = (0.05-0.01a)\,(t-6)_{+}^2 \exp(-0.1W_1-0.2aW_2), \quad t \leq 20,
\]
with administrative censoring of all remaining subjects at $t = 20$; about a third of the subjects are censored before $t = 12$. The sample size is $n = 1000$, and treatment is assigned with probability $\pr (A = 1 \mid W) =  1/\{1 + \exp(0.5+0.3W_1-0.5W_2-0.6W_3)\}$.

The targets are the counterfactual cumulative incidences of $\text{m}_2$ and $\text{d}$ at $t = 3, 6, 9, 12$ under the intervention with $a_{kg} = 1$ for $\text{o} \to \text{m}_1$ and $\text{m}_2 \to \text{m}_1$ and $a_{kg} = 0$ for all other transitions. This is a mixed regime in which the hazard of $\text{m}_1$ is borrowed from the treated arm and all other hazards from the control arm. The true values, $(0.117, 0.275, 0.387, 0.443)$ for $\text{m}_2$ and $(0.099, 0.281, 0.474, 0.636)$ for $\text{d}$, were obtained by simulating the hypothetical process for $4 \times 10^6$ subjects. The propensity score is estimated by logistic regression in $W$, the censoring hazards by arm-specific Cox models in $W$, and each of the three destination hazards by the Cox model (14), pooling the arms, with $z_{kg}(t)$ the indicators of the visited intermediate states and a common Breslow baseline. These working models are correctly specified. The one-step estimator \eqref{cond_2_est} and its standard error are computed on a grid of width $0.04$ by the backward recursion of Section E.2. The EIF-based estimator is evaluated with (1) all models correctly specified, (2) the true propensity score, (3) a misspecified propensity score, fitted by logistic regression on $W_1$ only, with correct hazard models, and (4) a correct propensity score with a misspecified hazard model. The plug-in estimator $\pn\{\widehat{F}_j^{\overline{a}}(t \mid W)\}$ is evaluated with (1) correctly specified and (2) misspecified hazard models. The misspecified hazard model omits the indicator of a preceding $\text{m}_1$ from the Cox model for the hazard of $\text{m}_2$, that is, it fixes $\delta_{12}^a = 1$ in both arms. The hazards of $\text{o} \to \text{m}_2$ and $\text{m}_1 \to \text{m}_2$ are then both misspecified, while the hazards of $\text{m}_1$ and $\text{d}$ remain correct.

\begin{figure}
\centering
\includegraphics[width=0.9\textwidth]{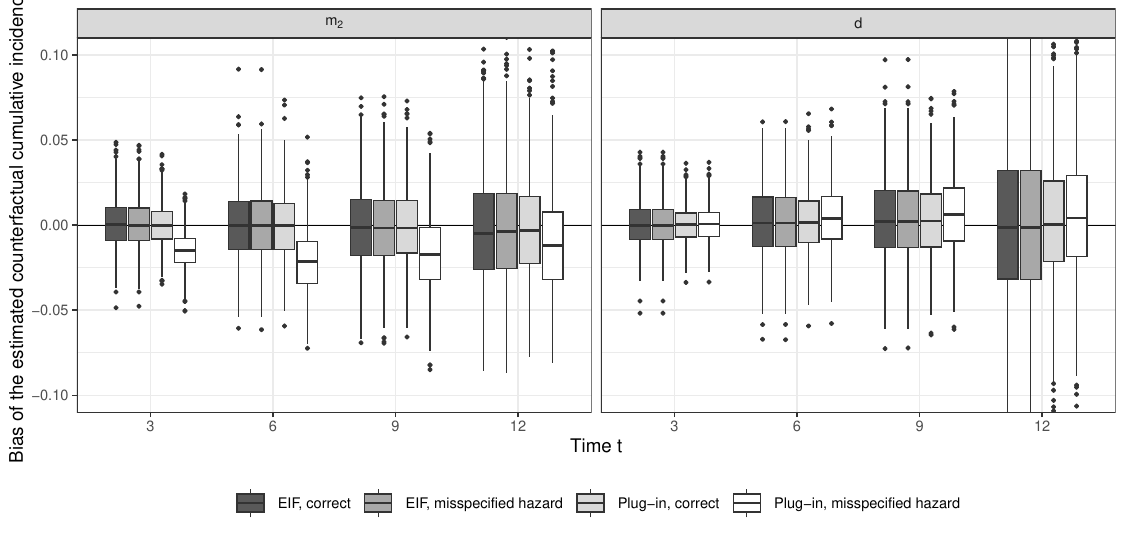}
\caption{Boxplots of the bias of the estimated counterfactual cumulative incidences of $\text{m}_2$ and $\text{d}$ at $t=3,6,9,12$ over 1000 replications: EIF-based (one-step) and plug-in estimators, with all working models correctly specified and with the misspecified hazard model for $\text{m}_2$. The vertical axis is truncated at $\pm 0.1$, which hides 0.5\% of the values, all at $t = 12$.} \label{fig:simu_bias}
\end{figure}

Confidence intervals are the point estimates $\pm$ 1.96 standard errors. Over 1000 replications, Figure \ref{fig:simu_bias} shows the bias of the two estimators and Table \ref{tab:simu} the average bias, standard deviation (SD), average standard error (SE), and coverage rate (CR) of nominal 95\% confidence intervals for $F_j^{\overline{a}}(t)$, $j = \text{m}_2, \text{d}$. SE and CR are reported for the EIF-based estimator only; a standard error for the plug-in estimator would require the bootstrap, which we do not compute.

\begin{table}
\centering
\caption{Average bias, standard deviation (SD), average standard error (SE), and coverage rate (CR) of nominal 95\% confidence intervals of the EIF-based estimator (EIF) and the plug-in estimator (Reg) for $F_j^{\overline{a}}(t)$ at selected time points. For the propensity score model (PS) and the hazard models (Haz), \checkmark\ denotes a correctly specified model, $\times$ a misspecified model, and T the true propensity score.} \label{tab:simu}
\begin{tabular}{llcccccccccc}
  \toprule
 & & \multicolumn{2}{c}{Event} & \multicolumn{4}{c}{$\text{m}_2$} & \multicolumn{4}{c}{d} \\
  \cmidrule(lr){5-8} \cmidrule(lr){9-12}
  & & \multicolumn{2}{c}{Time $t$} & 3 & 6 & 9 & 12 & 3 & 6 & 9 & 12 \\ 
  \midrule
  & Method & PS & Haz \\
  Bias & EIF & \checkmark & \checkmark & 0.001 & 0.000 & $-$0.001 & $-$0.002 & 0.001 & 0.002 & 0.003 & 0.003 \\
   & EIF & T & \checkmark & 0.001 & 0.000 & $-$0.001 & $-$0.001 & 0.001 & 0.002 & 0.003 & 0.003 \\
   & EIF & $\times$ & \checkmark & 0.001 & 0.000 & $-$0.001 & $-$0.001 & 0.001 & 0.002 & 0.003 & 0.003 \\
   & EIF & \checkmark & $\times$ & 0.000 & 0.000 & $-$0.001 & $-$0.001 & 0.001 & 0.002 & 0.003 & 0.003 \\
   & Reg & -- & \checkmark & 0.000 & 0.000 & $-$0.001 & $-$0.002 & 0.000 & 0.002 & 0.003 & 0.002 \\
   & Reg & -- & $\times$ & $-$0.015 & $-$0.021 & $-$0.017 & $-$0.011 & 0.001 & 0.004 & 0.007 & 0.006 \\
  SD & EIF & \checkmark & \checkmark & 0.015 & 0.021 & 0.024 & 0.036 & 0.014 & 0.021 & 0.024 & 0.053 \\
   & EIF & T & \checkmark & 0.015 & 0.021 & 0.024 & 0.036 & 0.014 & 0.021 & 0.024 & 0.053 \\
   & EIF & $\times$ & \checkmark & 0.015 & 0.021 & 0.023 & 0.035 & 0.013 & 0.021 & 0.024 & 0.052 \\
   & EIF & \checkmark & $\times$ & 0.015 & 0.021 & 0.024 & 0.036 & 0.014 & 0.021 & 0.024 & 0.053 \\
   & Reg & -- & \checkmark & 0.012 & 0.019 & 0.023 & 0.030 & 0.010 & 0.018 & 0.022 & 0.036 \\
   & Reg & -- & $\times$ & 0.011 & 0.019 & 0.022 & 0.030 & 0.010 & 0.018 & 0.022 & 0.036 \\
  SE & EIF & \checkmark & \checkmark & 0.014 & 0.020 & 0.023 & 0.031 & 0.013 & 0.021 & 0.024 & 0.046 \\
   & EIF & T & \checkmark & 0.014 & 0.020 & 0.023 & 0.031 & 0.013 & 0.021 & 0.024 & 0.046 \\
   & EIF & $\times$ & \checkmark & 0.014 & 0.020 & 0.023 & 0.031 & 0.013 & 0.020 & 0.024 & 0.045 \\
   & EIF & \checkmark & $\times$ & 0.014 & 0.020 & 0.023 & 0.031 & 0.013 & 0.021 & 0.024 & 0.046 \\
  CR & EIF & \checkmark & \checkmark & 0.938 & 0.945 & 0.945 & 0.929 & 0.942 & 0.950 & 0.947 & 0.915 \\
   & EIF & T & \checkmark & 0.934 & 0.945 & 0.942 & 0.925 & 0.941 & 0.949 & 0.941 & 0.915 \\
   & EIF & $\times$ & \checkmark & 0.938 & 0.944 & 0.942 & 0.927 & 0.940 & 0.943 & 0.941 & 0.916 \\
   & EIF & \checkmark & $\times$ & 0.939 & 0.946 & 0.940 & 0.923 & 0.942 & 0.950 & 0.946 & 0.915 \\
\bottomrule
\end{tabular}
\end{table}

Both estimators are nearly unbiased when all models are correctly specified. The plug-in estimator then has a smaller standard deviation, most visibly at $t = 12$, where censoring is heavy. When the hazard model for $\text{m}_2$ is misspecified, the plug-in estimator is biased for the incidence of $\text{m}_2$ by $-0.011$ to $-0.021$, between about 0.4 and 1.4 standard deviations. The EIF-based estimator shows no detectable bias under misspecification of either the propensity score or the hazard model. This is the case covered by condition (2) of Theorem 2. The misspecified transitions, $\text{o} \to \text{m}_2$ and $\text{m}_1 \to \text{m}_2$, are assigned to the control arm, and the transitions in $\mathcal{U}_k$ (Section 3.3) that the regime borrows from the treated arm, $\text{o} \to \text{m}_1$ and $\text{m}_2 \to \text{m}_1$, have correctly specified hazards in both arms. The weights are therefore correct, and the one-step estimator remains consistent under this mixed regime, as the simulation confirms. Coverage of the EIF-based confidence intervals is close to 95\% at $t \leq 9$ and slightly below at $t = 12$ (0.92--0.93). There the standard error underestimates the standard deviation by 13--14\%, because the censoring weights are large and variable late in follow-up. Estimating the propensity score and using its true value give the same standard deviation here. The misspecified propensity score does not affect the bias, as condition (1) of Theorem 2 implies.

\section{Additional data analysis results}

\subsection{Data pre-processing}

Each state is entered at most once. Same-day events were ordered as follows. A nonfatal EMACE recorded on the day of a CV death is part of the fatal EMACE (7 patients). Any other nonfatal event on the day of death is placed immediately before death, and the adjudicated cause is kept (4 nonfatal EMACE before a non-CV death, 3 on O$\to$E$_{\text{n}}$ and 1 on O$\to$M$\to$E$_{\text{n}}$; 6 MVE, of which 3 on O$\to$M and 2 on O$\to$E$_{\text{n}}$$\to$M preceded a non-CV death and 1 on O$\to$M a CV death). An EMACE and an MVE on the same day are ordered with the MVE first (3 patients). Under these rules, each of the 9,340 patients has exactly one trajectory. Table \ref{tab:trans} gives the observed transitions by arm.

\begin{table}[!htbp]
\centering
\caption{Observed transitions by treatment arm in LEADER. E$_{\text{n}}$: nonfatal first EMACE; E$_{\text{f}}$: fatal first EMACE; D$_{\text{c}}$: CV death after a nonfatal EMACE; D$_{\text{n}}$: non-CV death.} \label{tab:trans}
\begin{tabular}{lrr}
\toprule
Transition & Placebo & Liraglutide \\
\midrule
O $\to$ E$_{\text{n}}$ (nonfatal first EMACE) & 827 & 761 \\
O $\to$ M & 358 & 311 \\
O $\to$ E$_{\text{f}}$ (fatal first EMACE) & 167 & 134 \\
O $\to$ D$_{\text{n}}$ & 96 & 105 \\
E$_{\text{n}}$ $\to$ M after E$_{\text{n}}$ & 58 & 44 \\
E$_{\text{n}}$ $\to$ D$_{\text{c}}$ & 84 & 68 \\
E$_{\text{n}}$ $\to$ D$_{\text{n}}$ & 49 & 35 \\
M $\to$ E$_{\text{n}}$ after M (nonfatal first EMACE) & 52 & 45 \\
M $\to$ E$_{\text{f}}$ (fatal first EMACE) & 16 & 8 \\
M $\to$ D$_{\text{n}}$ & 13 & 16 \\
M after E$_{\text{n}}$ $\to$ D$_{\text{c}}$ & 5 & 6 \\
M after E$_{\text{n}}$ $\to$ D$_{\text{n}}$ & 5 & 4 \\
E$_{\text{n}}$ after M $\to$ D$_{\text{c}}$ & 6 & 3 \\
E$_{\text{n}}$ after M $\to$ D$_{\text{n}}$ & 6 & 2 \\
\midrule
Total & 1742 & 1542 \\
\bottomrule
\end{tabular}
\end{table}

\subsection{Cumulative incidences and total effects}

\begin{figure}[!htbp]
\centering
\includegraphics[width=0.96\textwidth]{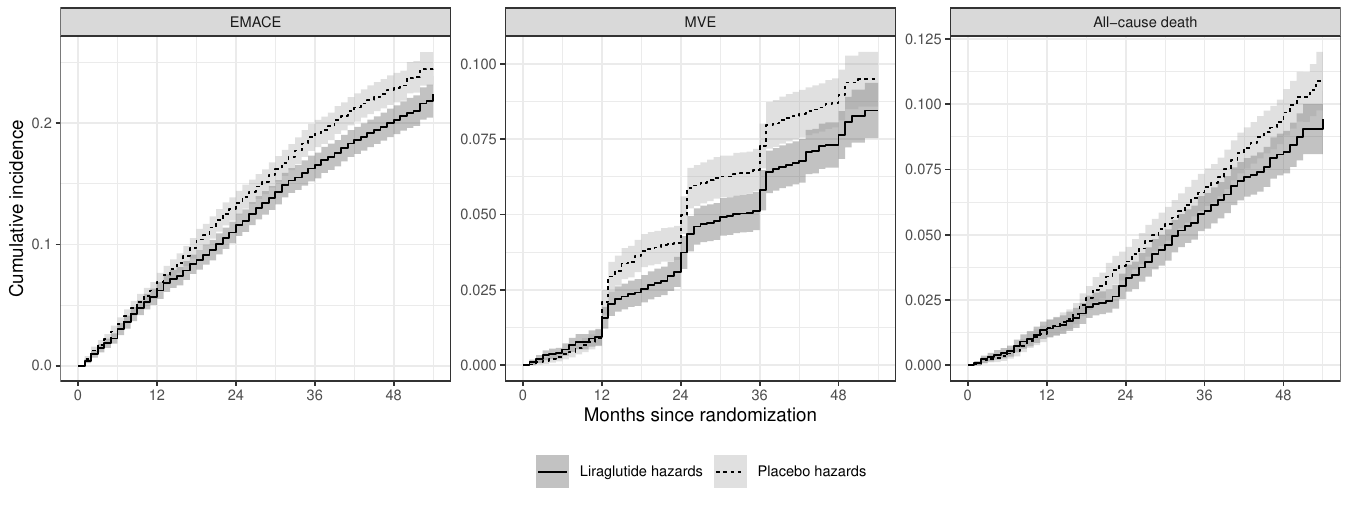} \\
\includegraphics[width=0.96\textwidth]{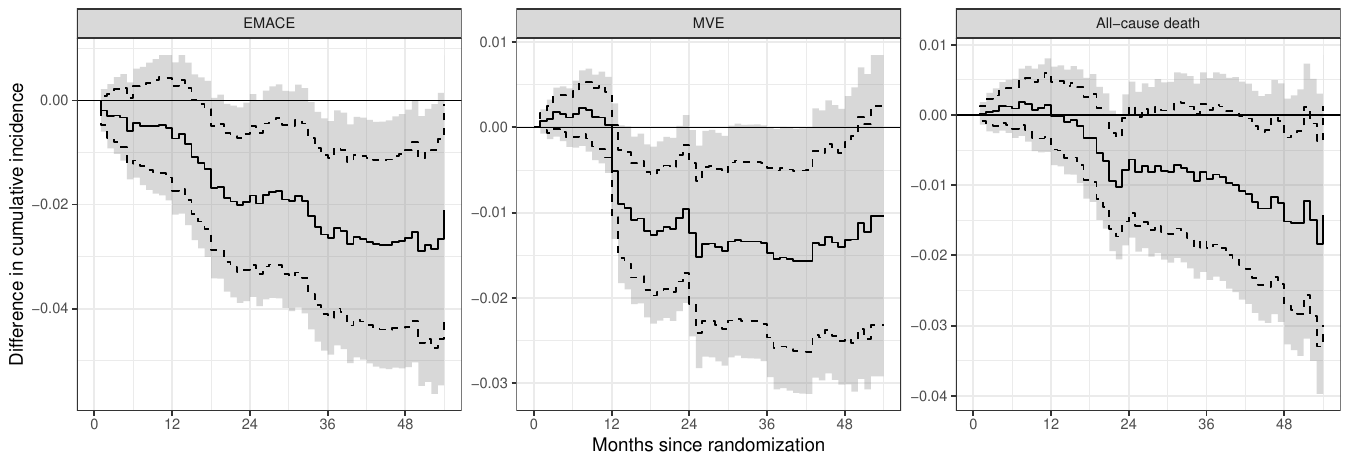}
\caption{Top: estimated cumulative incidences of first EMACE, MVE, and all-cause death under the all-placebo and all-liraglutide hazards, with pointwise 95\% confidence intervals (shaded). Bottom: total effects (difference in cumulative incidence), with pointwise 95\% confidence intervals (dashed) and 95\% simultaneous confidence bands over 1 to 54 months (shaded).} \label{fig:ate_total2}
\end{figure}

Figure \ref{fig:ate_total2} shows the estimated counterfactual cumulative incidences of first EMACE, MVE, and all-cause death under the all-placebo and all-liraglutide hazards, with pointwise 95\% confidence intervals, and the total effects with pointwise confidence intervals and 95\% simultaneous bands over 1 to 54 months. The simultaneous bands exclude zero over part of follow-up for first EMACE and MVE but not for death. By 54 months, the all-liraglutide hazards gained 0.95 months of EMACE-free time (SE 0.30, $P = 0.002$), 0.53 months of MVE-free time (SE 0.19, $P = 0.005$), and 0.36 months of survival (SE 0.18, $P = 0.043$) relative to the all-placebo hazards.

\subsection{Path-targeted effects over time}

\begin{figure}[!htbp]
\centering
\includegraphics[width=0.96\textwidth]{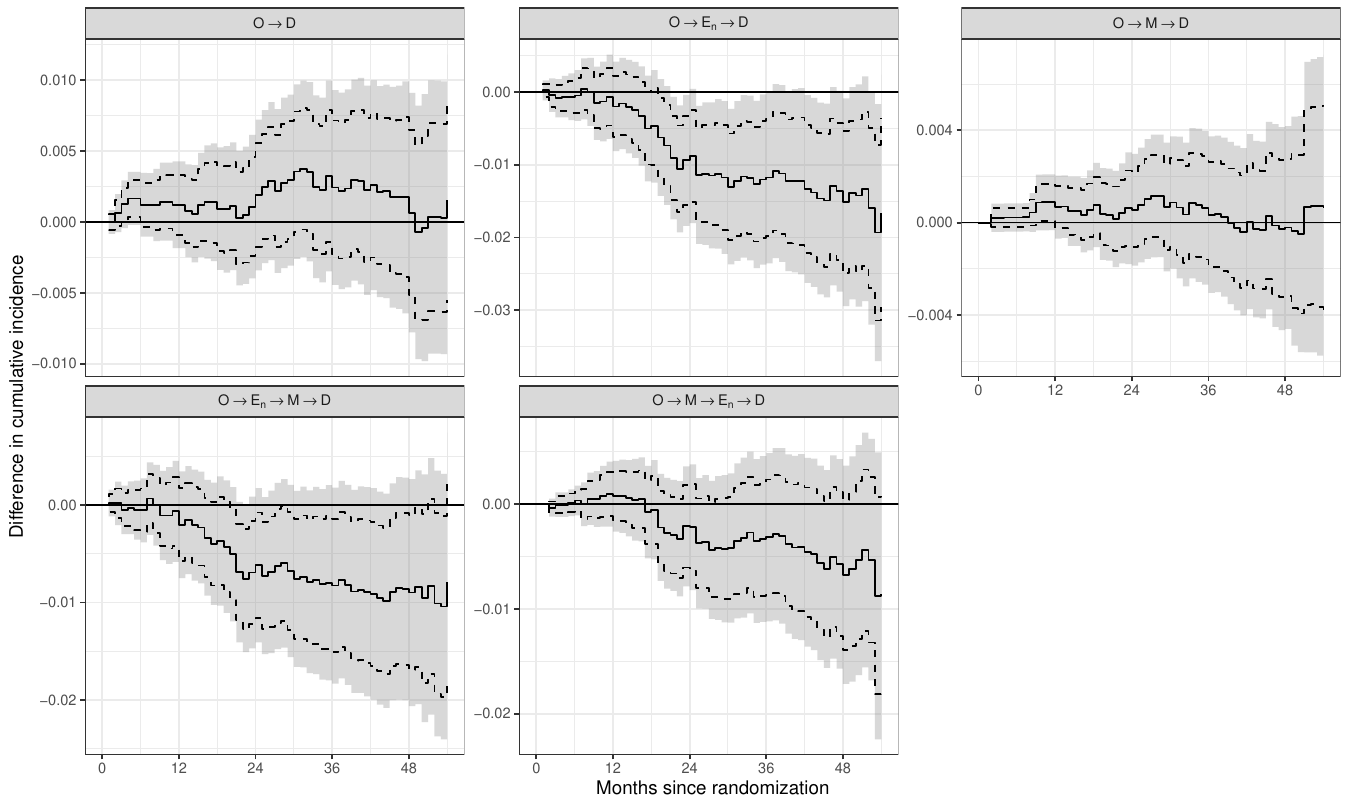}
\caption{Path-targeted effects on the cumulative incidence of death: difference between the regime in which the transitions along the path have liraglutide-arm hazards (all others placebo) and the all-placebo regime, with pointwise 95\% confidence intervals (dashed) and 95\% simultaneous bands over 1 to 54 months (shaded).} \label{fig:ate_path}
\end{figure}

Figure \ref{fig:ate_path} shows the path-targeted effects on the cumulative incidence of all-cause death (placebo reference). As explained in Section 2.3, these effects are not additive and do not apportion the total effect among paths; the path-targeted Shapley effects in Table 1 do. The simultaneous band excludes zero over part of follow-up only for O$\to$E$_{\text{n}}$$\to$D.

\subsection{Comparisons of dynamic treatment regimes} \label{sec:dtr}

The comparisons in this section are interpretable as comparisons of implementable treatment regimes only under the additional conditions stated in the Discussion of the main text: no treatment-induced or unmeasured confounding of transitions given the observed history, and no carry-over of earlier treatment. These conditions cannot be checked with the LEADER data, and the comparisons are reported for completeness rather than as recommendations. A regime is indexed by the treatment given while in O, after a nonfatal EMACE, and after an MVE. For example, O1E0M0 assigns liraglutide until the first event and placebo thereafter; a patient with both events receives the level of the most recently entered state. Table \ref{tab:dtr} reports the counterfactual cumulative incidences at 54 months, the restricted mean time lost to death by 54 months, and selected comparisons. The regimes differ mainly through the treatment given before the first event.

\begin{table}[!htbp]
\centering
\caption{Counterfactual cumulative incidences (\%) of first EMACE, MVE, and all-cause death at 54 months and restricted mean time lost to death by 54 months (months) under the eight regimes, and comparisons of restricted mean time lost (time gained by the first regime, survival for death and EMACE-free time for EMACE; standard error). For EMACE, the comparison O1M0 versus O0M1 is between O1E0M0 and O0E0M1.} \label{tab:dtr}
\begin{tabular}{lcccc}
\toprule
Regime & EMACE & MVE & Death & Time lost to death \\
\midrule
O1 E1 M1 & 22.42 & 8.45 & 9.43 & 2.30 \\
O0 E1 M1 & 24.47 & 9.29 & 10.23 & 2.57 \\
O1 E0 M1 & 22.42 & 8.65 & 10.00 & 2.39 \\
O0 E0 M1 & 24.47 & 9.49 & 10.91 & 2.69 \\
O1 E1 M0 & 22.51 & 8.45 & 9.33 & 2.26 \\
O0 E1 M0 & 24.51 & 9.29 & 10.12 & 2.53 \\
O1 E0 M0 & 22.51 & 8.65 & 9.96 & 2.37 \\
O0 E0 M0 & 24.51 & 9.49 & 10.86 & 2.66 \\
\midrule
Comparison (outcome) & & & Time gained & $P$ \\
\midrule
O1E0M0 vs O0E1M1 (D) &  &  & 0.208 (0.178) & 0.242 \\
O1E1M1 vs O0E1M1 (D) &  &  & 0.276 (0.149) & 0.064 \\
O1E0M0 vs O0E0M0 (D) &  &  & 0.294 (0.149) & 0.049 \\
O0E1M1 vs O0E0M0 (D) &  &  & 0.085 (0.105) & 0.416 \\
O1M0 vs O0M1 (E) &  &  & 0.886 (0.304) & 0.004 \\
\bottomrule
\end{tabular}
\end{table}

\subsection{Estimator based on the EIF under extended Markovness} \label{sec:extmarkov_res}

The working models of the main analysis make every transition hazard depend on the history only through the visited states and the time since the last transition, capped at 12 months. On the grid of width $h = 0.25$ months used for computation, this is a finite partition of the histories. The classes are the current state and the set of visited states, which the working models do not distinguish by order, crossed with 49 bins of the time since the last transition; the last bin collects sojourns of at least 12 months. The working models therefore satisfy Assumption \ref{ass:extmarkov} with this partition, and the one-step estimator \eqref{cond_h_est} can be computed from the same fitted models. We use the pooled Cox models of the main analysis, with treatment, treatment-by-covariate, and treatment-by-$z_{kg}(t)$ interactions and a common Breslow baseline, without refitting. Only the weight changes, from the history-specific density ratio $\omega_k^{\overline{a}}$ to the ratio $\widehat{P}_k^{\overline{a}} / \widehat\pi_k^{a_{kg}}$ of the hypothetical to the observed occupancy of the class, and the increment from $\widehat\Delta_{kg}^{\overline{a}}$ to its class average $\widehat{D}_{kg}^{\overline{a}}$. The plug-in estimates are identical to those of the main analysis.

\begin{figure}[!htbp]
\centering
\includegraphics[width=0.96\textwidth]{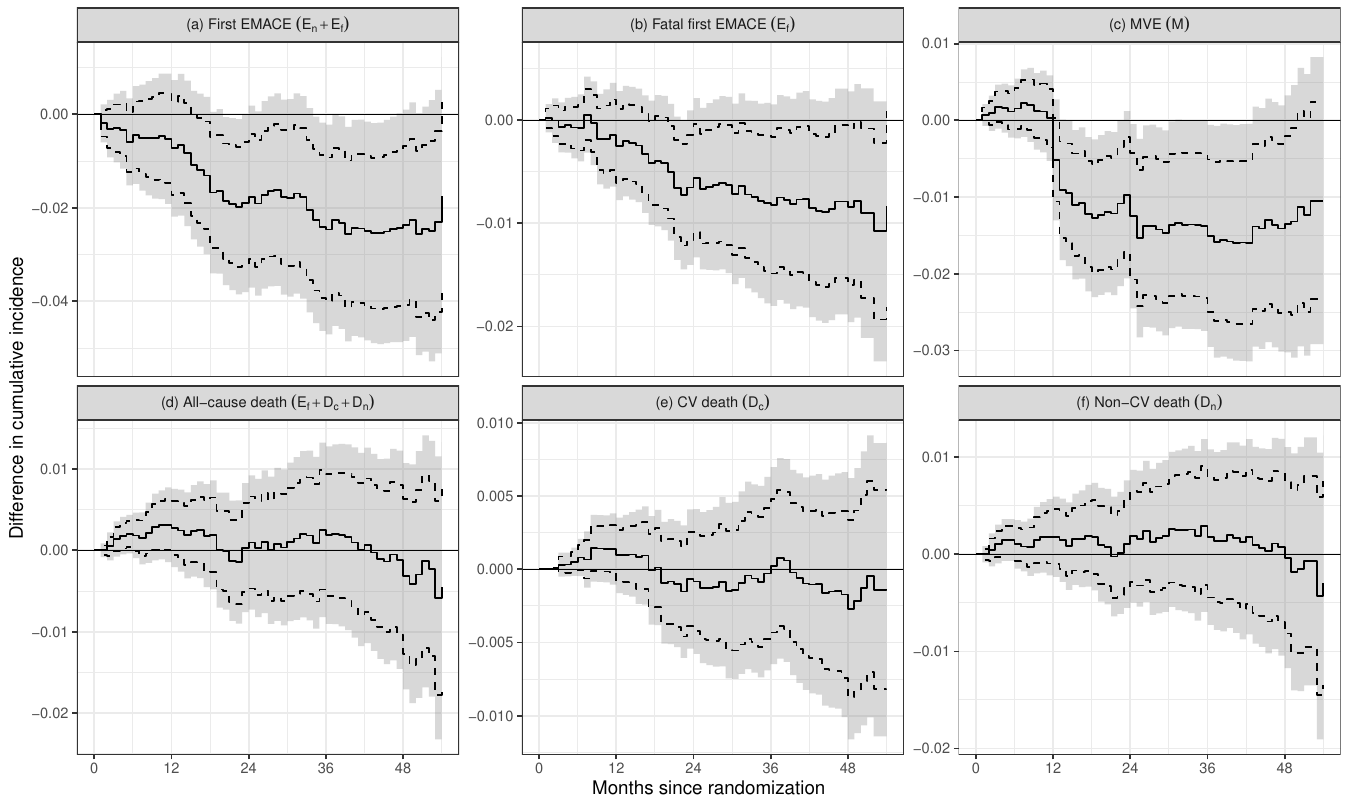}
\caption{Event-targeted effects estimated with the one-step estimator based on the EIF under extended Markovness (Lemma \ref{lem:extmarkov}), with the working models of the main analysis, in the layout of Figure 5. Solid: estimate; dashed: pointwise 95\% confidence intervals; shaded: 95\% simultaneous confidence band over 1 to 54 months.} \label{fig:extmarkov}
\end{figure}

Figure \ref{fig:extmarkov} shows the six event-targeted effects of Figure 5, and Table \ref{tab:extmarkov} compares the restricted-mean contrasts with those of the main analysis. The two estimators differ by at most 0.01 months, and their standard errors differ by less than 1\%. By Theorem \ref{thmS1}, the estimator based on the projected EIF is efficient in the submodel, and its asymptotic variance is no larger than that of the estimator of the main text. The near equality of the standard errors indicates that, in LEADER, $\omega_k^{\overline{a}} \Delta_{kg}^{\overline{a}}$ varies little among the histories within a class, so little efficiency is lost by not imposing the partition. The simultaneous bands exclude zero over part of follow-up for first EMACE and MVE, as in the main analysis. The conclusions of the main text are therefore not sensitive to the choice between the two estimators.

\begin{table}[!htbp]
\centering
\caption{Event-targeted and total effects estimated with the EIF of the main text (nonparametric) and with the EIF under extended Markovness (Lemma \ref{lem:extmarkov}), with the same working models: time free of the outcome gained by 54 months (months; standard error). Placebo reference; the $P$-values are for the estimator under extended Markovness.} \label{tab:extmarkov}
\begin{tabular}{lccc}
\toprule
Effect (outcome) & Main analysis & Extended Markov & $P$ \\
\midrule
\multicolumn{4}{l}{\textit{Event-targeted, into E$_{\text{n}}$ and E$_{\text{f}}$}} \\
~~~First EMACE & 0.895 (0.303) & 0.885 (0.303) & 0.004 \\
~~~Fatal first EMACE & 0.294 (0.124) & 0.296 (0.124) & 0.017 \\
\multicolumn{4}{l}{\textit{Event-targeted, into M}} \\
~~~MVE & 0.534 (0.187) & 0.534 (0.187) & 0.004 \\
\multicolumn{4}{l}{\textit{Event-targeted, into D$_{\text{c}}$ and D$_{\text{n}}$}} \\
~~~All-cause death & $-$0.023 (0.132) & $-$0.030 (0.132) & 0.819 \\
~~~CV death after a nonfatal EMACE & 0.029 (0.079) & 0.023 (0.079) & 0.773 \\
~~~Non-CV death & $-$0.057 (0.112) & $-$0.058 (0.113) & 0.605 \\
\multicolumn{4}{l}{\textit{Total}} \\
~~~First EMACE & 0.950 (0.303) & 0.941 (0.303) & 0.002 \\
~~~MVE & 0.527 (0.187) & 0.528 (0.187) & 0.005 \\
~~~All-cause death & 0.361 (0.178) & 0.366 (0.178) & 0.040 \\
\bottomrule
\end{tabular}
\end{table}

\subsection{Results of the frailty-indexed sensitivity analysis} \label{sec:frailty_res}

\begin{figure}[!htbp]
\centering
\includegraphics[width=0.96\textwidth]{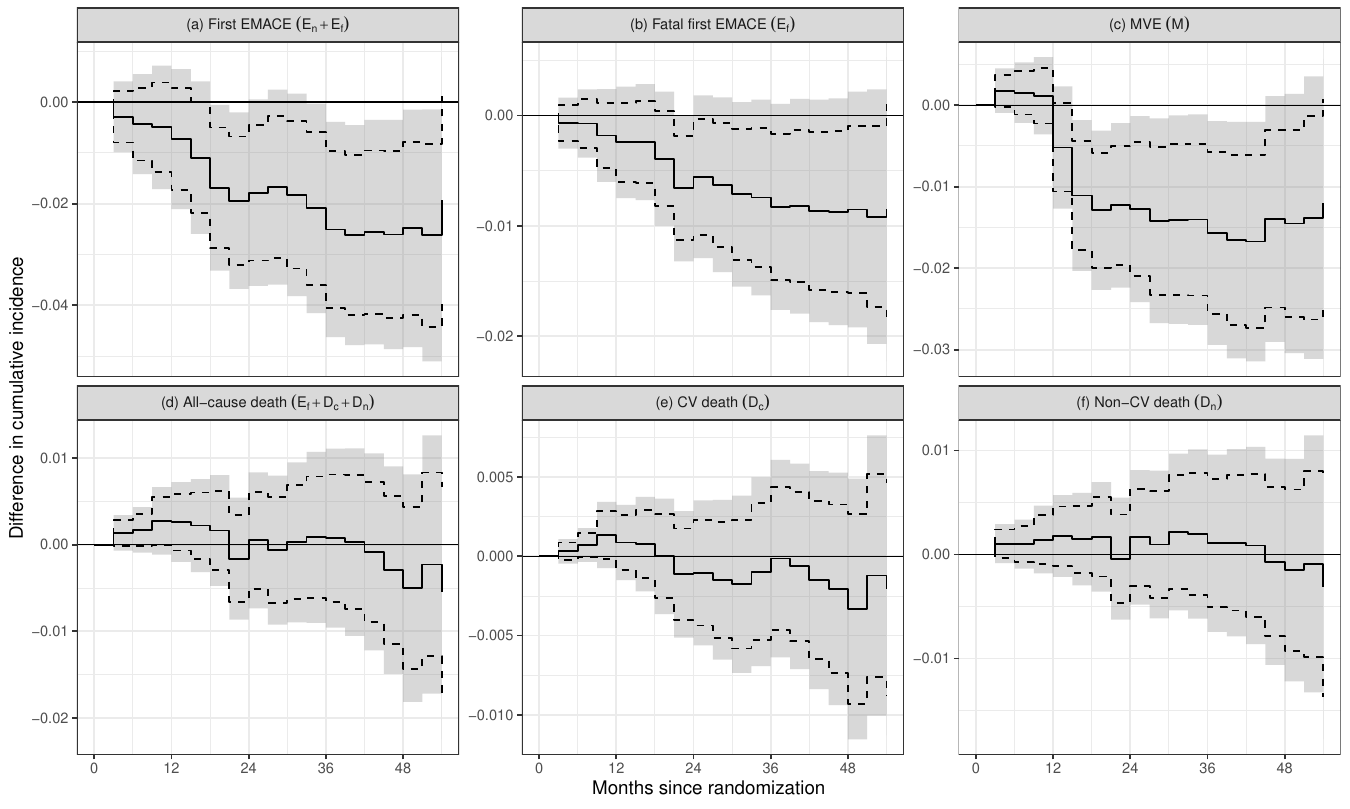}
\caption{Frailty-indexed event-targeted effects with gamma frailty of estimated variance $\widehat\nu = 0.63$, in the layout of Figure 5. Solid: one-step estimate; dashed: pointwise 95\% confidence intervals; shaded: 95\% simultaneous confidence band over the grid of every 3 months up to 54 months.} \label{fig:frailty_hat}
\end{figure}

\begin{figure}[!htbp]
\centering
\includegraphics[width=0.96\textwidth]{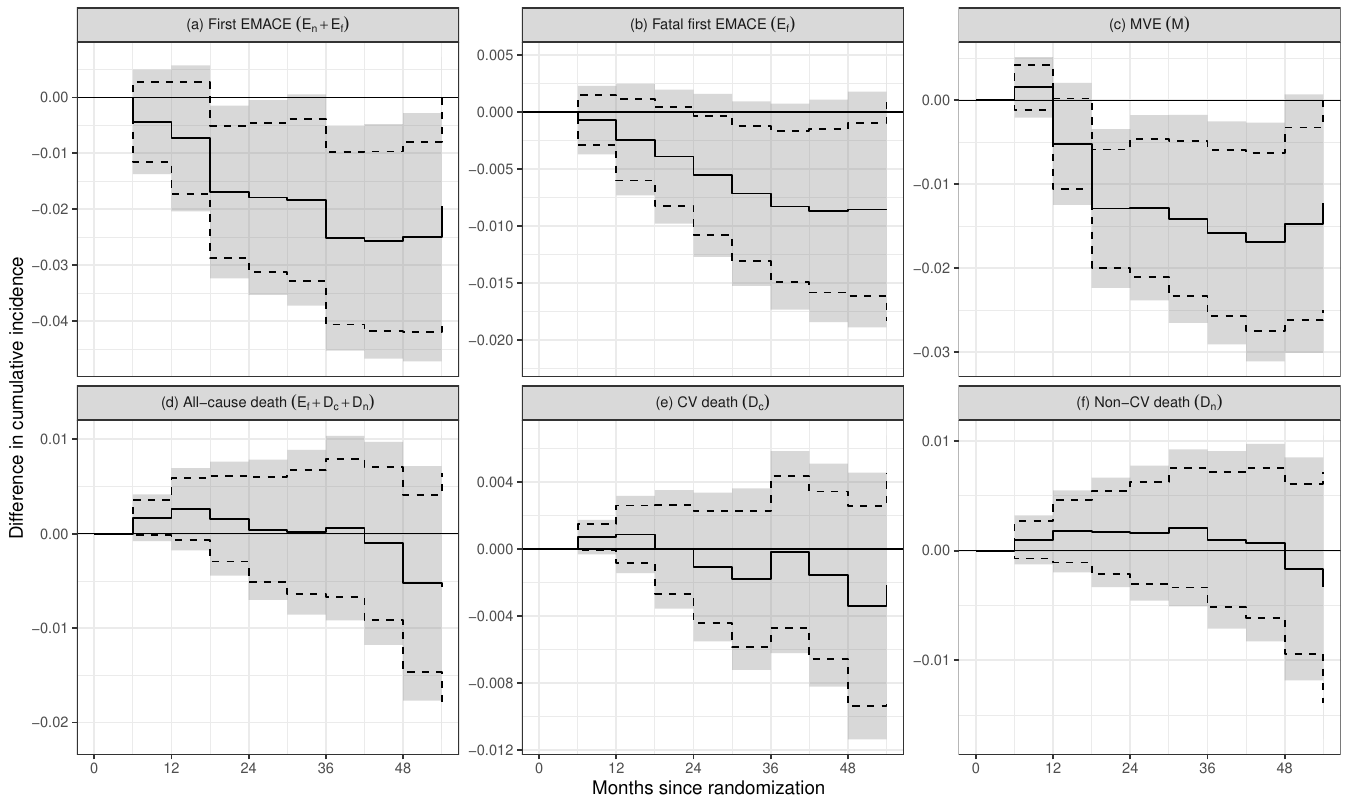}
\caption{Frailty-indexed event-targeted effects with gamma frailty of variance $\nu = 0.75$, in the layout of Figure 5. Solid: one-step estimate; dashed: pointwise 95\% confidence intervals; shaded: 95\% simultaneous confidence band over the grid of every 6 months up to 54 months.} \label{fig:frailty_075}
\end{figure}

Table \ref{tab:frailty} reports the one-step estimates of the frailty-indexed estimands of Section \ref{sec:density_est} with gamma frailty. Figures \ref{fig:frailty_hat} and \ref{fig:frailty_075} show the event-targeted effects over time for the estimated variance $\widehat\nu = 0.63$ and for $\nu = 0.75$, in the layout of Figure 5; the figures for $\nu = 0.25$ and $0.5$ are similar. The row without frailty is the main analysis ($\nu = 0$). As required by Section \ref{sec:fr_ident}, the placebo-regime cumulative incidence of death and the total effect do not change with $\nu$ beyond rounding. The mixed-regime effects change by at most 0.03 months over $\nu \in [0, 0.75]$, much less than their standard errors. The event-targeted effects on first EMACE and MVE remain significant, the event-targeted effect on death of switching the transitions into D$_{\text{c}}$ and D$_{\text{n}}$ remains close to zero, and the transition-targeted effect of $(\text{O}, \text{E}_{\text{n}})$ and the path-targeted effects of O$\to$E$_{\text{n}}$$\to$D and O$\to$E$_{\text{n}}$$\to$M$\to$D are essentially unchanged. The simultaneous bands exclude zero over part of follow-up for first EMACE and MVE for every $\nu$, as in the main analysis. Holding a shared latent frailty fixed under the intervention therefore does not change the conclusions in this range. The analysis indexes the estimand by the amount of latent heterogeneity held fixed and provides no evidence about the presence or absence of unmeasured confounding.

\begin{table}[!htbp]
\centering
\caption{Frailty-indexed one-step estimates with gamma frailty of variance $\nu$ (standard errors in parentheses): placebo-regime cumulative incidence of death at 54 months (\%), and total, event-targeted, transition-targeted, and path-targeted effects (time free of the outcome gained by 54 months, months; placebo reference). The event-targeted effects on first EMACE, MVE, and death switch the transitions into E$_{\text{n}}$ and E$_{\text{f}}$, into M, and into D$_{\text{c}}$ and D$_{\text{n}}$, respectively; the transition- and path-targeted effects are on all-cause death. The row without frailty is the main analysis.} \label{tab:frailty}
\resizebox{\textwidth}{!}{%
\begin{tabular}{lcccccccc}
\toprule
Frailty & CIF of death & Total, death & Into E$_{\text{n}}$, E$_{\text{f}}$ & Into M & Into D$_{\text{c}}$, D$_{\text{n}}$ & (O, E$_{\text{n}}$) & O$\to$E$_{\text{n}}$$\to$D & O$\to$E$_{\text{n}}$$\to$M$\to$D \\
\midrule
No frailty (main analysis) & 10.86 (0.57) & 0.36 (0.18) & 0.90 (0.30) & 0.53 (0.19) & $-$0.02 (0.13) & 0.36 (0.12) & 0.46 (0.15) & 0.30 (0.13) \\
$\nu = 0.25$ & 10.86 (0.57) & 0.36 (0.18) & 0.90 (0.30) & 0.54 (0.19) & $-$0.01 (0.13) & 0.36 (0.12) & 0.46 (0.15) & 0.30 (0.13) \\
$\nu = 0.5$ & 10.87 (0.57) & 0.36 (0.18) & 0.91 (0.30) & 0.55 (0.19) & 0.00 (0.13) & 0.35 (0.12) & 0.46 (0.15) & 0.29 (0.13) \\
$\widehat\nu = 0.63$ & 10.87 (0.57) & 0.36 (0.18) & 0.91 (0.30) & 0.56 (0.19) & 0.01 (0.13) & 0.35 (0.12) & 0.46 (0.15) & 0.29 (0.12) \\
$\nu = 0.75$ & 10.87 (0.57) & 0.36 (0.18) & 0.91 (0.30) & 0.56 (0.19) & 0.01 (0.13) & 0.35 (0.12) & 0.46 (0.15) & 0.28 (0.12) \\
\bottomrule
\end{tabular}}
\end{table}

\subsection{Secondary analysis: first occurrence of individual events} \label{sec:indiv}

We split the first event into nine mutually exclusive types: the six components of EMACE (nonfatal myocardial infarction, nonfatal stroke, cardiovascular death, coronary revascularization, heart failure, and unstable angina pectoris), the two components of MVE (nephropathy and retinopathy), and non-CV death. This gives a competing-risks model with the single transient state O and nine absorbing states. It is the special case of Lemma 1 in which the only transient state is O, so that the weight reduces to $P_{\text{O}}^{\overline{a}} / \{P_{\text{O}}^{a} G^{a}\}$. The target is the cumulative incidence of each event type as the \textit{first} event. An event that follows another event is not counted, so the composite EMACE and MVE here are first-event versions of those in the main text, which also count an EMACE after an MVE and vice versa. Same-day events follow the tie rules of Section D.1. In addition, nonfatal EMACE components recorded on the same day (133 patients, mostly myocardial infarction with revascularization) are ordered as myocardial infarction, stroke, heart failure, unstable angina pectoris, then revascularization. A revascularization performed on the day of a myocardial infarction or an episode of unstable angina is thus attributed to that event. With these rules the components add up to the composites, and the first-event counts of EMACE (including fatal first EMACE), MVE, and non-CV death agree with the transitions out of O in Table \ref{tab:trans}.

The total effect compares the all-liraglutide and all-placebo hazards of the nine first-event transitions. The transition-targeted effect of $(\text{O}, k)$ switches only the hazard of a first event of type $k$ to liraglutide, with the other eight at placebo. For a composite, all of its component hazards are switched, which is an event-targeted effect. Cause-specific hazards were fitted in each arm by Cox models with the seven baseline covariates. The censoring model, the propensity score, and the grid were those of the main analysis. Under the all-placebo and all-liraglutide hazards, the one-step estimates of the first-event cumulative incidences at 54 months agreed with the unadjusted Aalen--Johansen estimates in the corresponding arm to within 0.1 percentage points.

\begin{figure}[!htbp]
\centering
\includegraphics[width=\textwidth]{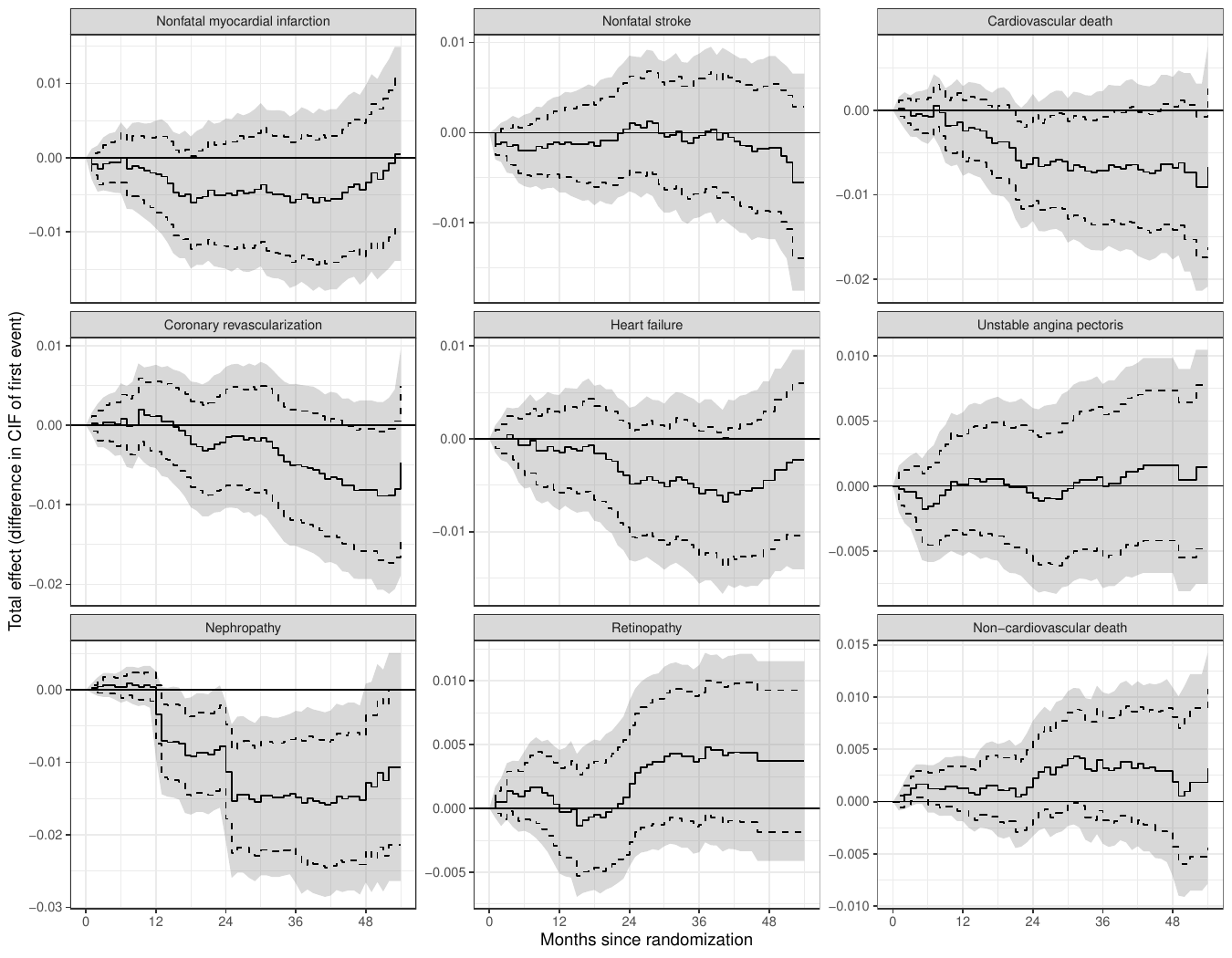}
\caption{Total effects on the cumulative incidence of each of the nine event types as the first event (all-liraglutide versus all-placebo hazards), with pointwise 95\% confidence intervals (dashed) and 95\% simultaneous bands over 1 to 54 months (shaded).} \label{fig:ate_total1}
\end{figure}

\begin{figure}[!htbp]
\centering
\includegraphics[width=\textwidth]{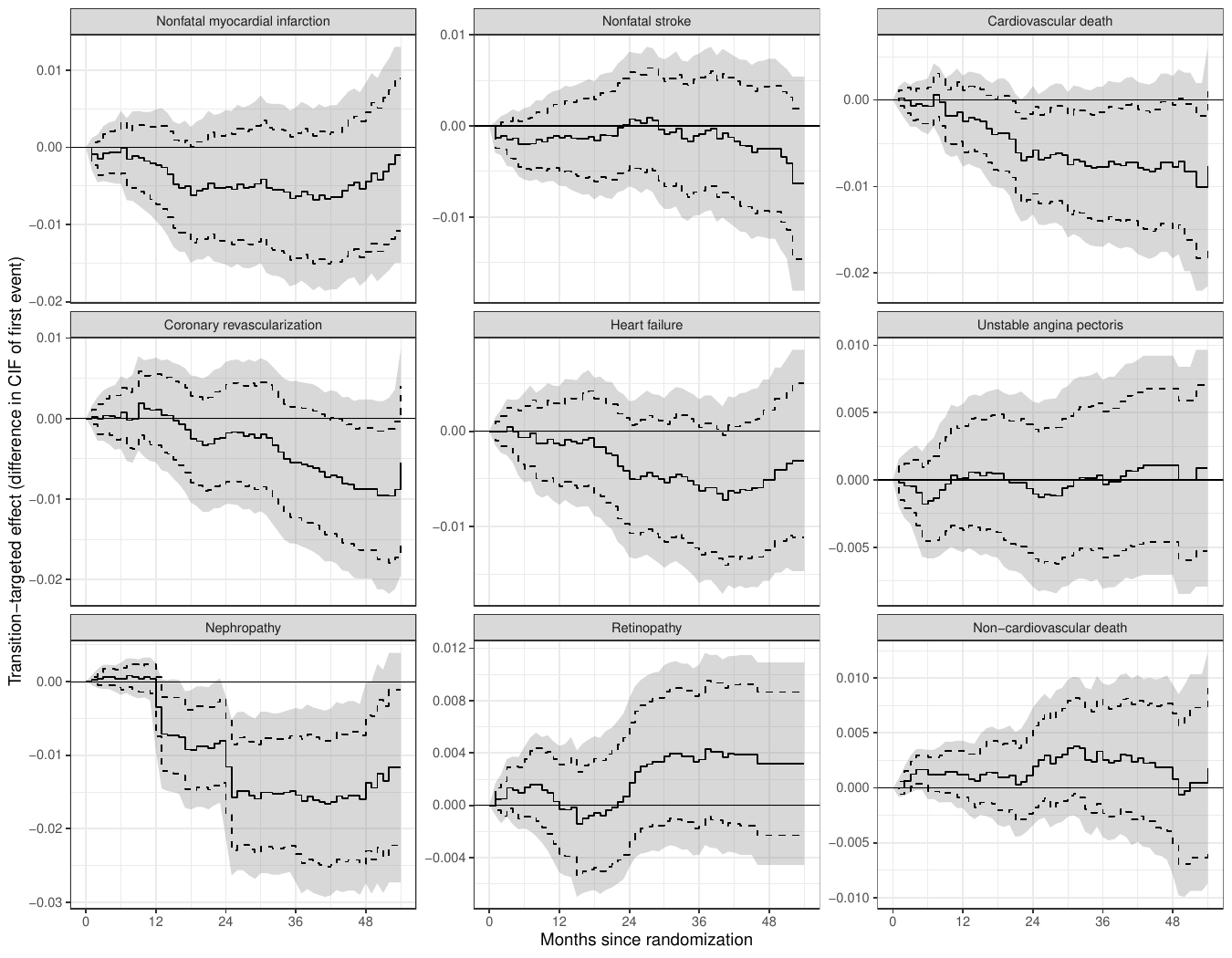}
\caption{Transition-targeted effects of $(\text{O}, k)$ on the cumulative incidence of event type $k$ as the first event: hazard of a first event of type $k$ from the liraglutide arm, all other first-event hazards from the placebo arm, versus all-placebo hazards. Pointwise 95\% confidence intervals (dashed) and 95\% simultaneous bands over 1 to 54 months (shaded).} \label{fig:ate2}
\end{figure}

Table \ref{tab:pvals} reports the counts and the restricted-mean contrasts, and Figures \ref{fig:ate_total1} and \ref{fig:ate2} show the effects over time. Total and transition-targeted effects differ by at most 0.05 months: at these event rates, switching the hazards of the other first events changes the incidence of a given first event very little. Among the individual events, the transition-targeted effects on nephropathy (0.55 months of nephropathy-free time gained by 54 months, SE 0.15, $P < 0.001$) and on cardiovascular death (0.29 months, SE 0.12, $P = 0.017$) had $P < 0.05$, in the direction of the LEADER reports \citep{marso2016liraglutide}. The 95\% simultaneous band excluded zero only for nephropathy. The estimates for the other EMACE components all favored liraglutide except unstable angina pectoris, with confidence intervals including zero. The estimates for retinopathy and non-CV death were in the opposite direction, also with intervals including zero.

\begin{table}[!htbp]
\centering
\caption{First occurrences of individual events by arm, and total and transition-targeted effects on the cumulative incidence of each event as the first event: time free of the event gained by 54 months (months; standard error), with $P$-values of the restricted-mean contrasts. Placebo reference. $^\dagger$For composites, the hazards of all components are switched (event-targeted effect).} \label{tab:pvals}
\resizebox{\textwidth}{!}{%
\begin{tabular}{lrrcccc}
\toprule
& & & \multicolumn{2}{c}{Total effect} & \multicolumn{2}{c}{Transition-targeted effect} \\
\cmidrule(lr){4-5} \cmidrule(lr){6-7}
Event of first occurrence & Liraglutide & Placebo & Time gained & $P$ & Time gained & $P$ \\
\midrule
EMACE$^\dagger$ & 895 & 994 & 0.849 (0.300) & 0.005 & 0.874 (0.299) & 0.003 \\
~~~MACE$^\dagger$ & 492 & 548 & 0.518 (0.229) & 0.024 & 0.563 (0.227) & 0.013 \\
~~~~~~Nonfatal myocardial infarction & 225 & 237 & 0.194 (0.163) & 0.235 & 0.222 (0.162) & 0.172 \\
~~~~~~Nonfatal stroke & 133 & 144 & 0.060 (0.124) & 0.630 & 0.080 (0.123) & 0.514 \\
~~~~~~Cardiovascular death & 134 & 167 & 0.265 (0.121) & 0.028 & 0.285 (0.120) & 0.017 \\
~~~Coronary revascularization & 153 & 185 & 0.172 (0.141) & 0.222 & 0.189 (0.140) & 0.178 \\
~~~Heart failure & 149 & 167 & 0.170 (0.136) & 0.209 & 0.186 (0.135) & 0.168 \\
~~~Unstable angina pectoris & 101 & 94 & $-$0.012 (0.116) & 0.916 & 0.000 (0.115) & 0.998 \\
MVE$^\dagger$ & 311 & 358 & 0.395 (0.179) & 0.028 & 0.428 (0.178) & 0.016 \\
~~~Nephropathy & 214 & 278 & 0.523 (0.148) & $<0.001$ & 0.545 (0.147) & $<0.001$ \\
~~~Retinopathy & 97 & 80 & $-$0.128 (0.107) & 0.232 & $-$0.113 (0.106) & 0.286 \\
Non-cardiovascular death & 105 & 96 & $-$0.116 (0.090) & 0.195 & $-$0.090 (0.088) & 0.305 \\
\bottomrule
\end{tabular}}
\end{table}

\subsection{Secondary analysis: CV and non-CV death as the two causes of death} \label{sec:cvsplit}

\begin{figure}[!htbp]
\centering
\includegraphics[width=0.85\textwidth]{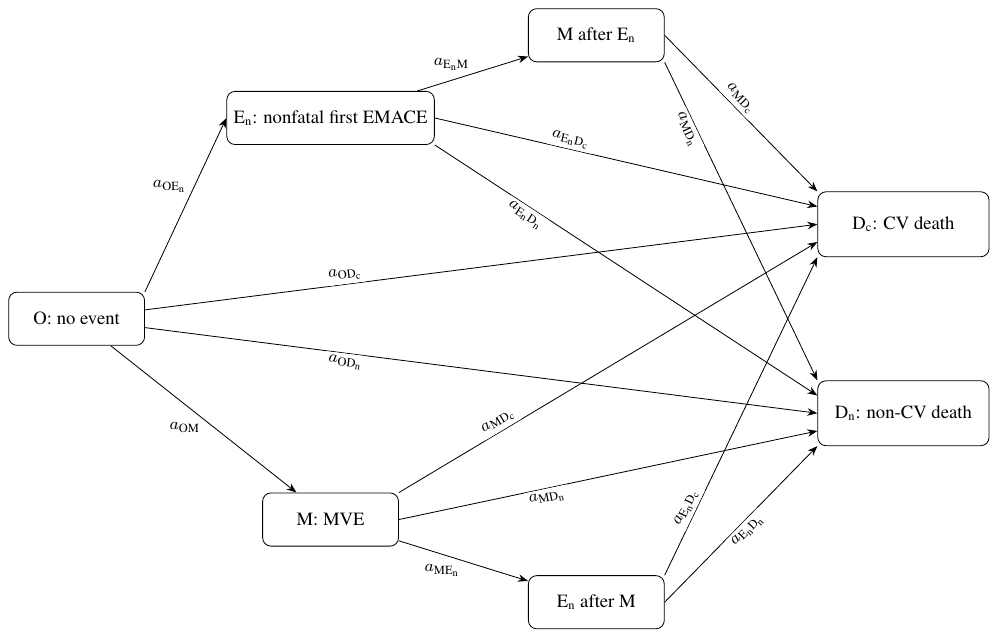}
\caption{State structure of the secondary analysis in which fatal first EMACE is merged into CV death. Each arrow is labeled by the component of $\overline{a}$ that sets its hazard.} \label{fig:structure_cv}
\end{figure}

In the main analysis, a fatal first EMACE (E$_{\text{f}}$) is a first EMACE, and its transition shares the component of $\overline{a}$ of the transition into E$_{\text{n}}$ from the same state. Here we instead merge E$_{\text{f}}$ into CV death. There are then two causes of death, CV death (D$_{\text{c}}$) and non-CV death (D$_{\text{n}}$), treated as different absorbing states (Figure \ref{fig:structure_cv}). The transient states and the fitted hazards are those of the main analysis. However, a CV death without a preceding nonfatal EMACE, from O or from M, is now a transition into D$_{\text{c}}$ with its own component of $\overline{a}$. The intervention vector therefore has ten components, for the transitions O$\to$E$_{\text{n}}$, O$\to$M, E$_{\text{n}}$$\to$M, M$\to$E$_{\text{n}}$, O$\to$D$_{\text{c}}$, O$\to$D$_{\text{n}}$, E$_{\text{n}}$$\to$D$_{\text{c}}$, E$_{\text{n}}$$\to$D$_{\text{n}}$, M$\to$D$_{\text{c}}$, and M$\to$D$_{\text{n}}$. As in the main analysis, the transitions out of ``M after E$_{\text{n}}$'' share the components of the transitions out of M, and those out of ``E$_{\text{n}}$ after M'' share the components of the transitions out of E$_{\text{n}}$.

\begin{figure}[!htbp]
\centering
\includegraphics[width=0.96\textwidth]{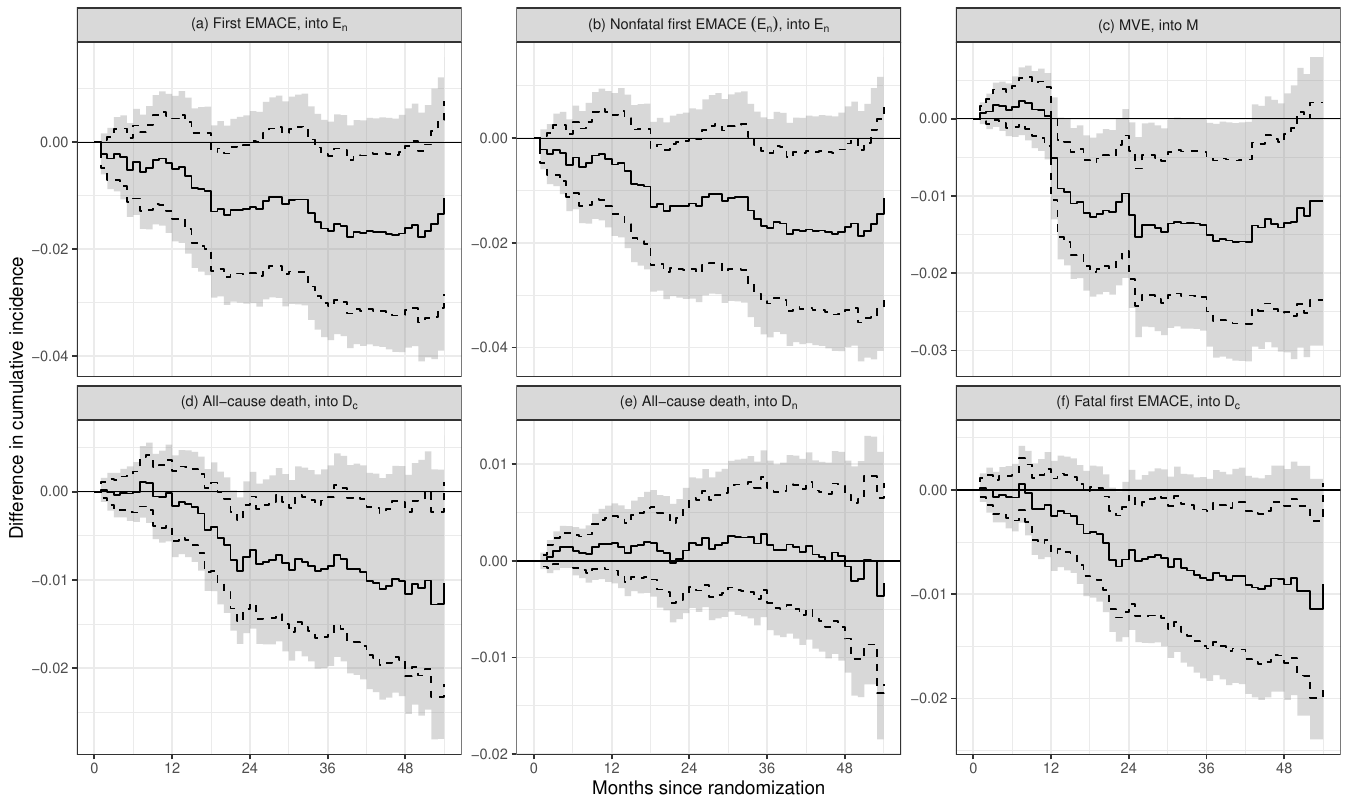}
\caption{Event-targeted effects in the secondary analysis in which fatal first EMACE is merged into CV death: difference in the counterfactual cumulative incidence between the regime in which the hazards of the designated transitions are taken from the liraglutide arm, with all other hazards at placebo, and the all-placebo regime. Solid: estimate; dashed: pointwise 95\% confidence intervals; shaded: 95\% simultaneous confidence band over 1 to 54 months.} \label{fig:secondary}
\end{figure}

Figure \ref{fig:secondary} and Table \ref{tab:secondary} show six event-targeted effects with the placebo reference: (a) on first EMACE, whose cumulative incidence is the sum of the sub-incidences along O$\to$E$_{\text{n}}$, O$\to$D$_{\text{c}}$, O$\to$M$\to$E$_{\text{n}}$, and O$\to$M$\to$D$_{\text{c}}$, switching the transitions into E$_{\text{n}}$; (b) on nonfatal first EMACE (E$_{\text{n}}$), switching the transitions into E$_{\text{n}}$; (c) on MVE, switching the transitions into M; (d) and (e) on all-cause death (D$_{\text{c}}$ or D$_{\text{n}}$), switching the transitions into D$_{\text{c}}$ and into D$_{\text{n}}$, respectively; and (f) on fatal first EMACE, the sum of the sub-incidences of D$_{\text{c}}$ along O$\to$D$_{\text{c}}$ and O$\to$M$\to$D$_{\text{c}}$, switching the transitions into D$_{\text{c}}$. Effect (c) coincides with the event-targeted effect on MVE in the main analysis, because the transitions into M are the same in both structures.

Switching the transitions into D$_{\text{c}}$, which now include fatal first EMACE, gained 0.33 months of survival by 54 months (SE 0.14, $P = 0.018$), most of it through fatal first EMACE (0.31 months, SE 0.12, $P = 0.011$), whereas switching the transitions into D$_{\text{n}}$ had no detectable effect on survival. Switching only the transitions into nonfatal first EMACE gained 0.60 months of EMACE-free time (SE 0.28, $P = 0.032$). This is less than the 0.90 months of the main analysis, in which the hazards of fatal first EMACE are switched as well. Together with the main analysis, these results are consistent with a survival effect that operates mainly through the hazards of a first EMACE, fatal or nonfatal, rather than through the hazards of death after a nonfatal event. The simultaneous bands exclude zero over part of follow-up for MVE, and only around 22 months for (d) and (f).

\begin{table}[!htbp]
\centering
\caption{Event-targeted effects in the secondary analysis in which fatal first EMACE is merged into CV death: time free of the outcome gained by 54 months (months; standard error) when the hazards of the switched transitions are taken from the liraglutide arm, with all other hazards at placebo.} \label{tab:secondary}
\begin{tabular}{lllcc}
\toprule
 & Outcome & Switched transitions & Time gained & $P$ \\
\midrule
(a) & First EMACE & Into E$_{\text{n}}$ & 0.603 (0.281) & 0.032 \\
(b) & Nonfatal first EMACE & Into E$_{\text{n}}$ & 0.627 (0.287) & 0.029 \\
(c) & MVE & Into M & 0.534 (0.187) & 0.004 \\
(d) & All-cause death & Into D$_{\text{c}}$ & 0.335 (0.141) & 0.018 \\
(e) & All-cause death & Into D$_{\text{n}}$ & $-$0.056 (0.107) & 0.599 \\
(f) & Fatal first EMACE & Into D$_{\text{c}}$ & 0.311 (0.123) & 0.011 \\
\bottomrule
\end{tabular}
\end{table}

\section{Proofs of theoretical results}

\subsection{Proof of Theorem 1} \label{sec:proof_thm1}

First, we show that the potential transition hazards are identifiable.
\begin{align*}
&\quad ~ \mathrm{d}\Lambda_{kg}^{a}(t \mid w,\overline{x}(t)) \\
&= \pr\{t< T^{a}_g<t+\mathrm{d}t \mid W=w, \overline{X}^{a}(t)=\overline{x}(t), X^{a}(t)=k\} \\
&= \pr\{t< T^{a}_g<t+\mathrm{d}t \mid W=w, \overline{X}^{a}(t)=\overline{x}(t), X^{a}(t)=k, A=a\} \\
&= \pr\{t< T^{a}_g<t+\mathrm{d}t \mid W=w, \overline{X}^a(t)=\overline{x}(t), X^a(t)=k, T_{\text{c}}^a \geq t, A=a\} \\
&= \pr\{t< T_g<t+\mathrm{d}t, \Delta_g = 1 \mid W=w, \overline{X}(t)=\overline{x}(t), X(t)=k, A=a\}.
\end{align*}
The second equality uses ignorability, the third random censoring, and the fourth consistency. Here $X(t)=k$ implies that censoring has not occurred by $t$, and positivity ensures that the conditioning event has positive probability (density) on the support of $\overline{X}^{a}(t)$ given $W$. This identifies $\mathrm{d}\Lambda_{kg}^{a}(t \mid w, \overline{x}(t))$ on that support only. By the last part of Assumption 3, the law of $\overline{X}^{\overline{a}}$ given $W$ is absolutely continuous with respect to that of $\overline{X}^{a}$ for $a = 0, 1$: the likelihood ratio of $\overline{X}^{\overline{a}}(t^*)$ to $\overline{X}^{a}(t^*)$ is a finite product of hazard ratios and survival ratios. Hence every hazard $\mathrm{d}\Lambda_{kg}^{a_{kg}}$ that enters the construction of $\overline{X}^{\overline{a}}$ is evaluated at a history at which it is identified.

Next, for a path $q\in\mathcal{Q}_j$, given the first $r-1$ transitions at times $t_1,\ldots,t_{r-1}$, the conditional density of the $r$-th transition at time $t_r$ is, by the competing-risks construction,
\begin{align*}
\mathrm{d}\pr_r(t_r \mid a_{q(r-1),\cdot}, w, q[\overline{t}_r])
&= \exp\left\{-\sum_{k\in\underline{\mathcal{G}}_{q(r-1)}} \Lambda_{q(r-1),k}^{a_{q(r-1),k}}(t_r \mid w,q[\overline{t}_r])\right\} \mathrm{d}\Lambda_{q(r-1),q(r)}^{a_{q(r-1),q(r)}}(t_r \mid w,q[\overline{t}_r]),
\end{align*}
where $q[\overline{t}_r]$ is the path history before $t_r$, that is, $\{(q(s),t_s): s=1,\ldots,r-1\}$ with $t_1 < \cdots < t_{r-1} < t_r$ and $x(t_r)=q(r-1)$. The sub-incidence of path $q$ at time $t$ is therefore
\begin{align*}
F_j^{\overline{a}}(t;q) = \int_{\mathcal{W}} \int_0^t \int_0^{t_{l(q)}} \cdots \int_0^{t_2} \prod_{r=1}^{l(q)} \mathrm{d} \pr_r(t_r \mid a_{q(r-1),\cdot}, w, q[\overline{t}_r]) \mathrm{d}P(w),
\end{align*}
and $F_j^{\overline{a}}(t) = \sum_{q\in\mathcal{Q}_j} F_j^{\overline{a}}(t;q)$. $\square$

\subsection{Proof of Lemma 1} \label{sec:proof_lemma1}

Throughout Sections E.2 and E.3 we fix $j \in \mathcal{G}$, the intervention $\overline{a}$, and $t \in [0, t^*]$, and abbreviate $F(t \mid w) = F_j^{\overline{a}}(t \mid w)$, $F(t \mid w; \overline{x}(s-)) = F_j^{\overline{a}}(t \mid w; \overline{x}(s-))$, and $\Delta_{kg}(s; w, \overline{x}(s-)) = \Delta_{kg}^{\overline{a}}(t \mid w; \overline{x}(s-))$, the quantities defined in Section 3.1. As in the main text, the cumulative hazards are taken to be absolutely continuous in calendar time, with rates $\lambda_{kg}^{a}(s \mid w, \overline{x}(s-))$, so that transition densities exist; continuity of the cumulative hazards suffices for the martingale arguments below. The cumulative hazard $\Lambda_{kg}^{a}(s \mid w, \overline{x}(s-))$ is accumulated over the current sojourn, as in Theorem 1. Let $\mathcal{E}^* = \mathcal{E} \cup \{(k, \text{c}): k \in \mathcal{G}\}$, let $N_{k\text{c}}(s)$ count censoring from state $k$ by time $s$, and let $\mathrm{d}\Lambda_{k\text{c}}^{a}(s \mid w, \overline{x}(s-))$ be the censoring hazard in arm $a$. The submodel parameter is denoted by $v$ to avoid confusion with the positivity constant $\epsilon$.

\paragraph{Step 1: likelihood and tangent space.}
Let $\mathcal{F}_s = \sigma\{A, W, \overline{X}(s)\}$ be the observed-data filtration. By Assumptions 1--4 (Section E.1), the $\mathcal{F}$-intensity of $N_{kg}$ is $Y_k(s)\,\mathrm{d}\Lambda_{kg}^{A}(s \mid W, \overline{X}(s-))$ for $(k,g) \in \mathcal{E}$, where $\Lambda_{kg}^{a}$ is the potential transition hazard, and that of $N_{k\text{c}}$ is $Y_k(s)\,\mathrm{d}\Lambda_{k\text{c}}^{A}(s \mid W, \overline{X}(s-))$. Hence
\[
M_{kg}(s) = N_{kg}(s) - \int_0^s Y_k(u)\,\mathrm{d}\Lambda_{kg}^{A}(u \mid W, \overline{X}(u-)), \qquad (k,g) \in \mathcal{E}^*,
\]
are square-integrable $\mathcal{F}$-martingales with continuous compensators. Since no two counting processes jump at the same time, $\langle M_{kg}, M_{k'g'}\rangle \equiv 0$ for $(k,g) \neq (k',g')$, and $\langle M_{kg}\rangle(s) = \int_0^s Y_k\,\mathrm{d}\Lambda_{kg}^{A}$. By Jacod's formula \citep{jacod1975multivariate, andersen2012statistical}, the likelihood of $O$ factorizes into $p(W)$, $\pr(A \mid W)$, and one factor per hazard in $\mathcal{E}^*$. The nonparametric model $\mathcal{M}$ leaves $p(w)$, $\pr(a \mid w)$, and every hazard unrestricted. Consider the one-dimensional submodels
\begin{equation}\label{eq:submodel}
\begin{aligned}
p_v(w) &= \{1 + v\, b_W(w)\}\, p(w), \qquad \pr_v(a \mid w) = \{1 + v\, b_A(a, w)\}\,\pr(a \mid w), \\
\mathrm{d}\Lambda_{kg,v}^{a}(s \mid w, \overline{x}(s-)) &= \{1 + v\, b_{kg}(s; a, w, \overline{x}(s-))\}\,\mathrm{d}\Lambda_{kg}^{a}(s \mid w, \overline{x}(s-)), \qquad (k,g) \in \mathcal{E}^*,
\end{aligned}
\end{equation}
where $b_W$, $b_A$, and $b_{kg}$ are bounded, $\E\, b_W(W) = 0$, $\E\{b_A(A,W) \mid W\} = 0$, and $b_{kg}(s; a, w, \overline{x}(s-))$ is a predictable function of the history. The score at $v = 0$ is
\begin{equation}\label{eq:score}
S(O) = b_W(W) + b_A(A, W) + \sum_{(k,g)\in\mathcal{E}^*} \int_0^{t^*} b_{kg}(s; A, W, \overline{X}(s-))\,\mathrm{d}M_{kg}(s).
\end{equation}
The tangent space $\dot{\mathcal{P}}$, the closed linear span of these scores in $L_2(P)$, is the orthogonal direct sum
\[
\dot{\mathcal{P}} = \dot{\mathcal{P}}_W \oplus \dot{\mathcal{P}}_A \oplus \bigoplus_{(k,g)\in\mathcal{E}^*} \dot{\mathcal{P}}_{kg},
\]
where $\dot{\mathcal{P}}_W = \{b(W): \E\,b(W) = 0\}$, $\dot{\mathcal{P}}_A = \{b(A,W): \E\{b(A,W) \mid W\} = 0\}$, and
\[
\dot{\mathcal{P}}_{kg} = \left\{\int_0^{t^*} b\,\mathrm{d}M_{kg}: b \text{ predictable}, \E\int_0^{t^*} b^2 Y_k\,\mathrm{d}\Lambda_{kg}^{A} < \infty\right\}.
\]
Orthogonality holds because $\E\{\int b\,\mathrm{d}M_{kg} \mid A, W\} = 0$, because $\E\{b_A(A,W) \mid W\} = 0$, and because $\E\{\int b\,\mathrm{d}M_{kg}\int b'\,\mathrm{d}M_{k'g'}\} = \E\int b\, b'\,\mathrm{d}\langle M_{kg}, M_{k'g'}\rangle = 0$ for $(k,g) \neq (k',g')$. Moreover, by the martingale representation theorem for marked point processes \citep{jacod1975multivariate, bremaud1981point}, every mean-zero square-integrable function of $O$ has the form \eqref{eq:score}, so $\dot{\mathcal{P}} = L_2^0(P)$ and the model is locally nonparametric. Consequently, suppose that $\psi \in L_2^0(P)$ satisfies
\begin{equation}\label{eq:gradient}
\frac{\partial}{\partial v} F_{j,v}^{\overline{a}}(t)\Big|_{v=0} = \E\{\psi(O)\, S(O)\}
\end{equation}
for every submodel \eqref{eq:submodel}, where $F_{j,v}^{\overline{a}}(t)$ denotes the target functional evaluated at $P_v$. Then $\psi$ is the unique influence function of $F_j^{\overline{a}}(t)$ and hence its efficient influence function \citep{bickel1993efficient, van2013weak}. We show that the function $\psi$ displayed in Lemma 1 satisfies \eqref{eq:gradient}.

\paragraph{Step 2: the target as a functional of the hazards.}
Given $W = w$, the hypothetical trajectory $\overline{X}^{\overline{a}}$ is, by construction, a marked point process on $\mathcal{G}$ whose intensity of a transition from $k$ to $g$ is $Y_k^{\overline{a}}(s)\,\mathrm{d}\Lambda_{kg}^{a_{kg}}(s \mid w, \overline{X}^{\overline{a}}(s-))$, where $Y_k^{\overline{a}}(s) = \mathds{1}\{X^{\overline{a}}(s-) = k\}$ and $N_{kg}^{\overline{a}}$ counts the hypothetical transitions from $k$ to $g$. Let $\pr^{\overline{a}}(\cdot \mid w)$ denote its law and $\E^{\overline{a}}(\cdot \mid w)$ the corresponding expectation. Theorem 1 states that $F(t \mid w) = \pr^{\overline{a}}(T_j^{\overline{a}} \leq t \mid w)$. The density of the history $\overline{X}^{\overline{a}}(s-)$ under $\pr^{\overline{a}}(\cdot \mid w)$, with respect to the product of counting measure on state sequences and Lebesgue measure on transition times, is the $p^{\overline{a}}(s, \overline{x}(s-) \mid w)$ defined in the main text. Under the submodel \eqref{eq:submodel} the hypothetical process has intensities $Y_k^{\overline{a}}(s)\{1 + v\, b_{kg}(s; a_{kg}, w, \overline{X}^{\overline{a}}(s-))\}\,\mathrm{d}\Lambda_{kg}^{a_{kg}}(s \mid w, \overline{X}^{\overline{a}}(s-))$; write $\pr_v^{\overline{a}}(\cdot \mid w)$ for its law and $F_v(t \mid w) = \pr_v^{\overline{a}}(T_j^{\overline{a}} \leq t \mid w)$. Then $F_{j,v}^{\overline{a}}(t) = \int F_v(t \mid w)\,p_v(w)\,\mathrm{d}w$ and, by the product rule,
\begin{equation}\label{eq:product_rule}
\frac{\partial}{\partial v} F_{j,v}^{\overline{a}}(t)\Big|_{v=0} = \int F(t \mid w)\,b_W(w)\,p(w)\,\mathrm{d}w + \int \frac{\partial}{\partial v}F_v(t \mid w)\Big|_{v=0}\, p(w)\,\mathrm{d}w,
\end{equation}
provided the derivative inside the second integral exists and is bounded, which we show next. Note that $F_v(t \mid w)$ depends neither on $b_A$ nor on the censoring directions $b_{k\text{c}}$.

\paragraph{Step 3: derivative with respect to the hazards.}
Write $b_{kg}(s) = b_{kg}(s; a_{kg}, w, \overline{X}^{\overline{a}}(s-))$ and $\mathrm{d}\Lambda_{kg}(s) = \mathrm{d}\Lambda_{kg}^{a_{kg}}(s \mid w, \overline{X}^{\overline{a}}(s-))$. For $|v| < 1/\sup|b_{kg}|$, the laws $\pr_v^{\overline{a}}(\cdot \mid w)$ and $\pr^{\overline{a}}(\cdot \mid w)$ are mutually absolutely continuous on $\sigma\{\overline{X}^{\overline{a}}(t)\}$ with likelihood ratio
\[
L_v(t) = \prod_{(k,g)\in\mathcal{E}} \prod_{s \leq t}\{1 + v\, b_{kg}(s)\}^{\mathrm{d}N_{kg}^{\overline{a}}(s)} \exp\left\{-v\int_0^t Y_k^{\overline{a}}(s)\,b_{kg}(s)\,\mathrm{d}\Lambda_{kg}(s)\right\}
\]
\citep{jacod1975multivariate, andersen2012statistical}. Since each state is visited at most once, $\overline{X}^{\overline{a}}$ has at most $|\mathcal{G}| - 1$ jumps, so $L_v(t)$ and $\partial_v L_v(t)$ are bounded uniformly in $v$ near zero. Let $\Phi = \mathds{1}(T_j^{\overline{a}} \leq t)$. By dominated convergence,
\begin{equation}\label{eq:score_method}
\frac{\partial}{\partial v}F_v(t \mid w)\Big|_{v=0} = \frac{\partial}{\partial v}\E^{\overline{a}}\{\Phi\,L_v(t) \mid w\}\Big|_{v=0} = \E^{\overline{a}}\left\{\Phi \sum_{(k,g)\in\mathcal{E}}\int_0^t b_{kg}(s)\,\mathrm{d}M_{kg}^{\overline{a}}(s) \Bigm| w\right\},
\end{equation}
where $M_{kg}^{\overline{a}}(s) = N_{kg}^{\overline{a}}(s) - \int_0^s Y_k^{\overline{a}}\,\mathrm{d}\Lambda_{kg}$ are the martingales of the hypothetical process under $\pr^{\overline{a}}(\cdot \mid w)$.

\paragraph{Step 4: martingale representation of $\Phi$.}
We claim that
\begin{equation}\label{eq:mart_rep}
\Phi = F(t \mid w) + \sum_{(k,g)\in\mathcal{E}}\int_0^t \Delta_{kg}(s; w, \overline{X}^{\overline{a}}(s-))\,\mathrm{d}M_{kg}^{\overline{a}}(s).
\end{equation}
For $s \leq t$ let $U(s) = \pr^{\overline{a}}\{T_j^{\overline{a}} \leq t \mid w, \overline{X}^{\overline{a}}(s)\}$. Because the hypothetical process is Markov on the space of histories, $U(s) = F(t \mid w; \overline{X}^{\overline{a}}(s))$, where $F(t \mid w; \overline{x}(s))$ denotes the counterfactual cumulative incidence given the history up to and including time $s$ (if a transition occurs at $s$, it is included). The process $U$ is a bounded martingale with $U(0) = F(t \mid w)$ and $U(t) = \mathds{1}\{j \in \overline{X}^{\overline{a}}(t)\} = \Phi$. Fix a history $\overline{x}(s-)$ with current state $k$, entered at time $t_{r-1} < s$, and suppose $j \notin \overline{x}(s-)$. Write $\overline{x}(s-)\oplus(g,u)$ for the history in which the sojourn in $k$ is extended to $u$ and followed by a transition to $g$ at $u$. Under the hypothetical hazards, the first transition out of $k$ after $s$ occurs to state $g$ at time $u$ with density $\exp\{-\sum_{g'\in\underline{\mathcal{G}}_k}[\Lambda_{kg'}^{a_{kg'}}(u \mid w, \overline{x}) - \Lambda_{kg'}^{a_{kg'}}(s \mid w, \overline{x})]\}\,\lambda_{kg}^{a_{kg}}(u \mid w, \overline{x})$. Conditioning on this transition gives the backward equation
\begin{equation}\label{eq:backward}
\begin{aligned}
F(t \mid w; \overline{x}(s-)) &= \sum_{g\in\underline{\mathcal{G}}_k}\int_s^t \exp\left\{-\sum_{g'\in\underline{\mathcal{G}}_k}\left[\Lambda_{kg'}^{a_{kg'}}(u \mid w, \overline{x}) - \Lambda_{kg'}^{a_{kg'}}(s \mid w, \overline{x})\right]\right\} \\
&\qquad\qquad \times F(t \mid w; \overline{x}(s-)\oplus(g,u))\,\lambda_{kg}^{a_{kg}}(u \mid w, \overline{x})\,\mathrm{d}u,
\end{aligned}
\end{equation}
because the event of no transition before $t$ contributes zero. (If $j \in \overline{x}(s-)$, then $F \equiv 1$ and $\Delta_{kg} \equiv 0$ for every later transition.) Differentiating \eqref{eq:backward} with respect to $s$ shows that, for a fixed sequence of past states and transition times, $s \mapsto F(t \mid w; \overline{x}(s-))$ is absolutely continuous during the sojourn in $k$, with
\begin{equation}\label{eq:backward_diff}
\begin{aligned}
\frac{\partial}{\partial s}F(t \mid w; \overline{x}(s-)) &= \sum_{g'\in\underline{\mathcal{G}}_k}\lambda_{kg'}^{a_{kg'}}(s \mid w, \overline{x})\,F(t \mid w; \overline{x}(s-)) \\
&\quad - \sum_{g\in\underline{\mathcal{G}}_k}\lambda_{kg}^{a_{kg}}(s \mid w, \overline{x})\,F(t \mid w; \overline{x}(s-)\oplus(g,s)) \\
&= -\sum_{g\in\underline{\mathcal{G}}_k}\Delta_{kg}(s; w, \overline{x}(s-))\,\lambda_{kg}^{a_{kg}}(s \mid w, \overline{x}).
\end{aligned}
\end{equation}
Along a realization of $\overline{X}^{\overline{a}}$, the process $U$ is therefore absolutely continuous between transitions with derivative \eqref{eq:backward_diff} evaluated at the current history, and jumps by $\Delta_{kg}(s; w, \overline{X}^{\overline{a}}(s-))$ at a transition from $k$ to $g$ at time $s$. Summing the continuous and the jump parts over $[0,t]$,
\[
U(t) - U(0) = \sum_{(k,g)\in\mathcal{E}}\left[\int_0^t \Delta_{kg}(s; w, \overline{X}^{\overline{a}}(s-))\,\mathrm{d}N_{kg}^{\overline{a}}(s) - \int_0^t \Delta_{kg}(s; w, \overline{X}^{\overline{a}}(s-))\,Y_k^{\overline{a}}(s)\,\mathrm{d}\Lambda_{kg}(s)\right],
\]
which is \eqref{eq:mart_rep}.

\paragraph{Step 5: the derivative under the hypothetical law.}
Substitute \eqref{eq:mart_rep} into \eqref{eq:score_method}. The martingale terms have mean zero, $\langle M_{kg}^{\overline{a}}, M_{k'g'}^{\overline{a}}\rangle \equiv 0$ for $(k,g) \neq (k',g')$ because the hypothetical counting processes have no common jumps, and $\E^{\overline{a}}\{\int b\,\mathrm{d}M_{kg}^{\overline{a}}\int \Delta\,\mathrm{d}M_{kg}^{\overline{a}} \mid w\} = \E^{\overline{a}}\{\int b\,\Delta\,\mathrm{d}\langle M_{kg}^{\overline{a}}\rangle \mid w\}$ for bounded integrands. Hence
\begin{equation}\label{eq:deriv_hyp}
\begin{aligned}
&\frac{\partial}{\partial v}F_v(t \mid w)\Big|_{v=0} \\
&= \sum_{(k,g)\in\mathcal{E}}\E^{\overline{a}}\left\{\int_0^t b_{kg}(s; a_{kg}, w, \overline{X}^{\overline{a}}(s-))\,\Delta_{kg}(s; w, \overline{X}^{\overline{a}}(s-))\,Y_k^{\overline{a}}(s)\,\lambda_{kg}^{a_{kg}}(s \mid w, \overline{X}^{\overline{a}}(s-))\,\mathrm{d}s \Bigm| w\right\}.
\end{aligned}
\end{equation}
This is the derivative of the g-formula with respect to the hazards, expressed as an expectation under the hypothetical process. It is bounded by $\sup|b_{kg}|$ times $\sum_{(k,g)}\E^{\overline{a}}\{N_{kg}^{\overline{a}}(t) \mid w\} \leq |\mathcal{G}| \sup|b_{kg}|$, so the exchange of differentiation and integration in \eqref{eq:product_rule} is justified. Steps 2--5 do not use Assumption 5.

\paragraph{Step 6: change of measure.}
Fix $s \in [0,t]$, $a \in \{0,1\}$, and $k \in \mathcal{G}$. By Assumptions 1--4, conditionally on $A = a$ and $W = w$ the observed process has event intensities $Y_k(s)\,\mathrm{d}\Lambda_{kg}^{a}(s \mid w, \overline{X}(s-))$ and censoring intensities $Y_k(s)\,\mathrm{d}\Lambda_{k\text{c}}^{a}(s \mid w, \overline{X}(s-))$, so by Jacod's formula the density of the event $\{\overline{X}(s-) \in \mathrm{d}\overline{x}(s-),\, T_{\text{c}} \geq s\}$ given $A = a$ and $W = w$ is
\[
p^{a}(s, \overline{x}(s-) \mid w)\,G^{a}(s \mid w, \overline{x}(s-)), \qquad G^{a}(s \mid w, \overline{x}(s-)) = \exp\left\{-\int_0^s \mathrm{d}\Lambda_{x(u-),\text{c}}^{a}(u \mid w, \overline{x}(u-))\right\},
\]
where $p^{a}$ is the density of the uncensored potential trajectory $\overline{X}^{a}(s-)$, and $G^{a}(s \mid w, \overline{x}(s-)) = \pr\{T_{\text{c}} \geq s \mid A = a, W = w, \overline{X}^{a}(s-) = \overline{x}(s-)\}$ by Assumption 2. Hence, for any bounded measurable function $\Phi_s$ of histories with current state $k$,
\begin{equation}\label{eq:change_measure}
\begin{aligned}
\E^{\overline{a}}\{Y_k^{\overline{a}}(s)\,\Phi_s(\overline{X}^{\overline{a}}(s-)) \mid w\} &= \int_{x(s-) = k}\Phi_s(\overline{x})\,p^{\overline{a}}(s, \overline{x} \mid w)\,\mathrm{d}\overline{x} \\
&= \E\{Y_k(s)\,\omega_k^{\overline{a}}(s; a, W, \overline{X}(s-))\,\Phi_s(\overline{X}(s-)) \mid A = a, W = w\},
\end{aligned}
\end{equation}
The second equality multiplies and divides the integrand by $p^{a}G^{a}$ and uses the definition of $\omega_k^{\overline{a}}$. Assumption 5 guarantees that $\omega_k^{\overline{a}}(s; a_{kg}, w, \cdot)$ is well defined and bounded by $C$ on the support of $p^{\overline{a}}$. Now apply \eqref{eq:change_measure} with $a = a_{kg}$ and $\Phi_s(\overline{x}) = b_{kg}(s; a_{kg}, w, \overline{x})\,\Delta_{kg}(s; w, \overline{x})\,\lambda_{kg}^{a_{kg}}(s \mid w, \overline{x})$ to each term of \eqref{eq:deriv_hyp}. Exchanging $\int_0^t \mathrm{d}s$ and the expectation (Fubini; all integrands are bounded), and using $\E\{\mathds{1}(A = a)\,f(W, \overline{X}) \mid W\} = \pr(A = a \mid W)\,\E\{f(W, \overline{X}) \mid A = a, W\}$, we obtain
\begin{equation}\label{eq:deriv_obs}
\begin{aligned}
&\int\frac{\partial}{\partial v}F_v(t \mid w)\Big|_{v=0}\,p(w)\,\mathrm{d}w \\
&= \sum_{(k,g)\in\mathcal{E}}\E\left\{\frac{\mathds{1}(A = a_{kg})}{\pr(A = a_{kg} \mid W)}\int_0^t b_{kg}(s; A, W, \overline{X}(s-))\,\omega_k^{\overline{a}}(s; a_{kg}, W, \overline{X}(s-))\right. \\
&\qquad\qquad\qquad \left. \times \Delta_{kg}(s; W, \overline{X}(s-))\,Y_k(s)\,\mathrm{d}\Lambda_{kg}^{A}(s \mid W, \overline{X}(s-))\right\},
\end{aligned}
\end{equation}
where we used that $a_{kg} = A$ on the event $\{A = a_{kg}\}$.

\paragraph{Step 7: verification of \eqref{eq:gradient}.}
Let
\[
H_{kg}(s; a, w, \overline{x}(s-)) = \frac{\mathds{1}(a = a_{kg})}{\pr(A = a_{kg} \mid w)}\,\omega_k^{\overline{a}}(s; a_{kg}, w, \overline{x}(s-))\,\Delta_{kg}(s; w, \overline{x}(s-)),
\]
which is predictable and bounded by $C/\epsilon$ under Assumptions 3 and 5 since $|\Delta_{kg}| \leq 1$, and let
\[
\psi(O) = F(t \mid W) - F_j^{\overline{a}}(t) + \sum_{(k,g)\in\mathcal{E}}\int_0^t H_{kg}(s; A, W, \overline{X}(s-))\,\mathrm{d}M_{kg}(s)
\]
be the function displayed in Lemma 1. Then $\psi \in L_2(P)$, $\E\,\psi(O) = 0$, and $\psi \in \dot{\mathcal{P}}_W \oplus \bigoplus_{(k,g)\in\mathcal{E}}\dot{\mathcal{P}}_{kg} \subset \dot{\mathcal{P}}$. For a score of the form \eqref{eq:score}, the orthogonality relations of Step 1 give
\begin{equation}\label{eq:psi_S}
\begin{aligned}
\E\{\psi(O)S(O)\} &= \E\left[\{F(t \mid W) - F_j^{\overline{a}}(t)\}\,b_W(W)\right] \\
&\quad + \sum_{(k,g)\in\mathcal{E}}\E\left\{\int_0^t H_{kg}(s; A, W, \overline{X}(s-))\,b_{kg}(s; A, W, \overline{X}(s-))\right. \\
&\qquad\qquad\qquad\qquad \left. \times Y_k(s)\,\mathrm{d}\Lambda_{kg}^{A}(s \mid W, \overline{X}(s-))\right\},
\end{aligned}
\end{equation}
Here we used three facts. The term $F(t \mid W) - F_j^{\overline{a}}(t)$ is orthogonal to $b_A$ and to every martingale term. The integral $\int H_{kg}\,\mathrm{d}M_{kg}$ is orthogonal to $b_W$, $b_A$, and to $\int b_{k'g'}\,\mathrm{d}M_{k'g'}$ for $(k',g') \neq (k,g)$, including the censoring directions. Finally, $\E\{\int H_{kg}\,\mathrm{d}M_{kg}\int b_{kg}\,\mathrm{d}M_{kg}\} = \E\int H_{kg}\,b_{kg}\,\mathrm{d}\langle M_{kg}\rangle$. The first term on the right-hand side of \eqref{eq:psi_S} is the first term of \eqref{eq:product_rule}, and the second is \eqref{eq:deriv_obs}, which equals the second term of \eqref{eq:product_rule} by \eqref{eq:deriv_hyp}. Thus \eqref{eq:gradient} holds for every submodel \eqref{eq:submodel}, and $\psi$ is the efficient influence function of $F_j^{\overline{a}}(t)$ in the nonparametric model. Neither the propensity directions $b_A$ nor the censoring directions $b_{k\text{c}}$ contribute to either side of \eqref{eq:gradient}, which is why the propensity score and the censoring hazards appear in $\psi$ only as weights. $\square$

\subsection{Proof of Lemma \ref{lem:extmarkov}} \label{sec:proof_lemmaS1}

Let $\mathcal{M}_h \subset \mathcal{M}$ be the model of Assumption \ref{ass:extmarkov}, in which $\mathrm{d}\Lambda_{kg}^{a}(s \mid w, \overline{x}(s-)) = \mathrm{d}\Lambda_{kg}^{a}(s \mid w, h(\overline{x}(s-)))$ for $(k,g) \in \mathcal{E}$, where $h(\overline{x}(s-))$ is the class of the history before $s$, the class-specific hazards being otherwise unrestricted, and the covariate distribution, the propensity score, and the censoring hazards unrestricted. We use the notation of the proof of Lemma 1 and the quantities $P_k^{\overline{a}}(s \mid w; c)$, $D_{kg}^{\overline{a}}(t \mid w; c, s)$, and $\pi_k^{a}(s \mid w; c)$ defined in Section \ref{sec:extmarkov}.

\paragraph{Tangent space.}
A submodel \eqref{eq:submodel} stays in $\mathcal{M}_h$ if and only if the directions $b_{kg}$, $(k,g) \in \mathcal{E}$, depend on the history only through its class, $b_{kg}(s; a, w, \overline{x}(s-)) = b_{kg}(s; a, w, h(\overline{x}(s-)))$. The tangent space of $\mathcal{M}_h$ is therefore
\[
\dot{\mathcal{P}}^h = \dot{\mathcal{P}}_W \oplus \dot{\mathcal{P}}_A \oplus \bigoplus_{(k,g)\in\mathcal{E}}\dot{\mathcal{P}}_{kg}^h \oplus \bigoplus_{k\in\mathcal{G}}\dot{\mathcal{P}}_{k\text{c}}, \qquad \dot{\mathcal{P}}_{kg}^h = \left\{\int_0^{t^*} b(s; A, W, h(\overline{X}(s-)))\,\mathrm{d}M_{kg}(s)\right\} \subset \dot{\mathcal{P}}_{kg},
\]
with the same orthogonality relations as in Step 1 of the proof of Lemma 1. The efficient influence function in $\mathcal{M}_h$ is the unique element $\psi_h \in \dot{\mathcal{P}}^h$ that satisfies \eqref{eq:gradient} for every submodel of $\mathcal{M}_h$ \citep{bickel1993efficient}.

\paragraph{The derivative in $\mathcal{M}_h$.}
Steps 2--5 of the proof of Lemma 1 do not use Assumption 5 and remain valid in $\mathcal{M}_h$. For class-measurable directions, group the histories in \eqref{eq:deriv_hyp} by class. Under Assumption \ref{ass:extmarkov} the rate $\lambda_{kg}^{a_{kg}}(s \mid w, \cdot)$ is constant on each class, and by the definitions of $P_k^{\overline{a}}$ and $D_{kg}^{\overline{a}}$,
\[
\E^{\overline{a}}\left\{\Delta_{kg}(s; w, \overline{X}^{\overline{a}}(s-))\,Y_k^{\overline{a}}(s)\,\mathds{1}\{\overline{X}^{\overline{a}}(s-) \in c\} \Bigm| w\right\} = P_k^{\overline{a}}(s \mid w; c)\,D_{kg}^{\overline{a}}(t \mid w; c, s),
\]
so that
\begin{equation}\label{eq:deriv_class}
\frac{\partial}{\partial v}F_v(t \mid w)\Big|_{v=0} = \sum_{(k,g)\in\mathcal{E}}\sum_{c}\int_0^t b_{kg}(s; a_{kg}, w, c)\,P_k^{\overline{a}}(s \mid w; c)\,D_{kg}^{\overline{a}}(t \mid w; c, s)\,\mathrm{d}\Lambda_{kg}^{a_{kg}}(s \mid w, c),
\end{equation}
the sum running over the finitely many classes $c$.

\paragraph{Verification.}
Let
\[
\bar{H}_{kg}(s; a, w, c) = \frac{\mathds{1}(a = a_{kg})}{\pr(A = a_{kg} \mid w)}\,\frac{P_k^{\overline{a}}(s \mid w; c)\,D_{kg}^{\overline{a}}(t \mid w; c, s)}{\pi_k^{a_{kg}}(s \mid w; c)},
\]
with the convention $\bar{H}_{kg} = 0$ when $P_k^{\overline{a}}(s \mid w; c) = 0$. By Assumption 3 and the positivity part of Assumption \ref{ass:extmarkov}, $|\bar{H}_{kg}| \leq \epsilon^{-2}$. Hence
\[
\psi_h(O) = F(t \mid W) - F_j^{\overline{a}}(t) + \sum_{(k,g)\in\mathcal{E}}\int_0^t \bar{H}_{kg}(s; A, W, h(\overline{X}(s-)))\,\mathrm{d}M_{kg}(s),
\]
which is the transition form of Lemma \ref{lem:extmarkov}, belongs to $\dot{\mathcal{P}}_W \oplus \bigoplus_{(k,g)\in\mathcal{E}}\dot{\mathcal{P}}_{kg}^h \subset \dot{\mathcal{P}}^h$. For a score $S$ of a submodel of $\mathcal{M}_h$, the orthogonality relations give, as in Step 7,
\begin{align*}
\E\{\psi_h(O)S(O)\} &= \E\left[\{F(t \mid W) - F_j^{\overline{a}}(t)\}\,b_W(W)\right] \\
&\quad + \sum_{(k,g)\in\mathcal{E}}\E\left\{\int_0^t \bar{H}_{kg}(s; A, W, h(\overline{X}(s-)))\,b_{kg}(s; A, W, h(\overline{X}(s-)))\right. \\
&\qquad\qquad\qquad\qquad \left. \times Y_k(s)\,\mathrm{d}\Lambda_{kg}^{A}(s \mid W, h(\overline{X}(s-)))\right\} \\
&= \E\left[\{F(t \mid W) - F_j^{\overline{a}}(t)\}\,b_W(W)\right] \\
&\quad + \sum_{(k,g)\in\mathcal{E}}\E\left\{\sum_{c}\int_0^t \bar{H}_{kg}(s; A, W, c)\,b_{kg}(s; A, W, c)\,\pi_k^{A}(s \mid W; c)\,\mathrm{d}\Lambda_{kg}^{A}(s \mid W, c)\right\} \\
&= \E\left[\{F(t \mid W) - F_j^{\overline{a}}(t)\}\,b_W(W)\right] \\
&\quad + \sum_{(k,g)\in\mathcal{E}}\E\left\{\sum_{c}\int_0^t b_{kg}(s; a_{kg}, W, c)\,P_k^{\overline{a}}(s \mid W; c)\,D_{kg}^{\overline{a}}(t \mid W; c, s)\right. \\
&\qquad\qquad\qquad\qquad \left. \times \mathrm{d}\Lambda_{kg}^{a_{kg}}(s \mid W, c)\right\}.
\end{align*}
The second equality conditions on $(A, W)$ inside the $\mathrm{d}s$-integral. It uses $\E[Y_k(s)\,\mathds{1}\{\overline{X}(s-) \in c\} \mid A, W] = \pi_k^{A}(s \mid W; c)$ and the fact that $\bar{H}_{kg}$, $b_{kg}$, and the class-specific hazard are functions of $(s, A, W, c)$ only. The third equality uses $\E\{\mathds{1}(A = a)\,f(W) \mid W\} = \pr(A = a \mid W)\,f(W)$, which cancels the propensity score and $\pi_k^{a_{kg}}$. By \eqref{eq:product_rule} and \eqref{eq:deriv_class}, the right-hand side equals $\partial_v F_{j,v}^{\overline{a}}(t)|_{v=0}$. Hence $\psi_h$ satisfies \eqref{eq:gradient} on $\dot{\mathcal{P}}^h$ and lies in $\dot{\mathcal{P}}^h$, so it is the efficient influence function in $\mathcal{M}_h$.

\paragraph{Relation to Lemma 1.}
When Assumption 5 holds as well, the function $\psi$ of Lemma 1 satisfies \eqref{eq:gradient} for every submodel of $\mathcal{M}_h \subset \mathcal{M}$, so it is an influence function in $\mathcal{M}_h$. Moreover, $\psi_h$ is the projection of $\psi$ onto $\dot{\mathcal{P}}^h$: $\psi_h \in \dot{\mathcal{P}}^h$, and $\psi - \psi_h = \sum_{(k,g)} \int_0^t (H_{kg} - \bar{H}_{kg})\,\mathrm{d}M_{kg}$ is orthogonal to $\dot{\mathcal{P}}^h$, because \eqref{eq:change_measure} with $\Phi_s(\overline{x}) = \Delta_{kg}(s; w, \overline{x})\,\mathds{1}\{\overline{x} \in c\}$ gives
\[
\E\left[H_{kg}(s; A, W, \overline{X}(s-))\,Y_k(s)\,\mathds{1}\{\overline{X}(s-) \in c\} \mid A = a_{kg}, W = w\right] = \pi_k^{a_{kg}}(s \mid w; c)\,\bar{H}_{kg}(s; a_{kg}, w, c),
\]
so that $\E\{\int (H_{kg} - \bar{H}_{kg})\,\mathrm{d}M_{kg} \int b\,\mathrm{d}M_{kg}\} = 0$ for every class-measurable $b$. The nonparametric efficient influence function thus remains a valid influence function under Assumption \ref{ass:extmarkov}, with variance no smaller than that of $\psi_h$. The variances are equal only if $\omega_k^{\overline{a}}\Delta_{kg}$ is constant across the histories within each class, because the class average in $\psi_h$ removes this variability. In particular, the one-step estimator of the main text, which is based on $\psi$, remains consistent and asymptotically normal in $\mathcal{M}_h$ with variance $\E(\psi^2)$, but is not efficient there. The argument uses the partition only through the class indicator and the class positivity. Lemma \ref{lem:extmarkov} therefore holds for any finite partition satisfying Assumption \ref{ass:extmarkov}, including partitions that record a discretized sojourn time. In that case the class of the history changes with $s$, and all class indicators are evaluated at each $s$ accordingly.

\paragraph{Path form.}
It remains to show that the transition form equals the path form in Lemma \ref{lem:extmarkov}. Fix $(k,g) \in \mathcal{E}$, $s \in [0,t]$, and a class $c$. By the definitions of $P_k^{\overline{a}}$ and $D_{kg}^{\overline{a}}$,
\begin{align*}
&P_k^{\overline{a}}(s \mid w; c)\,D_{kg}^{\overline{a}}(t \mid w; c, s) \\
&= \int_{x(s-) = k,\ \overline{x}(s-) \in c}\left\{F(t \mid w; \overline{x}(s-)\oplus(g,s)) - F(t \mid w; \overline{x}(s-))\right\}p^{\overline{a}}(s, \overline{x}(s-) \mid w)\,\mathrm{d}\overline{x}(s-).
\end{align*}
Expanding $F(t \mid w; \cdot)$ over the paths that complete the given history to $j$ by time $t$, and using the product form of $p^{\overline{a}}$ and of the sub-densities in Theorem 1, the first term equals
\begin{align*}
&\sum_{q\in\mathcal{Q}_j}\ \sum_{r:\, q(r-1) = k,\ q(r) = g}\ \int_{t_1 < \cdots < t_{r-1} < s < t_{r+1} < \cdots < t_{l(q)} \leq t}\mathds{1}\{h(q[\overline{t}_r]) = c\} \\
&\qquad\qquad \times \prod_{\ell \neq r}\mathrm{d}\pr_\ell(t_\ell \mid a_{q(\ell-1),\bcdot}, w, q[\overline{t}_\ell])\ \frac{\mathrm{d}\pr_r(s \mid a_{q(r-1),\bcdot}, w, q[\overline{t}_r])}{\mathrm{d}\Lambda_{q(r-1),q(r)}^{a_{q(r-1),q(r)}}(s \mid w, q[\overline{t}_r])},
\end{align*}
that is, the sub-density of path $q$ with its $r$-th transition at time $t_r = s$, divided by the hazard of that transition and integrated over the remaining transition times, where $h(q[\overline{t}_r])$ is the class of the history $q[\overline{t}_r]$ at time $s$. The second term equals
\[
\sum_{q\in\mathcal{Q}_j}\ \sum_{r:\, q(r-1) = k}\ \int_{t_1 < \cdots < t_{r-1} < s \leq t_r < \cdots < t_{l(q)} \leq t}\mathds{1}\{h(q[\overline{t}_r]) = c\}\prod_{\ell=1}^{l(q)}\mathrm{d}\pr_\ell(t_\ell \mid a_{q(\ell-1),\bcdot}, w, q[\overline{t}_\ell]),
\]
the probability that the hypothetical trajectory follows $q$ to $j$ by time $t$ and is in state $k$, with a history in class $c$, just before $s$. Substitute these expressions into the transition form with $c = h(\overline{X}(s-))$. Each class records the current state $k$, and $g$ cannot have been visited while the unit is in $k$, so $\{\overline{X}(s-) \in c\}$ implies $\{X(s-) = k, T_g \geq s\}$; moreover $\{X(s-) = k\}$ implies $T_{\text{c}} \geq s$. Hence $\pi_k^{a_{kg}}(s \mid W; h(q[\overline{t}_r])) = \pr\{T_g \wedge T_{\text{c}} \geq s, \overline{X}(s-) \in h(q[\overline{t}_r]) \mid A = a_{kg}, W\}$ is the denominator in the definition of $\eta$. Now exchange the integral over $s$ with the integrals over the transition times. For a given $(q, r)$ with $q(r-1) = k$ and $q(r) = g$, the first term contributes
\[
\int_0^t\cdots\int_0^{t_2}\prod_{\ell=1}^{l(q)}\mathrm{d}\pr_\ell(t_\ell \mid a_{q(\ell-1),\bcdot}, W, q[\overline{t}_\ell])\ \frac{\mathrm{d}\eta_{q(r-1),q(r)}^{a_{q(r-1),q(r)}}(t_r \mid W, q[\overline{t}_r])}{\mathrm{d}\Lambda_{q(r-1),q(r)}^{a_{q(r-1),q(r)}}(t_r \mid W, q[\overline{t}_r])},
\]
and, after summing over $g \in \underline{\mathcal{G}}_{q(r-1)}$, the second term contributes for a given $(q, r)$
\[
-\int_0^t\cdots\int_0^{t_2}\prod_{\ell=1}^{l(q)}\mathrm{d}\pr_\ell(t_\ell \mid a_{q(\ell-1),\bcdot}, W, q[\overline{t}_\ell])\ \sum_{g\in\underline{\mathcal{G}}_{q(r-1)}}\eta_{q(r-1),g}^{a_{q(r-1),g}}(t_r \mid W, q[\overline{t}_r]),
\]
because $\eta_{q(r-1),g}^{a_{q(r-1),g}}(t_r \mid W, q[\overline{t}_r])$ integrates the weighted martingale increment over $s \in (t_{r-1}, t_r]$, with the class indicator $\mathds{1}\{\overline{X}(s-) \in h(q[\overline{t}_r])\}$ evaluated at the left limit at each $s$. The indicator inside $M_{q(r-1),g}$ restricts $N_g$ and $Y_{\bcdot g}$ to transitions out of $q(r-1)$, so that $\mathrm{d}M_{q(r-1),g}(s; a, W, q[\overline{t}_r]) = \mathds{1}\{\overline{X}(s-) \in h(q[\overline{t}_r])\}\,\mathrm{d}M_{q(r-1),g}(s)$ in the notation of Step 1. (Evaluating the class at $\overline{X}(s)$ instead would make the indicator vanish at the jump of $N_g$, since the history then records the new state, and would destroy predictability.) Summing over $r$ and over $q \in \mathcal{Q}_j$ gives the path form. $\square$

\subsection{Derivations for the illness--death example} \label{sec:idm}

\paragraph{The nonparametric EIF \eqref{eif_id}.}
Under the Markov property with censoring depending on $w$ only, the density of the history ``in O at $s-$'' under the intervention is $p^{\overline{a}}(s, \overline{x}(s-) \mid w) = P_{\text{O}}^{\overline{a}}(s \mid w)$, and in arm $a$ it is $p^{a}(s, \overline{x}(s-) \mid w) = P_{\text{O}}^{a}(s \mid w)$. By the definition of $\omega$ in Section 3.1, $\omega_{\text{O}}^{\overline{a}}(s; a, w) = P_{\text{O}}^{\overline{a}}(s \mid w) / \{P_{\text{O}}^{a}(s \mid w)\, G^{a}(s \mid w)\}$. The history ``entered I at $u$ and in I at $s-$'' has density $P_{\text{O}}^{\overline{a}}(u \mid w)\, \lambda_{\text{OI}}^{a_{\text{OI}}}(u \mid w)\, \exp\{-\Lambda_{\text{ID}}^{a_{\text{ID}}}(s \mid w) + \Lambda_{\text{ID}}^{a_{\text{ID}}}(u \mid w)\}$ under the intervention and $P_{\text{O}}^{a}(u \mid w)\, \lambda_{\text{OI}}^{a}(u \mid w)\, \exp\{-\Lambda_{\text{ID}}^{a}(s \mid w) + \Lambda_{\text{ID}}^{a}(u \mid w)\}$ in arm $a$. With $a = a_{\text{ID}}$ the exponential factors coincide, which gives the expression for $\omega_{\text{I}}^{\overline{a}}(s; a, w, u)$ in Section \ref{sec:extmarkov}. Because the hazards are Markov, the counterfactual cumulative incidence given a history depends on it only through the current state, so $\Delta_{kg}^{\overline{a}}(t \mid w; \overline{x}(s-)) = F_{\text{D}}^{\overline{a}}(t \mid w; g, s) - F_{\text{D}}^{\overline{a}}(t \mid w; k, s)$. The conditional incidences $F_{\text{D}}^{\overline{a}}(t \mid w; k, s)$ of Section \ref{sec:extmarkov} follow from Theorem 1 started from the current state; the expression for $F_{\text{D}}^{\overline{a}}(t \mid w; \text{O}, s)$ is the backward equation \eqref{eq:backward}. Substituting into Lemma 1 gives \eqref{eif_id}. With $P_{\text{I}}^{\overline{a}}(s \mid w)$ the occupancy of I under the intervention, as defined in Section \ref{sec:extmarkov}, the g-formula of Theorem 1 reads
\[
F_{\text{D}}^{\overline{a}}(t \mid w) = \int_0^t P_{\text{O}}^{\overline{a}}(s \mid w) \, \mathrm{d}\Lambda_{\text{OD}}^{a_{\text{OD}}}(s \mid w) + \int_0^t P_{\text{I}}^{\overline{a}}(s \mid w) \, \mathrm{d}\Lambda_{\text{ID}}^{a_{\text{ID}}}(s \mid w),
\]
the two terms being the sub-incidences along O$\to$D and O$\to$I$\to$D.

\paragraph{The projected EIF \eqref{eif_id_h}.}
Under the Markov property, the classes in Assumption \ref{ass:extmarkov} are the states. For the transition $(k,g)$, the class of the hypothetical history during the sojourn in $k$ is ``currently in $k$,'' so $P_k^{\overline{a}}(s \mid W; c) = P_k^{\overline{a}}(s \mid W)$ is the occupancy probability of $k$ under the intervention and $\pi_k^{a}(s \mid W; c) = \pi_k^{a}(s \mid W)$. Since the counterfactual cumulative incidence given a history depends on the history only through the current state, $F_{\text{D}}^{\overline{a}}(t \mid W; \overline{x}(s-)) = F_{\text{D}}^{\overline{a}}(t \mid W; x(s-), s)$, so $\Delta_{kg}^{\overline{a}}(t \mid W; \overline{x}(s-)) = F_{\text{D}}^{\overline{a}}(t \mid W; g, s) - F_{\text{D}}^{\overline{a}}(t \mid W; k, s)$ for every history in the class, and $D_{kg}^{\overline{a}}(t \mid W; c, s)$ equals the same difference. Substituting into the transition form of Lemma \ref{lem:extmarkov} gives \eqref{eif_id_h}.

\subsection{Proof of Theorem 2} \label{sec:proof_thm2}

Throughout this section and the next we fix $j$, $\overline{a}$, and $t$, abbreviate as in Section E.2, and write $e(a \mid w) = \pr(A = a \mid w)$. Let $\p$ be the true distribution and $\pn$ the empirical measure. A set of nuisance functions $\p'$ consists of a propensity score $e'$, censoring survival functions $G'^{a}$, and transition hazards $\Lambda'^{a}_{kg}$ for $(k,g) \in \mathcal{E}$ and $a \in \{0,1\}$. From $\p'$ we compute $F'(t \mid w)$, $F'(t \mid w; \overline{x}(s-))$, and $\Delta'_{kg}(s; w, \overline{x}(s-))$ by Theorem 1 and the definition of $\Delta$ in Section 3.1, using the hazards $\Lambda'^{a_{kg}}_{kg}$. The density ratio $\rho'_k(s; a, w, \overline{x}(s-)) = p'^{\overline{a}}(s, \overline{x}(s-) \mid w) / p'^{a}(s, \overline{x}(s-) \mid w)$ is computed by the product formula of Equation (13), using the hazards of both arms, and the weight is $\omega'_k = \rho'_k / G'^{a}$. Let
\[
\psi(\p')(O) = F'(t \mid W) + \sum_{(k,g)\in\mathcal{E}} \frac{\mathds{1}(A = a_{kg})}{e'(a_{kg} \mid W)} \int_0^t \omega'_k(s; a_{kg}, W, \overline{X}(s-))\, \Delta'_{kg}(s; W, \overline{X}(s-))\, \mathrm{d}M'_{kg}(s),
\]
with $\mathrm{d}M'_{kg}(s) = \mathrm{d}N_{kg}(s) - Y_k(s)\,\mathrm{d}\Lambda'^{a_{kg}}_{kg}(s \mid W, \overline{X}(s-))$, so that $\widetilde{F}_j^{\overline{a}}(t) = \pn \psi(\widehat\p)$ and $\p\psi(\p) = F_j^{\overline{a}}(t)$, because the debiasing term has mean zero under the truth. Here and below, $\p$ integrates over a new observation with the nuisance functions held fixed.

\paragraph{An exact expression for the bias.}
Fix nuisance functions $\p'$ with $e' \geq \epsilon$ and bounded $\omega'_k$. We show that
\begin{equation}\label{eq:remainder}
\p\psi(\p') - F_j^{\overline{a}}(t) = \E \sum_{(k,g)\in\mathcal{E}} \E^{\overline{a}}\left[\int_0^t \left\{1 - r'_{kg}(s)\, \frac{\rho'_k(s)}{\rho_k(s)}\right\} \Delta'_{kg}(s)\, Y_k^{\overline{a}}(s)\, \mathrm{d}\{\Lambda'^{a_{kg}}_{kg} - \Lambda^{a_{kg}}_{kg}\}(s \mid W, \overline{X}^{\overline{a}}(s-)) \Bigm| W \right],
\end{equation}
where $\E^{\overline{a}}(\cdot \mid W)$ is the expectation under the law of the true hypothetical process given $W$ (Section E.2, Step 2), all functions are evaluated at $(s; a_{kg}, W, \overline{X}^{\overline{a}}(s-))$, and
\[
r'_{kg}(s) = \frac{e(a_{kg} \mid W)\, G^{a_{kg}}(s \mid W, \overline{X}^{\overline{a}}(s-))}{e'(a_{kg} \mid W)\, G'^{a_{kg}}(s \mid W, \overline{X}^{\overline{a}}(s-))}
\]
is the ratio of the true to the working propensity score and censoring survival.

\textit{Plug-in term.} Fix $W = w$ and let $U'(s) = F'(t \mid w; \overline{X}^{\overline{a}}(s))$ along a trajectory of the true hypothetical process. Then $U'(0) = F'(t \mid w)$ and $U'(t) = \mathds{1}\{j \in \overline{X}^{\overline{a}}(t)\}$, whatever the hazards from which $F'$ is computed. By the backward equation \eqref{eq:backward_diff} applied to the hazards $\Lambda'$, $U'$ is absolutely continuous between the transitions of the true process, with derivative $-\sum_{g \in \underline{\mathcal{G}}_k} \Delta'_{kg}\, \lambda'^{a_{kg}}_{kg}$ while the process is in $k$, and it jumps by $\Delta'_{kg}$ at a transition from $k$ to $g$. Hence
\[
U'(t) - U'(0) = \sum_{(k,g)\in\mathcal{E}} \int_0^t \Delta'_{kg}\, \mathrm{d}M_{kg}^{\overline{a}}(s) + \sum_{(k,g)\in\mathcal{E}} \int_0^t \Delta'_{kg}\, Y_k^{\overline{a}}(s)\, \mathrm{d}\{\Lambda^{a_{kg}}_{kg} - \Lambda'^{a_{kg}}_{kg}\}(s \mid w, \overline{X}^{\overline{a}}(s-)),
\]
where $M^{\overline{a}}_{kg}$ are the martingales of the true hypothetical process (Section E.2, Step 3). Taking expectations under $\pr^{\overline{a}}(\cdot \mid w)$,
\begin{equation}\label{eq:duhamel}
F'(t \mid w) - F(t \mid w) = \sum_{(k,g)\in\mathcal{E}} \E^{\overline{a}}\left[\int_0^t \Delta'_{kg}\, Y_k^{\overline{a}}(s)\, \mathrm{d}\{\Lambda'^{a_{kg}}_{kg} - \Lambda^{a_{kg}}_{kg}\}(s \mid w, \overline{X}^{\overline{a}}(s-)) \Bigm| w\right],
\end{equation}
which is Duhamel's equation for the hypothetical process; if the cumulative hazards $\Lambda'$ have jumps, as Breslow estimators do, it holds with product integrals in place of exponentials in $F'$ \citep[Section II.6]{andersen2012statistical}.

\textit{Debiasing term.} Given $A = a$ and $W$, the $\mathcal{F}$-intensity of $N_{kg}$ is $Y_k(s)\,\mathrm{d}\Lambda^{a}_{kg}(s \mid W, \overline{X}(s-))$ (Section E.2, Step 1), so for a bounded predictable $H$, $\E\{\int_0^t H\, \mathrm{d}M'_{kg} \mid A = a, W\} = \E\{\int_0^t H\, Y_k\, \mathrm{d}(\Lambda^{a}_{kg} - \Lambda'^{a}_{kg}) \mid A = a, W\}$. By \eqref{eq:change_measure} read from right to left, $\E\{Y_k(s)\, \Phi_s(\overline{X}(s-)) \mid A = a, W = w\} = \E^{\overline{a}}\{Y_k^{\overline{a}}(s)\, \Phi_s(\overline{X}^{\overline{a}}(s-)) / \omega_k^{\overline{a}}(s; a, w, \overline{X}^{\overline{a}}(s-)) \mid w\}$ for bounded $\Phi_s$, where $1/\omega_k^{\overline{a}} = p^{a} G^{a} / p^{\overline{a}}$ is well defined on the support of $p^{\overline{a}}$ by the last part of Assumption 3. With $a = a_{kg}$ and $\E\{\mathds{1}(A = a)\, f(W, \overline{X}) \mid W\} = e(a \mid W)\, \E\{f(W, \overline{X}) \mid A = a, W\}$,
\[
    \begin{aligned}
        & \E\{\psi(\p')(O) - F'(t \mid W) \mid W\} \\
        = & - \sum_{(k,g)\in\mathcal{E}} \E^{\overline{a}}\left[\int_0^t \frac{e(a_{kg} \mid W)}{e'(a_{kg} \mid W)}\, \frac{\omega'_k}{\omega_k}\, \Delta'_{kg}\, Y_k^{\overline{a}}(s)\, \mathrm{d}\{\Lambda'^{a_{kg}}_{kg} - \Lambda^{a_{kg}}_{kg}\}(s \mid W, \overline{X}^{\overline{a}}(s-)) \Bigm| W\right].
    \end{aligned}
\]
Since $\omega'_k / \omega_k = (\rho'_k / \rho_k)(G^{a_{kg}} / G'^{a_{kg}})$, adding \eqref{eq:duhamel} and integrating over $W$ gives \eqref{eq:remainder}.

\paragraph{Structure of the density ratio.}
By Equation (13), $\rho_k(s; a, w, \cdot)$ is a product of hazard ratios and survival ratios of the transitions in $\mathcal{U}_k$, and a factor equals one identically whenever the transition is assigned arm $a$, because the same hazard then appears in the numerator and in the denominator. Hence $\rho_k$, and likewise $\rho'_k$, involves only the hazards, in both arms, of the transitions in $\mathcal{U}_k$ with $a_{k'g'} \neq a_{kg}$; in particular it does not involve $\Lambda_{kg}^{a_{kg}}$. Consequently $\rho'_k = \rho_k$ whenever these hazards are correctly specified in both arms, and $\rho'_k = \rho_k = 1$ whatever the hazards when all components of $\overline{a}$ are equal.

\paragraph{Consistency.}
Let $\p^{*}$ denote the limits of the fitted nuisance functions. Since $\psi(\widehat\p)$ lies in a Glivenko--Cantelli class with probability tending to one and $\|\psi(\widehat\p) - \psi(\p^{*})\|_{L_2} = o_p(1)$, we have $\pn \psi(\widehat\p) - \p\psi(\p^{*}) = (\pn - \p)\psi(\p^{*}) + (\pn - \p)\{\psi(\widehat\p) - \psi(\p^{*})\} + \p\{\psi(\widehat\p) - \psi(\p^{*})\} = o_p(1)$. Hence $\widetilde{F}_j^{\overline{a}}(t)$ converges to $\p\psi(\p^{*})$, whose bias is given by \eqref{eq:remainder} with $\p' = \p^{*}$. Under condition (1) of the theorem, $\mathrm{d}\{\Lambda^{*a_{kg}}_{kg} - \Lambda^{a_{kg}}_{kg}\} = 0$ for every $(k,g)$, so every term vanishes. The hazards of the other arm, which enter only through $\rho^{*}_k$, and the propensity score and censoring model may be arbitrary. Under condition (2), $r^{*}_{kg} \equiv 1$. For a transition whose fitted hazard is correct, the integrator vanishes. For a misspecified transition $(k,g)$, the condition on the transitions in $\mathcal{U}_k$ borrowed from the other arm gives $\rho^{*}_k = \rho_k$ by the structure just described, so the coefficient vanishes. In either case $\p\psi(\p^{*}) = F_j^{\overline{a}}(t)$. A single misspecified transition hazard satisfies condition (2) under any $\overline{a}$, because every other hazard is correct. When all components of $\overline{a}$ are equal, $\rho^{*}_k = \rho_k = 1$ for every $k$, so condition (2) holds whatever the transition hazards. $\square$

\subsection{Proof of Theorem 3} \label{sec:proof_thm3}

Decompose
\begin{equation}\label{SemiEffDecom}
\sqrt{n} \{\widetilde{F}_j^{\overline{a}}(t) - F_j^{\overline{a}}(t)\} = \sqrt{n} (\pn-\p)\psi(\p) + \sqrt{n} (\pn-\p)\{\psi(\widehat\p)-\psi(\p)\} + \sqrt{n} \p\{\psi(\widehat\p)-\psi(\p)\}.
\end{equation}
The first term is asymptotically normal with mean zero and variance $\E[\EIF\{F_j^{\overline{a}}(t)\}]^2$, because $\psi(\p) - \p\psi(\p)$ is the efficient influence function of Lemma 1. The second is $o_{p}(1)$ because $\psi(\widehat\p)$ lies in a $\p$-Donsker class with probability tending to one (Condition S1(3)) and $\|\psi(\widehat\p)-\psi(\p)\|_{L_2} = o_p(1)$ by Condition S1(1), the fitted weights being bounded with probability tending to one \citep[Lemma 19.24]{van2000asymptotic}. The third term is identity \eqref{eq:remainder} with $\p' = \widehat\p$, which holds for any fitted nuisance functions. Its $(k,g)$-term is the integral of the coefficient $\widehat{e}_{kg} = \{1 - \widehat{r}_{kg}\, \widehat\rho_k / \rho_k\}\, \widehat\Delta_{kg}$ of Condition S1(2) against the error $\mathrm{d}\{\widehat\Lambda_{kg}^{a_{kg}} - \Lambda_{kg}^{a_{kg}}\}$, and it is $o_p(n^{-1/2})$ uniformly in $t \leq t^*$ by that condition. Hence $\sqrt{n}\, \p\{\psi(\widehat\p) - \psi(\p)\} = o_p(1)$, and by \eqref{SemiEffDecom}
\[
\sqrt{n}\{\widetilde{F}_j^{\overline{a}}(t) - F_j^{\overline{a}}(t)\} = \sqrt{n}(\pn - \p)\,\EIF\{F_j^{\overline{a}}(t)\} + o_p(1).
\]
Because $\EIF\{F_j^{\overline{a}}(t)\}$ is the efficient influence function in the nonparametric model $\mathcal{M}$ (Section E.2), $\widetilde{F}_j^{\overline{a}}(t)$ is regular in $\mathcal{M}$ and its asymptotic variance equals the efficiency bound in $\mathcal{M}$ \citep[Lemma 25.23]{van2000asymptotic}. The fitted Cox models enter only through $\widehat{\p}$, so the bound is that of $\mathcal{M}$ and not that of the Cox model, nor that of a submodel of the kind in Section \ref{sec:extmarkov}; these are submodels of $\mathcal{M}$ with bounds that are in general smaller. $\square$

\subsection{Proof of Theorem \ref{thmS1}} \label{sec:proof_thmS1}

Let $\psi_h(\p')(O) = F'(t \mid W) + \varphi'_h(O)$ denote the integrand of $\pn[\cdot]$ in \eqref{cond_h_est} evaluated at nuisance functions $\p'$. These consist of the propensity score, the censoring hazards, and the class-specific transition hazards, from which the occupancies and class averages are computed. Then $\widetilde{F}_{j,h}^{\overline{a}}(t) = \pn\psi_h(\widehat\p)$ and $\p\psi_h(\p) = F_j^{\overline{a}}(t)$. The decomposition \eqref{SemiEffDecom} applies with $\psi_h(\cdot)$ in place of $\psi(\cdot)$. The first term is asymptotically normal with variance $\E[\psi_h(\p)(O) - F_j^{\overline{a}}(t)]^2$, the variance stated in Theorem \ref{thmS1}, because $\psi_h(\p) - \p\psi_h(\p)$ is the efficient influence function of Lemma \ref{lem:extmarkov}. By Condition \ref{condh}(3) the convergence holds as a process on $[0, t^*]$. The second term is $o_p(1)$ by Conditions \ref{condh}(1) and (3), the fitted weights being bounded with probability tending to one by the class positivity \citep[Lemma 19.24]{van2000asymptotic}. It remains to show that the third term, $\sqrt{n}\, \p\{\psi_h(\widehat\p) - \psi_h(\p)\}$, is $o_p(1)$.

\paragraph{An exact expression for the remainder.}
Fix $W = w$, and let $\widehat{F}(t \mid w; \overline{x}(s-))$ and $\widehat\Delta_{kg}(s; w, \overline{x}(s-))$ be the counterfactual incidence given a history and its increment, computed from the fitted class-specific hazards. Along a trajectory of the true hypothetical process, let $\widehat{U}(s) = \widehat{F}(t \mid w; \overline{X}^{\overline{a}}(s))$. Then $\widehat{U}(0) = \widehat{F}(t \mid w)$ and $\widehat{U}(t) = \mathds{1}\{j \in \overline{X}^{\overline{a}}(t)\}$, whatever the hazards, and by the backward equation \eqref{eq:backward_diff} under the fitted hazards, exactly as in the proof of Theorem 2 (Section \ref{sec:proof_thm2}),
\[
\widehat{U}(t) - \widehat{U}(0) = \sum_{(k,g)\in\mathcal{E}} \int_0^t \widehat\Delta_{kg}\,\mathrm{d}M_{kg}^{\overline{a}}(s) + \sum_{(k,g)\in\mathcal{E}} \int_0^t \widehat\Delta_{kg}\, Y_k^{\overline{a}}(s)\,\mathrm{d}\left\{\Lambda_{kg}^{a_{kg}} - \widehat\Lambda_{kg}^{a_{kg}}\right\}(s \mid w, h(\overline{X}^{\overline{a}}(s-))).
\]
Taking expectations under $\pr^{\overline{a}}(\cdot \mid w)$ and grouping the histories by class,
\begin{equation}\label{eq:duhamel_h}
\widehat{F}(t \mid w) - F(t \mid w) = \sum_{(k,g)\in\mathcal{E}} \sum_{c} \int_0^t P_k^{\overline{a}}(s \mid w; c)\, \widetilde{D}_{kg}(s \mid w; c)\, \mathrm{d}\left\{\widehat\Lambda_{kg}^{a_{kg}} - \Lambda_{kg}^{a_{kg}}\right\}(s \mid w, c),
\end{equation}
where $\widetilde{D}_{kg}(s \mid w; c) = \E^{\overline{a}}\{\widehat\Delta_{kg}(s; w, \overline{X}^{\overline{a}}(s-)) \mid X^{\overline{a}}(s-) = k, \overline{X}^{\overline{a}}(s-) \in c, W = w\}$ averages the fitted increment over the true hypothetical law of the histories in class $c$. If the fitted cumulative hazards have jumps, as Breslow estimators do, \eqref{eq:duhamel_h} holds with product integrals in place of exponentials \citep[Section II.6]{andersen2012statistical}. For the debiasing term, the fitted weights are functions of $(s, W, c)$, and $\E\{\mathds{1}(A = a)\, Y_k(s)\, \mathds{1}\{\overline{X}(s-) \in c\}\, \mathrm{d}\widehat{M}_{kg}(s) \mid W\} = \pr(A = a \mid W)\, \pi_k^{a}(s \mid W; c)\, \mathrm{d}\{\Lambda_{kg}^{a} - \widehat\Lambda_{kg}^{a}\}(s \mid W, c)$. Hence, with $a = a_{kg}$ in each term,
\[
\p\widehat\varphi_h = -\E\sum_{(k,g)\in\mathcal{E}}\sum_{c}\int_0^t \widehat{P}_k^{\overline{a}}\, \widehat{D}_{kg}\, \widehat\rho_{kg}\, \mathrm{d}\left\{\widehat\Lambda_{kg}^{a} - \Lambda_{kg}^{a}\right\}, \qquad \widehat\rho_{kg}(s \mid W; c) = \frac{\pr(A = a \mid W)\, \pi_k^{a}(s \mid W; c)}{\widehat\pr(A = a \mid W)\, \widehat\pi_k^{a}(s \mid W; c)},
\]
all functions being evaluated at $(s \mid W; c)$. Adding \eqref{eq:duhamel_h}, integrated over $W$, gives
\begin{equation}\label{eq:remainder_h}
\p\{\psi_h(\widehat\p) - \psi_h(\p)\} = \E\sum_{(k,g)\in\mathcal{E}}\sum_{c}\int_0^t \left\{P_k^{\overline{a}}\widetilde{D}_{kg} - \widehat{P}_k^{\overline{a}}\widehat{D}_{kg}\widehat\rho_{kg}\right\}(s \mid W; c)\, \mathrm{d}\left\{\widehat\Lambda_{kg}^{a_{kg}} - \Lambda_{kg}^{a_{kg}}\right\}(s \mid W, c).
\end{equation}
The coefficient in braces vanishes when the nuisance functions are correct, since then $\widehat{P}_k^{\overline{a}} = P_k^{\overline{a}}$, $\widehat\rho_{kg} = 1$, and $\widehat{D}_{kg} = \widetilde{D}_{kg} = D_{kg}^{\overline{a}}$. Otherwise it is a bounded factor times the errors of the fitted propensity score, censoring survival, occupancies, and class averages. Each term of \eqref{eq:remainder_h} is therefore the integral of a nuisance error against the error $\mathrm{d}\{\widehat\Lambda_{kg}^{a_{kg}} - \Lambda_{kg}^{a_{kg}}\}$, and it is $o_p(n^{-1/2})$ uniformly in $t \leq t^*$ by Condition \ref{condh}(2). Hence $\sqrt{n}\, \p\{\psi_h(\widehat\p) - \psi_h(\p)\} = o_p(1)$ and
\[
\sqrt{n}\{\widetilde{F}_{j,h}^{\overline{a}}(t) - F_j^{\overline{a}}(t)\} = \sqrt{n}(\pn - \p)\{\psi_h(\p)(O) - F_j^{\overline{a}}(t)\} + o_p(1)
\]
uniformly in $t \leq t^*$. Because $\psi_h(\p) - F_j^{\overline{a}}(t)$ is the efficient influence function in the model $\mathcal{M}_h$ of Assumption \ref{ass:extmarkov} (Section \ref{sec:proof_lemmaS1}), $\widetilde{F}_{j,h}^{\overline{a}}(t)$ is regular in $\mathcal{M}_h$, and its asymptotic variance equals the efficiency bound in $\mathcal{M}_h$ \citep[Lemma 25.23]{van2000asymptotic}. This variance is no larger than $\E[\psi(\p)(O) - F_j^{\overline{a}}(t)]^2$, because $\psi_h(\p) - F_j^{\overline{a}}(t)$ is the projection of $\psi(\p) - F_j^{\overline{a}}(t)$ onto the tangent space of $\mathcal{M}_h$ (Section \ref{sec:proof_lemmaS1}). Unlike the identity \eqref{eq:remainder} for the history-specific weight, the coefficient in \eqref{eq:remainder_h} involves the fitted hazard of $(k,g)$ itself through $\widehat{P}_k^{\overline{a}}$, $\widehat\pi_k^{a}$, and $\widehat{D}_{kg}$, because the class occupancies integrate the survival factor of $(k,g)$ over the entry times into $k$. The one-step correction therefore removes the first-order bias of the plug-in estimator but does not yield the exact consistency of Theorem 2. $\square$

\subsection{Proof of Lemma \ref{lem:frailty}} \label{sec:proof_frailty}

We use the notation of Sections \ref{sec:density_est} and \ref{sec:proof_lemma1}. By Section \ref{sec:fr_ident}, the observed law is parameterized, as in Step 1 of the proof of Lemma 1, by $p(w)$, $\pr(a \mid w)$, the observed-history event hazards $\lambda^{a}_{kg}$, and the censoring hazards, without restriction. The model is locally nonparametric, with scores \eqref{eq:score}, so an influence function is unique and efficient. The target \eqref{eq:fr_estimand} depends on the observed law through $p(w)$ and, via \eqref{eq:fr_demix}, through the event hazards of both arms; the propensity and censoring directions do not contribute. The direction $b_W$ gives the term $F_j^{\overline{a}}(t \mid W; \nu) - F_j^{\overline{a}}(t; \nu)$ as in \eqref{eq:product_rule}. Fix $W = w$ and consider the direction $b^{a}_{kg}(s; \overline{x}(s-)) := b_{kg}(s; a, w, \overline{x}(s-))$, which perturbs the observed-history hazards to $\lambda^{a}_{kg, v} = (1 + v b^{a}_{kg}) \lambda^{a}_{kg}$.

\paragraph{Step 1: perturbation of the conditional hazards.}
Along a history $\overline{x}$ in arm $a$, \eqref{eq:fr_demix} gives $\lambda^{a}_{kg, v}(s; \nu) = \lambda^{a}_{kg, v}(s) / \mu^{a}_{v}(s)$, where $\mu^{a}_{v}(s) = m_\nu\{N(s-), \Lambda^{a}_{\bullet, v}(s; \nu)\}$. Differentiating at $v = 0$,
\begin{align*}
    &\beta^{a}_{kg}(s) := \frac{\partial}{\partial v} \log \lambda^{a}_{kg, v}(s; \nu)\Big|_{v = 0} = b^{a}_{kg}(s) - \kappa^{a}(s)\, D^{a}(s), \\
    &D^{a}(s) := \frac{\partial}{\partial v} \Lambda^{a}_{\bullet, v}(s; \nu)\Big|_{v = 0} = \int_0^s \sum_{g} \lambda^{a}_{x(u-), g}(u; \nu)\, \beta^{a}_{x(u-), g}(u)\, \mathrm{d}u,
\end{align*}
so that $D^{a}$ solves the linear equation $\mathrm{d}D^{a}(s)/\mathrm{d}s = \sum_{g} \lambda^{a}_{x(s-), g}(s; \nu)\, b^{a}_{x(s-), g}(s) - \kappa^{a}(s) \lambda^{a}_{\bullet}(s; \nu) D^{a}(s)$, with $\lambda^{a}_{\bullet}(s; \nu) = \sum_{g} \lambda^{a}_{x(s-), g}(s; \nu)$ and $D^{a}(0) = 0$. Its solution is
\begin{equation}\label{eq:fr_D}
    D^{a}(s) = \int_0^s \Pi^{a}(u, s)\, \sum_{g} \lambda^{a}_{x(u-), g}(u; \nu)\, b^{a}_{x(u-), g}(u)\, \mathrm{d}u,
\end{equation}
with $\Pi^{a}$ the propagator \eqref{eq:fr_prop}. Within a sojourn, $\kappa^{a}(s) \lambda^{a}_{\bullet}(s; \nu) = \mathrm{d} \log m_\nu\{n, \Lambda^{a}_{\bullet}(s)\} / \mathrm{d}s$, which gives the closed form of $\Pi^{a}$ stated after \eqref{eq:fr_prop}. Since $m_\nu$ is decreasing in $L$, $\Pi^{a} \geq 1$, and $\Pi^{a}$ is bounded on $[0, t^*]$ because the conditional cumulative hazards are finite there.

\paragraph{Step 2: derivative of the target given $\xi$.}
For fixed $\xi$, the hypothetical process has hazards $\xi \lambda^{a_{kg}}_{kg}(\cdot; \nu)$, which under the submodel are perturbed in the directions $\beta^{a_{kg}}_{kg}$. Steps 3--5 of the proof of Lemma 1, applied to this process, give
\begin{equation}\label{eq:fr_deriv_xi}
\begin{aligned}
    \frac{\partial}{\partial v} F_j^{\overline{a}}(t \mid w, \xi; \nu)\Big|_{v=0} &= \sum_{(k,g) \in \mathcal{E}} \E^{\overline{a}}_{\xi}\Bigg[ \int_0^t \beta^{a_{kg}}_{kg}(s; \overline{X}(s-))\, \Delta_{kg}(t \mid w, \xi; \overline{X}(s-)) \\
    &\qquad\qquad \times Y_k^{\overline{a}}(s)\, \xi \lambda^{a_{kg}}_{kg}(s \mid w, \overline{X}(s-); \nu)\, \mathrm{d}s \Biggm| w, \xi \Bigg],
\end{aligned}
\end{equation}
where $\beta^{a_{kg}}_{kg}(s; \overline{X}(s-)) = b^{a_{kg}}_{kg} - \kappa^{a_{kg}} D^{a_{kg}}$, with $D^{a_{kg}}$ given by \eqref{eq:fr_D}, is evaluated along the hypothetical history. In particular, $D^{a_{kg}}$ accumulates the arm-$a_{kg}$ conditional hazards along a history that the regime may generate with other arms' hazards. This is why $\Lambda^{a}_{\bullet}$ and $\Pi^{a}$ are defined for every history and arm.

\paragraph{Step 3: collecting the coefficient of $b^{a}_{k'g'}(u)$.}
Substitute $\beta = b - \kappa D$ and \eqref{eq:fr_D} into \eqref{eq:fr_deriv_xi} and exchange the order of integration. The coefficient of $b^{a}_{k'g'}(u; \overline{X}(u-))$ consists of a direct term, present when $a = a_{k'g'}$, and an indirect term collecting all transitions $(k, g)$ with $a_{kg} = a$ and all $s > u$. Conditioning on $\overline{X}(u-) = \overline{x}$, the indirect term is $-\Gamma^{a}(u \mid w, \xi; \overline{x})$ of \eqref{eq:fr_gamma}. Writing the expectation over the history at $u-$ as an integral against $p^{\overline{a}}_\xi(u, \overline{x} \mid w)$,
\[
\begin{aligned}
    \frac{\partial}{\partial v} F_j^{\overline{a}}(t \mid w, \xi; \nu)\Big|_{v=0} &= \sum_{a} \sum_{(k',g')} \int_0^t \int_{x(u-) = k'} b^{a}_{k'g'}(u; \overline{x})\, \lambda^{a}_{k'g'}(u \mid w, \overline{x}; \nu) \\
    &\qquad \times \left\{ \mathds{1}(a = a_{k'g'})\, \xi\, \Delta_{k'g'}(t \mid w, \xi; \overline{x}) - \Gamma^{a}(u \mid w, \xi; \overline{x}) \right\} p^{\overline{a}}_\xi(u, \overline{x} \mid w)\, \mathrm{d}\overline{x}\, \mathrm{d}u.
\end{aligned}
\]

\paragraph{Step 4: integration over the frailty and change of measure.}
Integrating against $\phi_\nu(\xi)\, \mathrm{d}\xi$ gives $\partial_v F_j^{\overline{a}}(t \mid w; \nu)$. Given $A = a$ and $W = w$, the density of the event $\{\overline{X}(u-) \in \mathrm{d}\overline{x}, T_{\text{c}} \geq u\}$ is $p^{a}(u, \overline{x} \mid w)\, G^{a}(u \mid w, \overline{x})$ with $p^{a} = \int p^{a}_\xi \phi_\nu\, \mathrm{d}\xi$, and by \eqref{eq:fr_demix} the observed-history hazard is $\lambda^{a}_{k'g'}(u \mid w, \overline{x}) = \mu^{a}(u \mid w, \overline{x}) \lambda^{a}_{k'g'}(u \mid w, \overline{x}; \nu)$ with $\mu^{a} p^{a} = \int \xi p^{a}_\xi \phi_\nu\, \mathrm{d}\xi$. Multiplying and dividing the integrand by $p^{a} G^{a} \lambda^{a}_{k'g'}(u \mid w, \overline{x})$, as in Step 6 of the proof of Lemma 1,
\[
\begin{aligned}
    \frac{\partial}{\partial v} F_j^{\overline{a}}(t \mid w; \nu)\Big|_{v=0} &= \sum_{a} \sum_{(k',g')} \E\Bigg[ \frac{\mathds{1}(A = a)}{\pr(A = a \mid W)} \int_0^t b^{a}_{k'g'}(u; \overline{X}(u-))\, H^{a}_{k'g'}(u \mid W; \overline{X}(u-)) \\
    &\qquad\qquad \times Y_{k'}(u)\, \mathrm{d}\Lambda^{a}_{k'g'}(u \mid W, \overline{X}(u-)) \Biggm| W = w \Bigg],
\end{aligned}
\]
with $H^{a}_{k'g'}$ as in \eqref{eq:fr_H}; the factor $\lambda^{a}_{k'g'}(\cdot; \nu)$ cancels between the numerator and the observed-history hazard, which is why $\Gamma^{a}$ does not depend on $(k', g')$. The weights are bounded by Assumption 5 applied conditionally on $\xi$ and the boundedness of $\Pi^{a}$ and $\kappa^{a}$.

\paragraph{Step 5: verification.}
Step 7 of the proof of Lemma 1 applies with $H^{a}_{kg}(s \mid W; \overline{X}(s-))$ in place of $\omega_k^{\overline{a}} \Delta_{kg}^{\overline{a}}$ and with both arms contributing. By the martingale covariance,
\[
    \begin{aligned}
        & \E\left\{\frac{\mathds{1}(A = a)}{\pr(A = a \mid W)} \int_0^t H^{a}_{kg}\, \mathrm{d}M_{kg}(\cdot; a, W, \overline{X}) \int_0^{t^*} b_{kg}\, \mathrm{d}M_{kg}(\cdot; A, W, \overline{X})\right\} \\
        = & \E\left\{\frac{\mathds{1}(A = a)}{\pr(A = a \mid W)} \int_0^t H^{a}_{kg}\, b^{a}_{kg}\, Y_k\, \mathrm{d}\Lambda^{a}_{kg}\right\},
    \end{aligned}
\]
so $\E\{\psi_\nu(O) S(O)\}$ equals $\partial_v F_j^{\overline{a}}(t; \nu)$ for every score \eqref{eq:score}. Hence $\psi_\nu$ is the efficient influence function. $\square$


\end{document}